\documentclass[12pt]{article}
\usepackage{amsmath,amssymb}

\newcommand{\nc}{\newcommand}
\nc{\op}{\operatorname}
\def\basispi#1#2{{\partial\over\partial \pi^{#1}_{#2}}}
\def\basisx#1{{\partial\over\partial x^{#1}}}
\def\basisv#1{{\partial\over\partial v^{#1}}}

\def\ginverse#1#2{{(g^{-1})^{#1}_{#2}}}
\def\vertical#1#2{{E^{*}{}^{#1}_{#2}}}

\def\r{\mathbb R}
\def\rne{{\mathbb R}^{n}}
\def\rk{{\mathbb R}^{k}}

\def\rndual{{\mathbb R}^{n*}}

\def\canonicalj{{\cal J}}
\def\car{Carath\'eodory}
\def\blob{\quad\vrule height6pt width3pt}

\def\jh#1{{J^{(h)}_{#1}}}
\def\hdown#1#2{{h_{#1#2}}}
\def\hhatdown#1#2{{\hat h_{#1#2}}}

\def\rn{{\r^n}}
\nc\jetpi{\ensuremath{J^1\pi}}
\def\cojetpi{\ensuremath{J^1\pi^{*}}}
\def\lpie{\ensuremath{L_\pi E}}

\nc{\Gln}[1][n]{\ensuremath{\op{GL}({#1})}}
\nc{\Gv}{\ensuremath{\op{G}_{\op{v}}}}

\nc{\LE}{\ensuremath{\op{LE}}}

 \def\hook{\, \hbox to 15pt{\vbox{\vskip 6pt\hrule width 8pt height 1pt}
         \kern -5pt\vrule height 8pt width 1pt\hfil}}
\nc{\inv}{\ensuremath{^{-1}}}

\nc{\fr}{\frac}
\nc{\frb}[2]{\left.\frac{#1}{#2}\right|}
\nc{\ra}{\rightarrow}
\nc{\pa}{\partial}
\nc{\of}{\circ}
\nc{\iso}{\cong}

\def\xhatf{{X_{\hat f}}}

\def\xhatg{{X_{\hat g}}}
\def\vertical#1#2{{E^{*#1}_{#2}}}

\nc\bld{\bfseries}

\def\lag{\ensuremath{\operatorname{L}}}
\def\lagrangian{{\ensuremath{\cal L}}}

\def\car{Carath\'{e}odory}
\def\tn{\tau(n)}
\nc\jet{\ensuremath{J^1\pi}}
\def\cojet{\ensuremath{J^{1*}\pi}}

\def\basisz#1{\frac{\partial}{\partial z^{#1}}}
\def\multialpha#1{\alpha_1 \alpha_2\dots \alpha_{#1}}
\def\example{{\noindent{\bf Example:}\ }}
\def\vectorclass#1{{\lbrack\!\lbrack   {\hat X}_{\hat {#1}}\rbrack\!\rbrack}}
\def\vectorclassindicesinside #1#2#3{{\lbrack\!\lbrack {X}_{\hat
#1}^{#2}{#3}\rbrack\!\rbrack}}
\def\rhat#1{{{\hat r}_{#1}}}
\def\pb#1#2{\{\hat #1,\hat #2\}}

\def\npbblank{\{\ ,\ \}}

\def\pbnorighthat#1#2{\{\hat #1 , #2 \}}
\def\multibeta#1{\beta_1 \beta_2\dots \beta_{#1}}

\def\multiRbasisalpha#1{{\hat r}_{\alpha_1\alpha_2\dots \alpha_{ {#1}}} }

\def\multiRbasisbeta#1{{\hat r}_{\beta_1\beta_2\dots \beta_{ {#1}}} }

\def\popo#1#2{\frac{\partial #1}{\partial #2}}

\newtheorem{thm}{Theorem}[section]
\newtheorem{lem}[thm]{Lemma}
\newtheorem{prop}[thm]{Proposition}
\newtheorem{cor}[thm]{Corollary}
\newtheorem{definition}[thm]{Definition}

\def\rem{\bigskip\noindent{\bf Remark}$\;\;$}
\def\proof{\noindent{\bf Proof}$\;\;$}
\def\blob{\quad\rule{8pt}{8pt}}

\nc{\ba}{\begin{array}}
\nc{\ea}{\end{array}}
\def\basispi#1#2{{\frac{\partial}{\partial \pi^{#1}_{#2}}}}
\def\basisv#1#2{{\frac{\partial}{\partial v^{#1}_{#2}}}}
\def\basisu#1#2{{\frac{\partial}{\partial u^{#1}_{#2}}}}
\def\basisx#1{{\frac{\partial}{\partial x^{#1}}}}
\def\basisy#1{{\frac{\partial}{\partial y^{#1}}}}
\def\jet{j^1_x\phi}
\def\mat#1#2#3#4{\left( \begin{array}{cc} #1 & #2 \\
\noalign{\medskip} #3 & #4 \end{array} \right)}

\def\xhatf{{X_{\hat f}}}

\def\xhatg{{X_{\hat g}}}
\def\vertical#1#2{{E^{*#1}_{#2}}}

\def\lag{\ensuremath{\operatorname{L}}}
\def\lagrangian{{\ensuremath{\cal L}}}

\def\car{Carath\'{e}odory}

\def\qsubl{\ensuremath{Q_{\lag}}} 

\title{GEOMETRIC STRUCTURES ON FIELD THEORY}

\begin{document}

\normalsize
 \begin{center}
{\Large\bf GEOMETRIC STRUCTURES IN FIELD THEORY}
\end{center}
 \bigskip

\begin{center}
{\bf Manuel de Le\'on}

{\small IMFF (C.S.I.C.), Serrano 123, 28006, Madrid, Spain}

\underline{E-Mail: mdeleon@fresno.csic.es}
\end{center}

\begin{center}
{\bf Michael McLean and Larry K. Norris}

{\small Department of Mathematics, North Carolina State Univerisity  }

{\small  Box 8205, Raleigh, North Carolina, USA}

\underline{E-Mail: mamclean@unity.ncsu.edu, lkn@math.ncsu.edu }
\end{center}

\begin{center}
{\bf Angel Rey Roca and Modesto Salgado}

{\small Departamento de Xeometr\'{\i}a e Topolox\'{\i}a, University of Santiago de Compostela}
 
{\small Facultad de Matematicas, 15706 Santiago de Compostela, Spain }

\underline{E-Mail: modesto@zmat.usc.es}
\end{center}

\thispagestyle{empty}
\bigskip
\pagenumbering{roman}
\begin{abstract}
This review paper is concerned with the generalizations to field theory of
the tangent and cotangent structures and bundles that play fundamental roles
in the Lagrangian and Hamiltonian formulations of classical mechanics.  
The paper reviews, compares and constrasts the various generalizations in order to bring
some unity to the field of study.   The generalizations seem to fall into two categories.
In one direction  some have   generalized the geometric structures of the  bundles,
arriving at the various axiomatic systems such as $k$-symplectic and  $k$-tangent
structures. The other direction was to fundamentally extend the bundles themselves
 and to then explore the natural geometry of the extensions.  This latter direction gives 
us the multisymplectic geometry on  jet and cojet  bundles and
$n$-symplectic geometry on    frame bundles.

\end{abstract}

\newpage
\tableofcontents

\newpage

\section{Introduction}
\pagenumbering{arabic}

 This review paper is inspired  by the
geometric formulations of the Lagrangian and Hamiltonian descriptions of
classical mechanics.  The mathematical arenas of these well-known 
  formulations are respectively, the tangent and cotangent bundles of the
configuration space.  Over the years many have sought to study classical
field theory in  analogous ways, using various generalizations or
extensions of the tangent and cotangent bundles and/or their structures. No one has yet achieved a
perfect formalism, but there are   beautiful and useful results in
many arenas.

The generalizations seem to fall into two categories. In one direction,
some have   generalized the geometric structures of these bundles,
sometimes arriving at a formalism pertinent to field theory.  Another
direction was to fundamentally extend the bundles themselves and then
explore the natural geometry of the extensions.  The former gives us the
various axiomatic systems such as $k$-symplectic and  $k$-tangent structures. 
The latter gives us the multisymplectic geometry on  jet and cojet 
bundles and
$n$-symplectic geometry on    frame bundles.

\subsection{Cotangent-like structures}

The first step in this direction of generalization was
the development of symplectic geometry \cite{GS}.
Later, around 1960, Bruckheimer \cite{b} introduced
the notion of almost cotangent
structures.  These were futher investigated
by Clark and Goel \cite{cgc} in 1974.
In both cases the canonical $2$-form
became the model from which axioms were designed.

Between 1987 and 1991, several independent and closely related
generalizations were developed.  Polysymplectic geometry \cite{gun},
almost $k$-cotangent structures \cite{mc1,mc2}, and $k$-symplectic
geometry \cite{aw1,aw2} were based around the natural structure of the
$k$-cotangent bundle. This bundle, which can be thought of as the
fiberwise product of the cotangent bundle $k$ times, has a $k$-tuple of
$1$-forms  with which one works.  Also, the development of
the $n$-symplectic geometry of the frame bundle and its $\r^n$-valued
soldering $1$-form $\theta$ began during this time period \cite{No1,No2, No3, No4}. While
the development of $k$-tangent structures and $k$-symplectic geometry had purely geometric
motivations, polysymplectic geometry was created to study field theory and
$m$-symplectic geometry sought to generalize Hamiltonian mechanics.

\subsection{Tangent-like structures}

Around 1960, the theory of almost tangent structures was developed by
Clark and Bruckheimer \cite{cb} and Eliopoulos \cite{e} separately.
Almost tangent structures are generalizations of the tangent bundle. The
canonical vector valued one-form $J$, viewed as the object of central
interest, was axiomatized.

Almost $k$-tangent structures \cite{mt1,mt2} arose around 1988 as a
generalization of the geometry of the $k$-tangent bundle. This bundle is,
among other interpretations, the fiberwise product of the tangent bundle
with itself $k$ times. A section of this bundle is equivalent to a
$k$-tuple of vector fields. The central geometric object becomes a
$k$-tuple of $J's$.

Another version of the tangent structure arises on the jet bundle
(see \cite{Saunders}). This is a
very broad level of generalization since the idea of the jet of a section
generalizes  and incorporates  the notions of tangent vectors, cotangent
vectors, $k$-tangent vectors, and $k$-cotangent vectors. Such geometry has clear
importance to field theory since  one can envision any type of field as a
section of a fiber bundle.  

We present here also a new tangent-like structure, namely a canonically defined set of
tensor fields $J^i$, $i=1...n$ on the bundle of frames $LM$ of a manifold $M$.  This
tangent-like structure will be shown to induce the tangent structure on $TM$.

\subsection{Interconnections, and plan of the paper}

 In this    review paper our goal is to identify and clarify important connections between
the various structures mentioned above.  We also will consider
relationships between  some of the formalisms built on top of these
structures.

The $k$-cotangent, $k$-symplectic, and polysymplectic structures are nested
generalizations with $k$-cotangent being the most specific.  The $n$-symplectic
geometry of the frame bundle is also an example of a polysymplectic structure.
Later in the paper, we will draw some interesting connections between the frame
bundle and the $k$-cotangent bundle.

The frame bundle is an interesting case since in addition to having
a cotangent-like structure, it also has a tangent-like structure.
Exploiting the natural correlation of frames and co-frames, we can define
an $n$-tuple of $J$s in addition to the $m$-tuple of $\theta$s mentioned earlier.
These objects acquire additional properties and relationships on the frame
bundle.

The vector valued one-form $S_\alpha$ on the jet bundle is  later shown
to be directly related to the other tangent structures in the special
cases where they are comparable. Additionally, using new results
regarding the adapted frame bundle  we show a similar relationship
between the $k$-tangent structure there  and the $S_\alpha$ on the jet
bundle.

Venturing into the realm of multi-symplectic geometry, we show how the canonical
multi-symplectic form on the cojet bundle is tied to the canonical $k$-symplectic
structures we discuss. Moreover we show how the Cartan-Hamilton-Poincar\'e $n$-form on
$\jetpi$ is induced from the $m$-symplectic structure on $\lpie$.

What we strive to do in this paper is to unify perspectives.  We show
similarities and differences among the approaches  and draw strong
correlations.  Since no one geometry has emerged as dominant, it is important
that everyone   be aware of the options.  We hope this work may
serve as a guidebook and  translation table for those desiring to explore other
formalisms.

All the manifolds are supposed to be smooth. The differential of a mapping 
$F:M \longrightarrow N$ at a point $x \in M$ will be   denoted by
$F_{*}(x)$ or $TF(x)$. The induced tangent mapping will be denoted as
$TF : TM \longrightarrow TN$.

The names of the various theories are different, yet two names are so similar that
we feel it necessary to introduce the following convention that will be
followed throughout the paper.

 \begin{itemize}
\item We use the term {\em $k$-symplectic geometry} to refer to the works of
Awane  and the works of de Le\'on, Salgado, et. al. 

\item We use the terms {\em $n$-symplectic geometry}  and/or {\em
$m$-symplectic geometry} to refer to the works of Norris et. al.   
\end {itemize}

\section{Spaces with tangent-like structures}

 In this section we first recall the definitions and main properties of
almost tangent and almost $k$-tangent structures. We describe the
canonical $n$-tangent structure of the frame bundle $LM$ of an
$n$-dimensional manifold $M$ in terms of the soldering form.

Secondly we recall Saunders's construction of the vector valued
$1$-form $S_\alpha$. This $1$-form    is  a generalization   to field theories defined on jet
bundles of  fibered manifolds, of the almost tangent
structure.

\subsection{Almost tangent structures and $TM$}

 An almost tangent structure $J$ on a $2n$-dimensional
manifold $M$ is   tensor field of type $(1,1)$ of constant rank $n$ such that  $J^2=0$.  The
manifold $M$ is then called an almost tangent manifold. Almost tangent
structures were introduced by Clark and Bruckheimer \cite{cb} and
Eliopoulos \cite{e} around $1960$ and have been studied by numerous authors
(see \cite{bc,cgt,ct,cram1,cram2,grif1,grif2,klein,ts}).

The canonical model of these structures is the tangent bundle $\tau_M: TM
\to M$ of an arbitrary manifold  $M$. Recall that for a vector $X_x$ at a
 point $x\in M$ its
vertical lift is the vector on $TM$ given by
\[X_x^V(v_x )= \displaystyle \frac{d}{dt}(v_x +tX_x)_{\vert_{t=0}}
  \in T_{v_x }(TM)\]
for all points $v_x \in TM$.

 The canonical tangent structure $J$  on $TM$ is defined by
$$J_{v_x }(Z_{v_x })=((\tau_M)_*(v_x )Z_{v_x })^V_{v_x}$$
for all vectors $Z_{v_x }\in T_{v_x }(TM)$, and it is locally given by
\begin{equation}\label{localj}
J=\frac{\displaystyle\partial}{\displaystyle\partial v^i}
  \otimes dx^i  
\end{equation}
with respect the bundle coordinates on $TM$.
 This tensor $J$ can be
regarded as the vertical lift of the identity tensor on $M$ to $TM$
\cite{mor}.

The integrability of these structures, which means the existence of local
coordinates such that the tensor field $J$ is locally given like as in
(\ref{localj}), is characterized as follows.
\begin{prop}
An almost tangent structure $J$ on $M$ is integrable
if and only if the Nijenhuis tensor $N_J$
of $J$ vanishes.  \blob
\end{prop}

Crampin and Thompson \cite{ct} proved  that an integrable almost  tangent
manifold $M$  satisfying  some natural global hypotheses is essentially the
tangent bundle of some differentiable manifold.

\subsection{Almost $k$-tangent structures and $T^1_kM$}

The almost $k$-tangent structures were introduced    as generalization of
the almost tangent structures   \cite{mt1,mt2}.
\begin{definition}
 An {\rm almost $k$-tangent structure} $J$ on a manifold $M$ of dimension $n+kn$ is
a  family $(J^1,\dots,J^k)$   of tensor fields   of type  $(1,1)$ such that
\begin{equation}\label{ec6160}
 J^A\circ J^B=J^B\circ J^A= 0,\qquad
  rank \,  J^A=n  , \qquad
  Im \, J^A\cap (\oplus_{B\neq A} Im \, J^B)=0,
\end{equation}

\noindent for $1\leq A,B\leq k$. In this case
the manifold $M$ is then called   an
{\rm almost $k$-tangent manifold}.
\end{definition}

The canonical model of these structures is the $k$-tangent vector bundle
$T^1_kM =J^1_0({\bf R}^k,M)$  of an arbitrary manifold $M$, that is  the
vector bundle with total space the manifold of $1$-jets of maps with
source at $0\in {\bf R}^k$ and with projection map $\tau
(j^1_0\sigma)=\sigma (0)$. 
 This  bundle is  also known as the {\em tangent bundle of $k^1$-velocities} of $M$
\cite{mor}.

 The manifold $T^{1}_{k}M$ can be canonically identified with the
Whitney sum of $k$ copies of $TM$, that is
\[
\begin{array}{ccc}
T^{1}_{k}M& \equiv & TM \oplus \dots \oplus TM, \\
j^1_0\sigma & \equiv & (j^1_0\sigma_1=v_1, \dots
,j^1_0\sigma_k=v_k)
\end{array}
\]
where $\sigma_A=\sigma(0,\dots,t,\dots,0)$ with $t\in {\bf R}$
at position $A$ and $v_A=
(\sigma_A)_*(0)(\frb{d}{dt}_0)$.

 If $(x^i)$ are local coordinates on $U \subseteq M$ then the induced local
coordinates $(x^i , v_A^i),\, 1\leq i \leq n,\, 1\leq A \leq k$,   on
$\tau^{-1}(U)\equiv T^1_kU$ are given by
$$ x^i(j^1_0\sigma)=x^i(\sigma (0)),\qquad
  v_A^i(j^1_0\sigma)=\displaystyle\frac{d}{dt}(x^i\circ
  \sigma_A)\vert_{{t=0}}=v_A(x^i) \, .$$

\begin{definition}
 For a vector  $X_x$ at $M$ we define its {\rm vertical $A$-lift}  \,
$(X_x)^A$ as the vector   on $T_k^1M$ given  by
  $$(X_x)^A(j^1_0\sigma ) = \frac{d}{dt} (
(v_1)_x,\dots,(v_{A-1})_x,(v_A)_x+tX_x,(v_{A+1})_x \dots,(v_k)_x
){\vert_{t=0}} \in T_{j^1_0\sigma}(T^1_kM)$$

\noindent for all points  $j^1_0\sigma \equiv ((v_1)_x,\dots,(v_k)_x)) \in
T^1_kM$.
\end{definition}
In local coordinates we have
\begin{equation}
(X_x)^A = \displaystyle\sum_{i=1}^n a^i \frac{\partial}{\partial v^i_A}
\end{equation}
for a vector $X_x = a^i \, \partial / \partial x^i.$

The {\em canonical vertical vector fields} on $T^1_kM$ are defined by
\begin{equation}\label{gen euler}
 C^A_B(x,X_1,X_2,\ldots,X_k) = (X_B)^A
\end{equation}
and are locally given by $C^A_B=v^i_B \basisv{i}{A}$.
 The {\em canonical $k$-tangent structure} $(J^1,\dots,J^k)$ on $T^1_kM$
is defined by
$$J^A(Z_{j^1_0\sigma})=
  (\tau_*
  (Z_{j^1_0\sigma}))^A$$
for all vectors $Z_{j^1_0\sigma}\in T_{j^1_0\sigma}(T^1_kM)$.
In local  coordinates we have
\begin{equation}\label{localJA}
J^A=\frac {\displaystyle\partial}{\displaystyle\partial v^i_A} \otimes
dx^i
\end{equation}

The tensors $J^A$ can be regarded as the  $(0,\dots,1_A,\dots,0)$-lift of the
identity tensor on $M$ to $T^1_kM$ defined in \cite{mor}.

We remark that an almost $1$-tangent structure is an almost
tangent structure.

In \cite{mt1,mt2} the almost $k$--tangent structures are  described as
$G$-structures, and the integrability of these structures, which is defined
as the existence of local coordinates  such that the tensor fields $J^A$ are
locally given as in (\ref{localJA}), is characterized by following proposition.

\begin{prop}
An almost $k$-tangent structure $(J^1,\dots,J^k)$ on $M$ is integrable
if and only if $\{J^A   ,J^B\}=0$ for all $1\leq A,B \leq k$,  where
$$\{J^A   ,J^B\}(X,Y)=[J^AX,J^BY]+J^AJ^B[X,Y]-J^A[X,J^BY]-J^B[J^AX,Y]\,
,$$ for any vector fields $X$, $Y$ on $M$.  \blob
\end{prop}

 In \cite{mt1,mt2} it is proved (in a way analogous to \cite{ct}) that an
integrable almost tangent manifold $M$  satisfying  some natural global
hypotheses is essentially the $k$-tangent bundle of some differentiable
manifold.

\subsection{The canonical $n$-tangent structure of $LM$}

We shall show that $LM$ has an intrinsic $n$-tangent structure
described in terms of the soldering form and fundamental
vertical vectors fields.

   Let $M$ be a $n$-dimensional
manifold and  $\lambda_M:LM\rightarrow M$
the principal fiber bundle of linear frames of $M$. A
point $u$ of $LM$ will be
denoted by the pair $(x,e_i)$ where $x\in M$ and
$(e_1,e_2,\dots,e_n)_x$ denotes
a linear frame at $x$. The projection map
$\lambda_M:LM\rightarrow M$  is defined by $\lambda_M(x,e_i)=x$.

 If $(U,x^i)$ is a chart on $M$
then we can introduce two different coordinates on $\lambda_M^{-1} (U)$.
First consider the {\em coframe} or {\em $n$-symplectic momentum}
coordinates $(x^i,\pi^i_j)$
on $\lambda_M^{-1} (U)$ defined by

\begin{equation}\label{momcoord}
x^i(u)=x^i(x) \ ,\qquad \pi^i_j(u)=e^i(\basisx j)  \, , 
\end{equation} 
where $(e^1,\dots,e^n)_x$ is the dual frame to $u=(e_1,\dots,e_n)_x$.

Secondly consider the {\em frame} or
{\em $n 
$-symplectic velocity} coordinates
$(x^i,v^i_j)$ on $\lambda_M^{-1} (U)$ defined by
\begin{equation}\label{velocoord}
 x^i(u)=x^i(x) \ ,\qquad v^i_j(u)=e_j(x^i) \, , 
\end{equation} 
  The relationship between the two coordinates systems on $LM$ is given by
 \begin{equation} \label{n-velocity coordinates}
   v^i_j(u) \pi^j_k(u)=\delta^i_k\ \ ,\qquad
   v^i_j(u) \pi^k_i(u)=\delta^k_j \, ,
   \end{equation}
for all $u$ in the domain of the $\pi^i_j$ momentum coordinates.

Denoting the standard basis of $gl(n,\mathbb{R})$ by $\{ E^i_j \}$, the
corresponding fundamental vertical vector fields $\vertical ij$
on $LM$ are given in momentum coordinates by
\begin{equation} \label{estar}
 \vertical ij = -\pi^i_k \basispi jk \, .
 \end{equation}
The bundle of linear frames $LM$ is an open and dense submanifold of the $n$-tangent bundle $T^
1_nM$, where $n=\dim M$. The general linear group $GL(n,\mathbb{R})$ acts naturally on both $LM$
and
$T^1_nM$. However, since each point in $LM$ is a linear frame, the action of $Gl(n,\mathbb{R})$ is
free on
$LM$ but not on $T^1_nM$. This reflects the fact that $LM$ has more intrinsic structure  than
$T^1_nM$.

On $LM$ we have an $\rne$-valued one-form,
the {\it soldering one-form}
$\hat{\theta} = \theta^i \, \hat{r}_i $. Here
$\hat{r}_i$ denotes the standard basis of $\r^n$. In momentum coordinates,
$\theta^i $ has the local  expression
\begin{equation}\label{localtitai}
\theta^i = \pi^i_j dx^j \, .
\end{equation}

$\hat{\theta}$ is the $n$-symplectic potential on $LM$.

Now the restriction of the $n$-tangent structure on $T^1_nM$ to $LM$ will yield an $n$-tangent
structure on $LM$. It is not difficult to show that the restriction of (\ref{localJA}) to $LM$ has,
in $n$-symplectic momentum coordinates, the form

\begin{equation}  \label{coordinate expression for Ji}
   J^i=-\pi^{i}_{a}\pi^{j}_{b}\basispi aj\otimes dx^{b}\, ,
   \end{equation}

 We will present now an alternative derivation of
this $n$-tangent structure on $LM$ that
is reminiscent of the geometric origins of other tangent-like structures.
We recall the formula

\begin{equation}
\xi^*(u)=\frac d{dt}(u\cdot \exp(t\xi))\vert_{t=0}
\end{equation}
for the value of the associated
fundamental vertical vector field $\xi^*$ on $LM$ defined at $u=(x,e_i)$
for each $\xi\in gl(n,\mathbb{Re eee})$.
These vector fields are smooth.
We define the vector-valued 1-forms $J^i$ by
\begin{equation}
(J^i)_u(X) = (E^i_j\theta^j_u(X))^*(u)\quad \forall\
X\in T_u(LM)
\end{equation}
This definition uses the group action  on $LM$ in a manner that parallels
the definition of the tangent structure on $TM$ and mixes in the canonical soldering $1$-forms in a
fundamental way. The difference is that the action of $GL(n,\mathbb{R})$ on $LM$ is global,
while the definition of
$J$ on $TM$ uses the fiberwise action of $T_nM$ on $T_nM$.

The mapping $\xi\to \xi^*$ is a linear mapping from the Lie algebra
$gl(n, \mathbb{R})$ to the Lie algebra of fundamental vertical vector
fields on $LM$. Hence
$$
(J^i)_u(X)= \theta^j_u(X)(E^i_j)^*(u)\quad \forall\
X\in T_u(LM)
$$
so that
\begin{equation}\label{m-tangent structure on LM}
J^i = \vertical ij\otimes\theta^j
\end{equation}

\noindent Substituting (\ref{estar}) and (\ref{localtitai}) into this formula yields the local
expression  (\ref{coordinate expression for Ji}).
This formula tell us that the canonical $n$-tangent structure on $T^1_nM$ is in fact another
representation of the soldering $1$-form $\hat{\theta}$. To see this explicity we note that the
mapping
$$ \hat{r}_i  \to E^j_i\otimes  \hat{r}_j \to E^{*j}_i\otimes  \hat{r}_i$$
is a linear representation of the basis vectors $\hat{r_i }$ of $\r^n$ in the space of $gl(n
,\mathbb{R})\otimes
\mathbb{R}^n$. Extending this representation to $\hat{\theta}= \theta^i \otimes \hat{r_i}$ we
obtain the $n$-tangent structure $\hat{J}$:
$$\hat{\theta}= \theta^i \otimes  \hat{r_i} \to   (E^{*i}_j\otimes \theta^j)   
\otimes   \hat{r_i} = \hat{J} \, .$$

\subsection{The vector-valued one-form $S_\alpha$ on $J^1\pi$}\label{vertendosect}

We now turn our attention to $1$-jets and review the tangent-like structure
present on $J^1\pi$ \, \cite{Saunders}.

Let $\pi:E\ra M$ be a fiber bundle where $M$ is $n$-dimensional and $E$ is
$m=(n+k)$-dimensional. Let
 $\tau_E\rfloor_{V \pi} :V\pi \rightarrow E$
be the vertical tangent bundle to $\pi$. We shall denote by
$\pi_{1,0}:\jetpi \to E$ the canonical projection and by $V\pi_{1,0}$ the
vertical distribution defined by $\pi_{1,0}$.

   Throughout this paper if $(x^i,y^A)$ are local fiber coordinates on $E$  we
take standard jet coordinates  $(x^i,y^A,y^A_i),\, 1\leq i\leq  n, 1\leq A\leq k,$
on the first jet bundle \jetpi\, the manifold of $1$-jets of sections of $\pi$.

\begin{definition}
Let $\phi:M\rightarrow E$ be a section of $\pi$, $x\in M$ and $y=\phi
(x)$. The {\rm vertical differential} of the section $\phi$ at the point $y\in
E$ is the map

\[
\begin{array}{ccccc}
d^V_y\phi & :& T_yE &\longrightarrow  & V_y \pi \\
          &  &    u  & \mapsto & u\, -\, (\phi \circ \pi)_*u
\end{array}
\]

\end{definition}
As $d^V_y\phi$ depends only on $j^1_x\phi$, the vertical differential can
be lifted to $J^1\pi$ in the following way.

\begin{definition}
   The {\rm canonical contact} $1$-form  $\omega^1$ on $J^1 \pi$ is the
 $V\pi$-valued $1$-form defined  by
 $$
\begin{array}{ccccc}
 {\bf \omega^1}(\jet)& :& T_{\jet}(J^1 \pi)& \longrightarrow & V_{\phi(x)}\pi \\
 & & \tilde{X}_{j^1_x\phi} &\mapsto &
 (d^V_y\phi)\left((\pi_{1,0})_*
 (\tilde{X}_{j^1_x\phi})\right)
\end{array}
$$
\end{definition}

\noindent In  coordinates,
\begin{equation}\label{coor3}
 {\bf \omega^1} \, = \, (dy^B - y^B_j \, dx^j) \otimes {\basisy B} \,
 \end{equation}

Next  let us  recall the definition of the vector-valued $1$-form
$S_\alpha$
 on $J^1 \pi$ where $\alpha$ is a $1$-form on $M$.
Given a point $j^1_x\phi\in J^1\pi$, a cotangent vector
 $\eta_x\in T^*_xM$ and a tangent vector
 $\xi \in V_{\phi(x)}\pi$,  there exists a well defined
 vector
  $\eta_x \odot_{\jet} \xi \in V_{j^1_x\phi}\pi_{1,0}$ called
    the {\em vertical lift}
  of $\xi$ to $V\pi_{1,0}$ by $\eta$.
 This vector is locally given by

 \begin{equation}\label{coor2}
 \eta_x \odot_{\jet} \xi_{\phi (x)} \quad = \quad
 \eta_i \, \xi^A \, {\basisv Ai} (\jet)\, .
 \end{equation}

\begin{definition}
   Let $\alpha \in \Lambda^1 M$ be any $1$-form on $M$. 
 The vector-valued $1$-form $S_\alpha$ on $J^1 \pi$ is defined by
 $$
\begin{array}{rcl}
 S_\alpha (\jet)  :
 T_{\jet} (J^1 \pi)
  & \longrightarrow & (V \pi_{1,0} )_{\jet} \\ \noalign{\medskip}
 \tilde{X}_{\jet}
  &  \rightarrow &
 S_\alpha (\jet) (\tilde{X}_{\jet}) \, = \,
 \alpha_x \odot_{\jet} {\bf \omega^1} 
  (\tilde{X}_{\jet}) \, .
\end{array}
$$
\end{definition}

 From (\ref{coor2}) and (\ref{coor3}) we have that in coordinates
\begin{equation}\label{soc}
 S_\alpha \, = \, \alpha_j \, (dy^A - y^A_i \, dx^i)
 \otimes {\basisv Aj} \, .
 \end{equation}

\noindent $S_\alpha$ can be considered a more general version of the canonical
tangent and $k$-tangent structures. This relationship is
 explored in section \ref{VertEndoVsKtangent}.
Note also that $S_\alpha$ plays an important role in the construction of
the Cartan-Hamilton-Poincar\'{e}  $n$-form  (see section \ref{pcformsect}).

\subsection{The adapted frame bundle $L_\pi E$}

An {\em adapted frame} at $e\in E$, $\pi:E  \to M$, is a frame where the  last $k$ basis vectors
are vertical with respect to $\pi$. The {\em adapted frame bundle} of $\pi$ \cite{FLN2,Lawson},
denoted by \lpie, consists of all adapted frames for $E$,
\[\lpie = \{(e_i,e_A)_e : e\in E,
\{e_i,e_A\} \mbox{ is a basis for } T_e E, \mbox{ and }
\pi_*(e)(e_A)=0\}\]
The canonical projection, $\lambda:\lpie\ra E$, is defined by
$\lambda(e_i,e_A)_e = e$.

\lpie\ is a reduced subbundle of $LE$, the frame bundle of $E$. As such it is a principal
fiber bundle over $E$.  Its structure group is \Gv\, the nonsingular block lower
triangular matrices
\begin{equation} \label{Gv defined}
\Gv = \left\{\left(\ba{cc}A&0\\C&B\ea\right): A\in Gl(n,\mathbb{R}), B\in
Gl(k,\mathbb{R}), C\in\mathbb{R}^{kn} \right\} \,
\end{equation}

\noindent \Gv\
 acts
on \lpie\ on the right by
\begin{equation}\label{actionlpie}
(e_i,e_A)_e \,{\mat A0CB}
= (A^i_j e_i+C^A_j e_A,B^A_B e_A)_e.
\end{equation}

 If $(x^i,y^A)$ are adapted coordinates on an open set $U\subseteq E$,
then one may induce several different coordinates on $\lambda^{-1} (U)$.
First consider the {\em coframe} or {\em $m$-symplectic momentum}
coordinates $(x^i,y^A,\pi^i_j,\pi^A_j,\pi^A_B)$
on $\lambda^{-1} (U)$ defined   in (\ref{momcoord}). Let us observe that  $\pi^i_A=0$ on 
$\lpie$.
 
We have as is customary retained the same symbols for the induced
horizontal coordinates.

Secondly consider the {\em frame} or {\em $m$-symplectic velocity} coordinates
$(x^i,y^A,v^i_j,v^A_j,v^A_B)$ on $\lambda^{-1} (U)$ defined  in (\ref{velocoord}).
Let us observe that  $v^i_A=0$ on 
$\lpie$.

The $v$ coordinates, viewed together as a block triangular matrix, form the
inverse of the $\pi$  coordinates defined  above.  The blocks have the following
relations:
\begin{equation*}
v^i_j \pi^j_s = \delta^i_s \qquad\qquad v^A_j \pi^j_s + v^A_B \pi^B_s = 0
\qquad\qquad v^A_B \pi^B_C = \delta^A_C
\end{equation*}

Lastly consider the following coordinates which
are constructed from the previous
two.\hfill\break Define
$(x^i,y^A,u^i_j,u^A_j,u^A_B)$ on $\lambda^{-1} (U)$   by
\begin{alignat*}{3}
x^i((e_i,e_A)_e)&=x^i(e) &\qquad\qquad  u^i_j&=\pi^i_j &\qquad\qquad
  u^A_j&=v^A_i \pi^i_j=-v^A_B \pi^B_j \\
y^A((e_i,e_A)_e)&=y^A(e) &\qquad\qquad  u^A_B &=\pi^A_B
\end{alignat*}

In Section 3.3
 we discuss the fact that $\lpie$ is an $H=Gl(n)\times Gl(k)$ principal bundle $\rho:\lpie \to
\jetpi$.
It will turn out that the $u^A_j$
coordinates are pull-ups under
$\rho$ of the standard jet coordinates on $\jetpi$.  As such, we  refer to these coordinates
 as {\em Lagrangian} coordinates.

 The fundamental vertical vector fields $\vertical {i}{j}$, $\vertical {A}{B}$ and 
$\vertical {i}{A}$,
on $\lpie$ are given, in Lagrangian coordinates, by
\begin{equation}\label{estar in lpie}
  \vertical ij = -u^i_s\basisu js\quad\quad
  \vertical AB = -u^A_C\basisu BC \quad \quad
  \vertical iA=u^i_sv_A^B\frac{\partial}{\partial u^B_s}
\end{equation}

On $L_\pi E$ we have also a $\r^{m+k}$-valued $1$-form, the {\it soldering
one-form}  $\hat{\theta} = {\theta}^i \, \hat{r}_i +
 \theta^A\, \hat{r}_A$, which is the restriction of the canonical soldering 1-form on $LE$ to
\lpie.   Here $\hat{r}_i, \hat{r}_A$ denotes the standard basis of $\r^{n+k}$.
 In Lagrangian coordinates, $\theta^i,\theta^A$
have the local  expression
\begin{equation}\label{local espression of titai and titaa}
\theta^i = u^i_j dx^j \, ,\quad \theta^A= u^A_B(dy^B-u^B_j dx^j)\, .
\end{equation}

 From   (\ref{m-tangent structure on LM}) we have that
the $(n+k)$-tangent structure on $LE$ is given by

$$J^i   = E^{*i}_j \otimes \theta^j +E^{*i}_B \otimes \theta^B,\quad
J^A = E^{*A}_j \otimes \theta^j +E^{*A}_B \otimes \theta^B$$

\noindent  Now considering its restriction to the principal
 fiber bundle $L_\pi E$
we have
$$
 (J^i)\vert_{L_\pi E}
\equiv J^i , \qquad
1\leq i \leq n,$$
$$ J^A\vert_{L_\pi E} \equiv
E^{*A}_j\vert_{L_\pi E} \otimes \theta^j +E^{*A}_B \otimes \theta^B
\qquad 1 \leq A \leq k\, .$$

\section{Relationships among the tangent-like structures}

In this section we show how the tangent, $k$-tangent, and similar
structures on various spaces are related.  We have already remarked in
Section 2.3 that the $n$-tangent structure on $LM$ and  the one on
$T^1_n M$ ($n=\dim M$) induce each other.  Now we complete the circle by
showing that the tangent structure on $TM$ induces the
$k$-tangent structure on $T^1_kM$ and that the $n$-tangent structure on
$LM$ induces the tangent structure on $TM$.

Secondly, we show that in the special cases where comparison makes sense
the vector valued one-form on $\jetpi$ is directly related to the
$k$-tangent structure on $T^1_kM$.  Furthermore, using recent results
relating the jet bundle and adapted frame bundle, we show a similar
relationship with the $(n+k)$-tangent structure on $\lpie$.

\subsection{Relationships among $TM$, $T^1_kM$, and $LM$}   %

\noindent {\underline {The $k$-tangent structure on $T^1_kM$ in terms of the
 tangent structure on $TM$}}
 \bigskip

One can induce $J^A$ on $T^1_k M$ from $J$ on $TM$.  We make use of the
{\em inclusion maps}
$$
\begin{array}{ccccr}
 i_A:& TM &\ra &T^1_kM &  \qquad 1\leq A \leq k\\
 & v_x  & \ra & (0,\ldots,0,v_x ,0,\ldots,0)
\end{array}
$$

 From (\ref{localj}), (\ref{localJA})   we obtain
\begin{prop}
$$J^A(u) = i_{A*}(\phi(u))  \of J_{\phi(u)} \of \phi_* (u)$$
\noindent  for all $u \in T^1_kM$, where $\phi:T^1_kM \ra TM$ is any $C^1$
bundle morphism over the identity on $M$({\em one of the $k$ projections
for example}).
\end{prop}

\bigskip
\noindent{\underline {The   tangent structure on $TM$ viewed from $LM$ }}
\bigskip

 \begin{lem}
Let $(J^1,\ldots,J^n)$ be the
canonical $n$-tangent
structure
of $LM$.
 For all vector fields $X$ on $LM$ we have
  \begin{equation}\label{transformation of Ji}
   J^i\of R_{g*}(X) = \ginverse ia R_{g*}\of
   J^a(X)
  \end{equation}
  \end{lem}

\noindent where $R_g$ denotes the right translation with respect to $g\in
GL(n,\mathbb{R})$.

 \proof \ \ It follows from (\ref{coordinate expression
for Ji}) and the identities
$$
R^{*}_{g}(\pi^{j}_{k})=(g^{-1})^{j}_{a}\pi^{a}_{k}, \quad
      R^{*}_{g}(d\pi^{j}_{k})=(g^{-1})^{j}_{a}d\pi^{a}_{k}, \quad
 R^{*}_{g}(dx^{i})=dx^{i} \, ,$$
$$
  R_{g*}(\basispi ij)= \ginverse ai \basispi aj , \quad
      R_{g*}(\basisx i)=\basisx i \, .$$
where $g$ is any element of $GL(n,\mathbb{R})$ . \blob

      Let  $\tilde TM$ denotes the manifold obtained from the
tangent bundle $TM$ by deleting the zero section. For a fixed, non-zero
element $\xi \in \rne$ let $\psi_{\xi}$ denote the mapping from $LM$ to
$\tilde TM$ defined as follows.  For each $u\in LM$ let
 \begin{equation}\label{def of psi}
   \psi_{\xi}(u)= [u,\xi]
     \end{equation}
 where we are identifying the tangent bundle $TM$ with the bundle
 associated to $LM$ and the standard action of $GL(n,\mathbb{R})$ on
 $\r^{n}$. The following lemma is easily verified for this map.

 \begin{lem}
 \begin{equation}\label{psixi}
(\psi_{\xi})_{*}(\vertical bc (u))=\xi^{b}v^{i}_{c}(u)
    \frb{\partial}{\partial y^i}_{[u,\xi]}
 \end{equation}
   \end{lem}  

 \rem In this case 
    Let $h=\hdown ij dx^{i}\otimes dx^{j}$ be any metric tensor
 field on the manifold $M$.  Then its associated covariant tensorial
function 
 on $LM$ is the $\rndual \otimes_{s}\rndual $ valued function with
 components
 \begin{equation}
   \hhatdown ij= (\hdown ab\circ\lambda)v^{a}_{i}v^{b}_{j}
   \label{hhatdown defined}
 \end{equation}
 (see \cite{KN}).
For simplicity we will drop the \, $\circ\lambda$ \,  notation and write simply
 $\hhatdown ij= \hdown ab  v^{a}_{i}v^{b}_{j}$.
 Moreover, we know that $(\hhatdown ab)$ obeys the transformation law
 \begin{equation}
   \hhatdown ab(u\cdot g)=g^{m}_{a}g^{n}_{b}\hhatdown mn(u)
   \label{transformation of hhatdown}
 \end{equation}
 for all $g\in GL(n)$.

  \begin{definition} Let $h$ be a fixed positive definite metric
 tensor field on $M$. The associated {\rm covariant
 $m$-tangent structure $(\jh i)$} {\rm based on $h$} is
 \begin{equation}\label{covariant j defined}
   \jh i= \displaystyle\sum_{j}^{m}\hhatdown ij J^j
     \end{equation}

 \end{definition}

 \begin{lem}
 \begin{equation}\label{tranformaion of jhdown}
   \jh{a}(u\cdot g)(R_{g*}(X)) = g^{b}_{a}  R_{g*}\left(
   \jh{b}(u)(X)\right)
  \end{equation}
  \end{lem}

 \proof \ \  The proof follows easily from (\ref{transformation of Ji})  and
 (\ref{transformation of hhatdown}).\quad \blob

\begin{thm} Let $h$ be an arbitrary positive definite metric tensor field on the
manifold $M$, and let $(\jh i)$ denote the covariant $m$-tangent structure
on $LM$ defined by $h$. For each point $[u,\xi]\in \tilde TM$ (note:
$\xi=(\xi^{i})$ is by assumption  non-zero)
 let   $\psi_{\xi}:LM\to \tilde TM$ be the map
defined  in (\ref{def of psi}) above.  Then the vector-valued 1-form
$\canonicalj$\ on $\tilde TM$ defined pointwise by
\begin{equation}\label{j on tm from jh on lm}
   X\longrightarrow \canonicalj ([u,\xi])(X)=
\frac{ \psi_{\xi*}\left( \xi^{i}\jh i(u)(\tilde X)\right)}{ \hhatdown ab(u)
     \xi^{a}\xi^{b}}  ,\qquad \forall\ X\in T_{[u,\xi]}(TM)
 \end{equation}

\noindent  is the canonical tangent structure on $\tilde TM$ given in local coordinates
by
\begin{equation}\label{canonical j in canonical coordinates}
   \canonicalj = \frac{\partial}{\partial y^i} \otimes dx^{i}
 \end{equation}
In equation (\ref{j on tm from jh on lm}) $\tilde X$
is any tangent vector at $u\in LM$ that projects to the same
vector at $\lambda_M(u)$ as does the vector $X\in T_{[u,\xi]}(TM)$; i.e.
$d\lambda_M(\tilde X)=d\tau(X)$.

\end{thm}

\proof\ \  We first show that the tangent vector
$\canonicalj([u,\xi])$ is well-defined.
Since $[u,\xi] = [u\cdot g, g\cdot \xi]$ we need to show that the
right-hand side of formula (\ref{j on tm from jh on lm}) remains
unchanged if we make the substitutions $u\to u\cdot g$ and
$\xi\to g\cdot\xi=(\ginverse ij \xi^{j})$.
Making the substitutions we have
\begin{equation}\label{initial step}
   \canonicalj ([u\cdot g, g\, \xi])(X)   =
   \frac{ \psi_{(g\, \xi)*}\left( (g\,
   \xi)^{i}\jh i(u\cdot g)(R_{g*}\tilde X)\right) } { \hhatdown
   ab(u\cdot g)(g\cdot \xi)^{a}(g\cdot \xi)^{b}}
 \end{equation}
Using $(g\,\xi^{i})=\ginverse im\xi^{m}$ and (\ref{tranformaion of
jhdown})
the numerator in this equation can be reduced as follows:
\begin{eqnarray*}
\psi_{(g\, \xi)*}\left( (g\,\xi)^{i}\jh i(u\cdot g)(R_{g*}\tilde X)\right)
&=&\psi_{(g\, \xi)*}\left( \ginverse im\xi^{m}  g^{b}_{i}  R_{g*}\left(
   \jh{b}(u)(X)\right)\right)  \\
&=& \psi_{(g\, \xi)*}\left( R_{g*}(\xi^{i} \jh i(u )(\tilde X)\right) \\
&=&(\psi_{(g\,\xi)}\circ R_{g})_{*}\left( \xi^{i} \jh i(u )(\tilde
X)\right)\\
&=& \psi_{\xi*}\left( \xi^{i} \jh i(u )(\tilde
X)\right)
\end{eqnarray*}
where the last  equality follows from the fact that $\psi_{g\,\xi }\circ R_{g}
=\psi_{\xi}$.

Similarly, using (\ref{transformation of hhatdown}) the denominator in
equation (\ref{initial step}) can be reduced as follows:
$$
\hhatdown ab(u\cdot g)(g\cdot \xi)^{a}(g\cdot \xi)^{b}
=\hhatdown  ab(u )  \xi^{a} \xi^{b}
$$
Substituting the last two results into (\ref{initial step}) we obtain
$$
 \frac{ \psi_{(g\cdot \xi)*}\left( (g\,
   \xi)^{i}\jh i(u\cdot g)(R_{g*}\tilde X)\right) }{ \hhatdown
   ab(u\cdot g)(g\cdot \xi)^{a}(g\cdot \xi)^{b}}
   = \frac{ \psi_{\xi*}\left( \xi^{i}\jh i(u)(\tilde X)\right)}
   { \hhatdown ab(u) \xi^{a}\xi^{b}}
$$
which proves that the mapping given in  (\ref{j on tm from jh on lm}) above
 is
well-defined.

We now calculate the numerator on the right-hand-side of the above
identity. From  (\ref{psixi}),
(\ref{covariant j defined}),   we obtain
\begin{eqnarray*}
\psi_{\xi*}\left( \xi^{i}\jh i(u)(\tilde X)  \right)
         & = & \left(\xi^{i}\hhatdown ij(u)\theta^{k}(u)(\tilde X)\right)\psi_{\xi*}
    \left( \vertical jk(u)   \right)  \\
 & = & \left(\xi^{i}\hhatdown ij(u)\pi^{k}_{l}(u)dx^{l}(\tilde X)\right)
 \left( \xi^{j}v^{a}_{k}(u)\frac{\partial}{\partial y^A}([u,\xi]) \right)\\
    & = & \left(\hhatdown ij(u)\xi^{i}\xi^{j}  \right)
     \left( \frac{\partial}{\partial y^A}([u,\xi]) dx^{a}(  X) \right)\\
    & = & \left(\hhatdown ij(u)\xi^{i}\xi^{j}  \right)
   \left( \frac{\partial}{\partial y^A}\otimes dx^{a} \right)([u,\xi]) (X)
\end{eqnarray*}
Since the metric $h$ is definite, the coefficient $\hhatdown
ij(u)\xi^{i}\xi^{j}$ is non-zero for all $u\in LM$. Hence we may
divide both sides of the last equation by this term and use linearity
of the mapping $\psi_{\xi}$ to obtain the desired result. \quad\blob

\subsection{The relationship between the vertical endomorphism on \jetpi\
and the canonical $k$-tangent structures}
\label{VertEndoVsKtangent}

 Now we shall describe the relationship between the vertical endomorphism on
\jetpi\ and the canonical $k$-tangent structure on $T^1_kM$ when $E$ is
the trivial bundle  $E=\rk\times M \rightarrow \rk$.  In this
case    $J^1 \pi$ is diffeomorphic to $   \rk \times T^1_kM$  via the
diffeomorphism  given by $j^1_t\phi \equiv (t,j^1_0\phi_t)$  where
$\phi_t (s)= \phi (t+s)$.
 In this case, (see (\ref{soc})),
 the vector valued $1$-form $S_\alpha$ is locally given by
$$
S_\alpha \, = \,
 \basisv iB
 \otimes
 \big(\alpha_B \,
 (dx^i -v^i_A \, dt^A)\big)
$$
with respect the
 coordinates
$(t^A,x^i,v_A^i)$ on $\rk \times T^1_kM $.

In the case $k=1$, we consider $\omega=dt$ and thus

$$
S_{dt}
= \,
 \frac{\partial}{\partial v^i}\,
 \otimes
 (dx^i - v^i \, dt)
= \,
 \frac{\partial}{\partial v^i}\otimes dx^i
 - \,  v^i \, \frac{\partial}{\partial v^i}\otimes \, dt $$

\noindent where  $(t,x^i,v^i)$ are the coordinates in $\rne \times TM$. Then
we have

   $$S_{dt} = J  \, - \, C\otimes \, dt$$

\noindent where $C$ denotes the canonical or Liouville vector field on $TM$ and
$J$ is the canonical tangent structure $J$ on $TM$.

In the general case, with $k$ arbitrary, if we fix $B$, $1\leq B\leq
k$, we have

%
$$
S_{dt^B}
=\,
 \frac{\partial}{\partial v^i_B}
 \otimes
 (dx^i -v^i_A \, dt^A)
\, = \,
\frac{\partial}{\partial v^i_B}\otimes dx^i
  - \,v^i_A \, \frac{\partial}{\partial v^i_B}\otimes \, dt^A
=J^B - C^B_A\otimes dt^B $$

\noindent where   $J^B$
is the canonical $k$-tangent structure on $T^1_kM$, and the $C^B_A$ are
the {\it canonical vertical  vector fields} defined in equation (\ref{gen
euler}).

\begin{prop} \label{trivial S case}
The relationship between the canonical $k$-tangent structure on $T^1_kM $
and the  vertical endomorphism $S_{dt^B}$, up to some obvious
identifications,  is given by
$$  J^B = S_{dt^B} \, + \, C^B_A\otimes dt^A $$
\end{prop}

\subsection{Strong relationships between $J^1\pi$
and $L_\pi E$}

   We shall consider two    ways of
describing 1-jets, each with its own charm:

1. Equivalence classes of local sections of $\pi$.
$$
J^1 \pi = \lbrace \jet : x \in M , \phi \in \Gamma_x(\pi) \rbrace
$$
where $\Gamma_x(\pi) $ denotes the set of sections of $\pi$ defined in a
neigboorhood of $x$.

2. Linear right-inverses to $\pi_* (e)$.
$$
J^1 \pi = \lbrace \tau_e : T_{\pi (e)} M  \rightarrow T_e E :
\pi_* (e) \circ \tau_e = id_{T_{\pi (e)} M} \rbrace
$$
We will use either  description of $J^1 \pi$ when it is convenient.

Let $H$ be the subgroup of \Gv\  isomorphic to $Gl(n)\times Gl(k)$ (defined in (\ref{Gv defined}))
given by
\[H =\left\{\left(\ba{cc}A&0\\0&B\ea\right):
            A\in Gl(n,\r),B\in Gl(k,\r)\right\}  \quad ,\]
and let ${\cal J}$ be the following subgroup of \Gv\
\[{\cal J} =\left\{\left(\ba{cc}I&0\\\xi&I\ea\right):
            \xi \in \mathbb{R}^{kn} \right\}   \quad .\]

   Although $H$ is a closed Lie subgroup of $G_V$, it is not
normal. As such $\Gv /H$ does not have a natural group structure; it is a
manifold with a left \Gv-action. For each coset $gH\in\Gv /H$, we select
the unique representative in ${\cal J}$.
\[\left(\ba{cc}A&0\\C&B\ea\right)\sim
\left(\ba{cc}A&0\\C&B\ea\right)\left(\ba{cc}A\inv&0\\0&B\inv\ea\right)
=\left(\ba{cc}I&0\\CA\inv&I\ea\right)\] By choosing these
representatives, we identify $\Gv /H$ with ${\cal J}$ and hence
$\mathbb{R}^{kn}$. These identifications are diffeomorphisms.

Consider how the left \Gv-action looks for our selected representatives.
\begin{equation}\label{action}
\left(\ba{cc}A&0\\C&B\ea\right) \left(\ba{cc}I&0\\ \xi&I\ea\right)
=\left(\ba{cc}A&0\\C+B\xi&B\ea\right)
\sim\left(\ba{cc}I&0\\CA\inv+B\xi A\inv&I\ea\right)
\end{equation}

\noindent So the \Gv-action appears {\em affine} when $\Gv /H$ is identified with
$\mathbb{R}^{kn}$. Therefore it is prudent to use this identification to
define an affine structure on $\Gv /H$ modeled on $ \mathbb{R}^{kn}$.
This \Gv-invariant structure will pass to the fibers of the associated
bundle discussed below, making it an affine bundle.

\begin{thm}\label{thjpiiso}\qquad
$\displaystyle \lpie\times_{\Gv}(\Gv/H) \iso \jetpi$
\end{thm}

\noindent {\bld Proof:}
The affine bundle isomorphism maps
each equivalence class $[(e_i,e_A)_e,(\xi^A_i)]$ to the
linear map $\phi:T_{\pi(e)}M\ra T_e E$ defined by
$\phi(\hat{e}_i) = e_i + \xi^A_i e_A$,
where we use the basis $\{\hat{e}_i\}$ where $\hat{e}_i=\pi_*(e)(e_i)$.
The inverse isomorphism
is given by
 $$
 \jet \longmapsto \lbrack \left( {\basisx i} ,
 {\basisy a} \right)_{\phi (x)},
 \left( \frac{\partial \phi^a}{\partial x^i} (x) \right) \rbrack
 $$

The following corollary, whose simple proof is made possible by the preceding
development, is a fundamental tool in lifting Lagrangian field theory to the
adapted frame bundle.

\begin{cor}
\lpie\ is a principal fiber bundle over \jetpi\ with structure group $H$.
\end{cor}
\par\noindent {\bld Proof:}
This fact follows directly from proposition 5.5 in reference \cite{KN}. \blob

We will denote the projection from \lpie\ to \jetpi\
by $\rho$. It is given by
  $$
 \begin{array}{rlcrlcl}
   \rho: & L_\pi E & \longrightarrow & J^1 \pi & & & \\
    & (e_i,e_A)_e & \longmapsto & \tau_e: & T_{\pi(e)} M & \rightarrow &
    T_e E \\
    & & & & \pi_*(e) (e_i) & \mapsto & e_i
 \end{array}
 $$

We  now show that
the $u^A_j$-coordinates defined in Section 2.5 are  the pull-ups of the jet
coordinates.
If $(x^i,y^A)$ are adapted coordinates on an open set $U\subseteq E$
and $u=(e_i,e_A)_e\in\lambda\inv(U)$ then
\begin{align*}
y^A_i\of\rho(u)
&= y^A_i(\tau_e) = (dy^A)_e \left( \tau_e (\frb{\pa}{\pa x^i}_{\pi(e)}) \right)
         = (dy^A)_e(\hat e^j(\frb{\pa}{\pa x^i}_{\pi(e)}) e_j) \\
        & = e^j(\frb{\pa}{\pa x^i}_{e}) (dy^A)_e(e_j)
= \pi^j_i   (u) v^A_j(u)= u^A_j(u)
\end{align*}

\subsection {The vector-valued $1$-form $S_\alpha$ on $J^1
\pi$ viewed from $L_\pi E$.}
   \lpie\ is a principal bundle over
\jetpi,   we shall establish in this subsection the relationship between
the vertical endomorphism $S_\alpha$  on \jetpi\ and the restriction of
the $(n+k)$-tangent structure of $LE$ to the vertical adapted bundle
\lpie. To be more precise, we show that $S_\alpha$ corresponds to the
tensors on $\lpie$:
 $$E^{*i}_B \otimes \theta^B  =
J^i - E^{*i}_j \otimes \theta^j,\quad 1\leq i\leq n\, .$$
Note the similarity to proposition \ref{trivial S case}.

\begin{prop}
 Let  $u = (e_i ,e_A)_e$ be a frame on \lpie\ and let us denote
by $u \cdot \xi$ the frame
$$
     u \cdot \xi  \quad = \quad (e_i , e_A)_e \,
   {\mat I0\xi I} \quad = \quad   (e_i + \xi^B_i \, e_B , e_A)_e \quad
      $$
   Let $\alpha$ be any $1$-form on $M$
and  $\lbrack u,\xi \rbrack =
\lbrack (e_i,e_A)_e, (\xi^A_i) \rbrack$ an element of $J^1 \pi$. Then
the relationship
 between $S_\alpha$ and the tensor fields $E^{*i}_B \otimes \theta^B$ is
given by
\begin{equation}\label{somega}
\displaystyle
 S_\alpha (\lbrack u , \xi \rbrack)
 (X_{\lbrack u, \xi \rbrack}) \,
 = \, \rho_* ( u \cdot \xi) \left(
 (\pi^* \alpha)_e (e_i) \,
 (E^{*i}_B \otimes \theta^B) (u \cdot \xi)
 ({\tilde{X}}_{u \cdot \xi }) \right)
\end{equation}
 where
 $$
    X_{\lbrack u, \xi \rbrack}  \in T_{\lbrack u, \xi \rbrack} (J^1 \pi)\, ,
\quad  {\tilde{X}}_{ u \cdot \xi }  \in
   T_{ u \, \xi } (L_\pi E)
 $$
 are vectors that project onto the same vector on $E$.
 \end{prop}

{\bf Proof :}
   First let us observe that, from the definition of $\rho$, we have
$$\rho (u\cdot \xi ) = \rho ((e_i + \xi^B_i \, e_B , e_A)_e) =
 \lbrack (e_i,e_A)_e,
 (\xi^A_i) \rbrack =\lbrack u,\xi \rbrack \quad .$$

 Now we shall prove that the right side of (\ref{somega}) does not depend on the
choice of the representative of the equivalence class $\lbrack (e_i,e_A)_e,
(\xi^A_i) \rbrack$. If
$$
\lbrack u , \xi \rbrack =\lbrack (e_i,e_A)_e, (\xi^A_i) \rbrack \, = \,
  \lbrack (\bar{e}_i,\bar{e}_A)_e, (\bar{\xi}^A_i) \rbrack \,  = \, \lbrack \bar{u}, \bar{\xi} \rbrack
\quad 
$$
we must prove that
 $$
   \rho_* (u \cdot \xi) \left(
   (\pi^* \alpha)_e (e_i) \,
   (E^{*i}_B \otimes \theta^B) (u \, \xi)
   (X_{u \cdot \xi}) \right) \,
   = \, \rho_* ((\bar{u} \cdot \bar{\xi})
   \left( (\pi^* \alpha)_e (\bar{e}_j) \, (E^{*j}_C \otimes \theta^C)
   ((\bar{u} \cdot \bar{\xi})
   (\bar{X}_{\bar{u} \cdot \bar{\xi}}) \right)
 $$
 for any vectors
$   X_{u \cdot \xi}  \in  T_{u \cdot \xi} (L_\pi E), \,
   \bar{X}_{\bar{u}\cdot \bar{\xi}}  \in
   T_{\bar{u} \cdot \bar{\xi}} (L_\pi E)
 $
 that project at the same vector on $E$.

But, in this case, we have from (\ref{actionlpie}) and (\ref{action})
\begin{equation}\label{ubarra}
\bar{u} =   (\bar{e}_j , \bar{e}_B)_e \,
     = (A^i_j \, e_i + C^A_j \, e_A\, ,B^A_B \, e_A ),\quad
   \bar{\xi}^B_j = -(B^{-1})^B_C \, C^C_j
 + (B^{-1})^B_C \, \xi^C_i \, A^i_j \quad .
\end{equation}

\noindent    Let us consider the frames
 $$
 \tilde{u} = (\tilde{e}_i, \tilde{e}_A)_e = u \cdot \xi =
  (e_i + \xi^B_i \, e_B , e_A)_e
 $$
$$
 \hat{u} = (\hat{e}_j, \hat{e}_B)_e = \bar{u} \cdot \bar{\xi} =
  (\bar{e}_j + \bar{\xi}^C_j \, \bar{e}_C , \bar{e}_B)_e
  = (A^i_j \, \tilde{e}_i ,
B^A_B \, \tilde{e}_A )_e $$
where the last identity comes from (\ref{ubarra}).
Then we deduce that the relationship between
the coordinates of $\tilde{u}$ and $\hat{u}$ are
 \begin{equation}\label{e2}
 \hat{v}^l_j = A^i_j \, \tilde{v}^l_i \, , \,
 \hat{v}^C_j = A^i_j \, \tilde{v}^C_i \, , \,
 \hat{v}^C_B = B^A_B \, \tilde{v}^C_A \, , \,
     \hat{u}^i_j = (A^{-1})^i_l \, \tilde{u}^l_j \, ,\,
   \hat{u}^A_l = \tilde{u}^A_l \,\, .
\end{equation}
On the other hand, the tensor fields
 $E^{*i}_B \otimes \theta^B$ are locally given by
\begin{equation}\label{eteta}
 E^{*i}_B \otimes \theta^B = u^i_j \, (dy^B - u^B_t \, dx^t) \otimes
 {\basisu Bj} \quad .
\end{equation}

\noindent From (\ref{eteta}),
 and  (\ref{e2}) we obtain
 \begin{equation}
   (E^{*j}_C \otimes \theta^C) (\hat{u}) =
  (A^{-1})^j_r \, {\tilde{u}}^r_l \, ((dy^A)_{\hat{u}} -
   {\tilde{u}}^A_t \, (dx^t)_{\hat{u}}) \otimes {\basisu Al} (\hat{u})
 \end{equation}

Since
 $  (\pi^* \alpha)_e (\bar{e}_j)  = A^i_j \,
   (\pi^* \alpha)_e (e_i)
 $ we deduce that
\begin{equation}\label{uhat}
\begin{array}{ccl}
 (\pi^* \alpha)_e (\bar{e}_j) \, (E^{*j}_C \otimes \theta^C) (\hat{u}) &
 = & (\pi^* \alpha)_e (e_i) \, {\tilde{u}}^i_l \, ((dy^A)_{\hat{u}} -
 {\tilde{u}}^A_t \, (dx^t)_{\hat{u}}) \otimes {\basisu al} (\hat{u})
\end{array}
\end{equation}
and
\begin{equation}\label{utilde}
(\pi^* \alpha )_e(e_i)  (E^{*i}_B \otimes \theta^B)(\tilde{u}) =
(\pi^* \alpha )_e(e_i) \, {\tilde{u}}^i_j \,
 ((dy^A)_{\tilde{u}} - {\tilde{u}}^A_l \, (dx^l)_{\tilde{u}})
 \otimes {\basisu Aj} (\tilde{u})
 \end{equation}

If the vectors
$ X_{u \cdot \xi}  \in  T_{u \cdot \xi} (L_\pi E), \,
   \bar{X}_{\bar{u}\cdot \bar{\xi}}  \in
   T_{\bar{u} \cdot \bar{\xi}} (L_\pi E)
 $
 project onto the same vector on $E$ then its components with respect the
coordinates $x^i$ and $y^A$ are equal and from (\ref{uhat})
 and (\ref{utilde})
we obtain that
$$\rho_*(\tilde{u}) \left( (\pi^* \alpha )_e(e_i)  (E^{*i}_B \otimes \theta^B)
(\tilde{u})(X_{\tilde{u}}) \right)=
\rho_*(\hat{u})\left((\pi^* \alpha )_e(\bar{e}_j)
(E^{*j}_C \otimes \theta^C) (\hat{u}) (\bar{X}_{\hat{u}})\right)$$
because $\rho(\tilde{u})=[u,\xi]=[\bar{u},\bar{\xi}]=\rho(\hat{u})$.

   Now we shall prove the identity (\ref{somega})
using the Theorem \ref{thjpiiso} . If \, $j^1_x \phi \equiv  \lbrack (e_i,e_A)_e , (\xi^A_i)
\rbrack$
\, and
\begin{equation}
 e_i = v^j_A    \, {\basisx j} (e) + v^B_i \, {\basisy B} (e) \quad ,
 \quad e_A = v^B_A\, {\basisy B} (e)
 \end{equation}
 then
$$
  \lbrack (e_i,e_A)_e , (\xi^A_i) \rbrack  =
  \lbrack ({\basisx j},{\basisy B})_e , {\mat {v^j_A   }0{v^B_i}{v^B_A}}
   \vert_e (\xi^A_i)  \rbrack
 $$

\noindent    From this  identity and (\ref{action}) we  deduce
that the coordinates of the 1-jet $\jet$,
 defined by the class
$\ \lbrack u , \xi \rbrack = \lbrack (e_i,e_A)_e , (\xi^A_i) \rbrack$,
 are
\begin{equation}\label{e5}
 \frac{\partial \phi^B}{\partial x^s} (\pi(e)) =
 u^r_s \, (v^B_r + v^B_A\, \xi^A_r)\vert_e
 \end{equation}
 and therefore from (\ref{soc}) we have
\begin{equation}\label{sotau}
 S_\alpha (\lbrack u,\xi \rbrack ) = \alpha_j(\pi(e)) \,
((dy^A)_{\lbrack u,\xi \rbrack } -
 u^r_i \, (v^A_r + v^A_B \, \xi^B_r)\vert_e \,
 (dx^i)_{\lbrack u,\xi \rbrack }) \otimes
 {\basisv Aj} (\lbrack u,\xi \rbrack )
 \end{equation}

\noindent The coodinates of the frame $\tilde{u}$
satisfy the identities
  $$
 \tilde{u}^i_j = u^i_j \quad , \quad
 \tilde{u}^A_t = \tilde{u}^l_t \, \tilde{v}^A_l =
 u^l_t \, (v^A_l + v^A_B \, \xi^B_l) \quad .
 $$
   Therefore, from (\ref{eteta}) we have
\begin{equation}\label{etu}
  (E^{*i}_B \otimes \theta^B) (\tilde{u})  =
u^i_j\vert_e \, ((dy^A)_{\tilde{u}} -
 u^l_t\vert_e \, (v^a_l + v^A_B \, \xi^B_l)\vert_e \, (dx^t)_{\tilde{u}}) \otimes
 {\basisu aj} (\tilde{u})
 \end{equation}
   If $\alpha = \alpha_r \, dx^r$, then
 $ (\pi^* \alpha)_e (e_i)   =
 \alpha_r (\pi(e)) \, v^r_i $
and from  (\ref{etu}) we obtain that
 $$
 (\pi^* \alpha)_e (e_i) \, (E^{*i}_B \otimes \theta^B) (\tilde{u}) =
 \alpha_j \, ((dy^A)_{\tilde{u}} - u^l_t\vert_e \,
 (v^a_l + v^A_B \, \xi^B_l)\vert_e \, (dx^t)_{\tilde{u}}) \otimes
 {\basisu Aj} (\tilde{u}))
 $$

\noindent    Now, since $\rho(\tilde{u}) =\lbrack u,\xi \rbrack $ ,
 from this last identity and  (\ref{sotau})
we  get the identity (\ref{somega}) taking into account
that $\rho^*y^A_j = u^A_j$. \blob

\section{Spaces with  cotangent-like structures}

In this section we shall define and give the main properties of  the almost
cotangent structure and its generalizations.

\subsection{Almost cotangent  structures and $T^*M$}

Almost cotangent structures were introduced by Bruckheimer \cite{b}. An
{\rm almost cotangent structure} on a $2m$-dimensional manifold $M$ consists
of a pair $(\omega , V)$ where $\omega $ is a symplectic form and $V$ is a
  distribution   such that
$$(i) \quad \omega\rfloor_{V\times V} =0,
      \qquad (ii) \quad   \ker\omega =\{0\}\, $$

The canonical model of this structure  is the cotangent
 bundle $\tau^*_M:T^*M  \to M$ of an
arbitrary manifold  $M$, where $\omega$ is the canonical symplectic form
$\omega_0=-d\theta_0$ on $T^*M$ and $V$ is the vertical distribution. Let
us recall the definition of the Liouville form $\theta_0$ in $T^*M$:
 $$\theta_0(\alpha)
(\tilde{X}_{\alpha})= \alpha
((\tau_M^*)_*(\alpha)(\tilde{X}_{\alpha})),$$ \noindent  for all
vectors $\tilde{X}_{\alpha}\in T_{\alpha}(T^*M) \, .$
In local coordinates $(x^i,p_i)$ on $T^*M$
\begin{equation}\label{locot}
\omega_0 = dx^i\wedge dp_i, \qquad  V=\langle \frac{\displaystyle\partial}
{\displaystyle\partial p_1}, \dots,
\frac{\displaystyle\partial}{\displaystyle\partial p_k}\rangle.
\end{equation}

Clark and Goel \cite{cgc} also investigated these structures, defining
them as a certain type of $G$-structure. They proved that the integrability
 of these structures, that is the existence of coordinates on the manifold
 such that
 $\omega_0$ and $V$ have the form of (\ref{locot}),  is
characterized by

\begin{prop}
An almost cotangent structure $(\omega,V)$ on $M$ is integrable
if and only if $\omega$ is closed and the distribution $V$
is involutive.  \blob
\end{prop}

Thompson \cite{ts,t} proved  that an integrable almost cotangent manifold
$M$ satisfiying  some natural global hypotheses is essentially the
cotangent bundle of some differentiable manifold.

\subsection{$k$-symplectic structures and $ (T^1_k)^*M $}

\begin{definition}\label{defksym}   \cite{aw1,aw2}
A {\rm $k$-symplectic structure}
on  a manifold $M$ of dimension $N=n+kn$ is a  family
$(\omega_A,V;1\leq A\leq k)$, where each  $\omega_A$ is a closed $2$-form and
$V$ is an $nk$-dimensional distribution on $M$ such that
 $$(i) \quad
\omega_{A_{\rfloor V\times V}}=0,\qquad (ii) \quad \displaystyle
{\cap_{A=1}^{k}} \ker\omega_A=\{0\}.$$
In this case $(M,\omega_A,V)$ is
called a {\rm $k$-symplectic manifold}.
\end{definition}

The canonical model of this structure  is the $k$-cotangent bundle
$(T^1_k)^*M =J^1(M, \rk)_0$  of an arbitrary manifold $M$, that is the
vector bundle with
 total space the manifold of $1$-jets of maps with target  at
$0\in {\bf R}^k$,  and projection    $\tau^*(j^1_{x,o}\sigma)=x$.

 The manifold $(T^{1}_{k})^*M$ can
be canonically identified with the
Whitney sum of $k$ copies of $T^*M$, say
\[
\begin{array}{ccc}
(T^{1}_{k})^*M & \equiv & T^*M \oplus \dots \oplus T^*M, \\ j_{x,0}\sigma
& \equiv & (j^1_{x,0}\sigma^1, \dots ,j^k_{x,0}\sigma^k)
\end{array}
\]
where $\sigma^A= \pi_A \circ \sigma:M \longrightarrow \mathbb{R}$ is the
 $A$-th component of $\sigma$.

The canonical $k$-symplectic
structure $(\omega_A,V;1\leq A\leq k)$,  on
$(T^1_k)^*M$
is defined by
\begin{gather*}
\omega_A = (\tau^*_A)^*(\omega_0)  \\ V(j^1_{x,0}\sigma) =
\ker(\tau^*)_*(j^1_{x,0}\sigma)
\end{gather*}
where $\tau^*_A = (T^1_k)^*M \rightarrow T^*M $
is the projection on the $A^{th}$-copy  $T^*M$
of $(T^1_k)^*M$, and $\omega_0$ is the canonical symplectic structure
of $T^*M$.

One can also define the $2$-forms $\omega_A$ by $\omega_A = -d\theta_A$
 where $\theta_A$ is
the $1$-form defined as follows
$$\theta_A(j^1_{x,0}\sigma)
(\tilde{X}_{\displaystyle j^1_{x,0}\sigma})= \sigma_{*}(x)(
(\tau_A^*)_*(j^1_{x,0}\sigma)\tilde{X}_{\displaystyle j^1_{x,0}\sigma})
 $$ for all vectors
$\tilde{X}_{\displaystyle j^1_{x,0}\sigma}\in
T_{j^1_{x,0}\sigma}(T^1_k)^*M \, .$

 If $(x^i)$ are local coordinates on $U \subseteq M$ then the induced  local
coordinates $(x^i , p^A_i),\, 1\leq i \leq n,\, 1\leq A \leq k$ on
$(T^1_k)^*U=(\tau^*)^{-1}(U)$ are given by

$$
  x^i(j^1_{x,0}\sigma)=x^i(x),\qquad
p^A_i(j^1_{x,0}\sigma)=d_x\sigma^A(\frb{\partial}{\partial x^i}_x )\, .
$$
Then the canonical $k$-symplectic structure
is locally given by
$$
\omega_A=\displaystyle \sum_{i =1}^{n}dx^i\wedge dp^A_{i}, \qquad
V=\langle\frac{\displaystyle\partial} {\displaystyle\partial p^1_{i}}, \dots,
\frac{\displaystyle\partial}{\displaystyle\partial p^k_{i}}\rangle \quad 1\leq
A \leq k\, .
$$

\begin{thm}\label{aw}\cite{aw1}
Let $(\omega_A,V;1\leq A\leq k)$ be a $k$-symplectic structure on $M$.
About every point of $M$ we can find a local coordinate system $(x^i ,
p^A_i),\, 1\leq i \leq n,\, 1\leq A \leq k$ such that
\begin{equation}\label{locksym}
\omega_A=  \sum_{i=1}^n dx^i\wedge dp^A_{i}, \quad  1\leq A \leq k
\end{equation}
\end{thm}

In \cite{gun} G\"{u}nther introduces the following definitions.
\begin{definition}
A closed non-degenerate $\r^n$-valued $2$-form
$$\bar{\omega} =\displaystyle \sum_{A=1}^n \omega_A \, {\hat{r}_A}$$
on a manifold $M$ of dimension $N$ is called a {\rm polysymplectic form}.
The pair $(M,\bar{\omega})$ is a {\rm polysymplectic manifold}.

 A polysymplectic form $\bar{\omega}$ on
a manifold $M$ is called {\rm standard} iff for every point of $M$ there exists
a local coordinate system such that $\omega_A$ is written locally as in
(\ref{locksym}). 
\end{definition}

>From  Theorem \ref{aw} it now follows
 that  the $k$-symplectic manifold structures
 coincide with the {\it standard polysymplectic}
 structures.

$\bar{\omega}$ is called by Norris \cite{No5} a general $n$-symplectic
structure. The difference in the formalism is that there exist natural
definitions of Poisson brackets in the $n$-symplectic theory of Norris.
See Section $9$ for a discussion of $n$-symplectic Poisson brackets in
the general case.

\subsection{Almost $k$-cotangent structures and $(T^1_k)^*M$}

In \cite{mc1} the almost $k$-cotangent structures were defined and
described as $G$-structures.
\begin{definition}
An {\rm almost $k$-cotangent structure} is a  family $(\omega_A,V_A;1\leq A\leq k)$,
where each  $\omega_A$ is a  $2$-form of constant rank $2n$ and $V_A$ is a
$n$-dimensional distribution on $M$, such that
$$(i)\quad  V_A \cap (\oplus_{B\neq A}V_B)=0, \quad
  (ii) \quad  \ker \omega_A=\oplus_{B\neq A} V_B, \quad
  (iii) \quad \omega_A\rfloor_{V_A\times V_A}=0$$
for all $1\leq A \leq k$.
\end{definition}

The canonical model of this structure  is   $(T^1_k)^*M$ with the $2$-forms
$\omega_A$, and  $V_A= \ker T\rho_A$ where $\rho_A:(T^1_k)^*M\ra
(T^1_{k-1})^*M$ is the projection given by
$$\rho_A(\alpha_1,\ldots,\alpha_k)=(\alpha_1,\ldots,\alpha_{A-1},
\alpha_{A+1},\ldots,\alpha_k).$$

The integrability of these structures is characterized by

\begin{prop}
An almost $k$-cotangent structure $(\omega_A,V_A;1\leq A\leq k)$ on $M$ is
integrable  if and only if the $2$-forms $\omega_A$ are closed and all
distributions   $V_{A_1}\oplus \dots  \oplus V_{A_k}$ are involutive.
\end{prop}
\rem  It can be proved  that an integrable almost $k$-cotangent structure
on a manifold $M$ is a $k$-symplectic structure on $M$ setting $V=
\displaystyle \oplus_{A=1}^k V_A$.

\subsection{The $n$-symplectic structure of $LM$}

The frame bundle $LM$ has a canonical $n$-symplectic structure   given
by
$\omega_i= -d\theta^i,\, V=\ker \lambda_M$ 
where $\theta^i$ are the components of the {\it soldering one-form}
and  $V$ is the vertical distribution. This structure was first introduced in
\cite{No1,No2} under the name  {\em generalized symplectic geometry on $LM$}, and later referred
to as $n$-symplectic geometry in \cite{No3}.
$n$-symplectic geometry is the generalized geometry that
one obtains on $LM$ when $d\hat \theta=d\theta^i \hat r_i$  is taken as a 
generalized symplectic 2-form.  The structure is rich enough to allow the definition of
generalized Poisson brackets and generalized Hamiltonian vector fields.  The ideas are
"generalized" in the sense that the observables of the theory are vector-valued on $LM$ rather
than $\r$-valued.  Moreover the generalized Hamiltonian vector fields are equivalence classes
of vector-valued vector fields. The details of this geometry in the more general case of a {\em
general $n$-symplectic manifold} are given in Section 9 of this paper.

The relationship between $n$-symplectic geometry on the bundle of linear
frames $LM$   and canonical symplectic geometry on the cotangent bundle
$T\sp *M$ has been developed in \cite{No3}, showing that the ordinary
symplectic geometry of $T\sp *M$ can be induced from the $n$-symplectic
geometry of $LM$ using the associated bundle construction.  This relationship will be discussed
further in Section 5.1.

  In \cite{FLN2} it is shown that $m$-symplectic geometry on frame bundles
can be viewed as a "covering theory" for the Hamiltonian formulation of
 field theory (multisymplectic manifolds). This relationship will be discussed
 in Section 7.4.  

Also in \cite{No4} it is shown that the 
Schouten-Nijenhuis brackets   of both symmetric and  antisymmetric
contravariant tensor fields have a natural geometrical interpretation in
terms of $n$-symplectic geometry on the bundle of linear frames $LM$.  Specifically, the
restriction of the $n$-symplectic Poisson bracket to the subspace of GL(n)-tensorial functions
is in fact the lift to $LM$ of the Schouten-Nijenhuis brackets.    See Section 9.4.

\subsection{$k$-cosymplectic structures
and $\r^k \times (T^1_k)^*M$}

   Let us  begin by recalling that a
    cosymplectic manifold is a triple $(M,\theta,\omega)$ consisting of a
   smooth $(2n+1)$-dimensional manifold $M$ with a closed $1$-form
   $\theta$ and a closed $2$-form $\omega$, such that $\theta \wedge
   \omega^n \neq0$. The standard example of a cosymplectic manifold is
   provided by  $(J^1(\r ,N)\equiv \r \times T^*N,dt,\pi^*\omega_0)$, with
   $t:\r \times T^*N \to \r$ and $\pi:\r \times T^*N \to  T^*N$ the
   canonical projections and $\omega_0$ the canonical symplectic form on
   $T^*N$.

\begin{definition}\label{deest}
Let $M$ be a differentiable manifold  of dimension $(k+1)n+k$. A family
$(\eta_A,\omega_A,V;1\leq A\leq k)$, where each $\eta_A$ is a closed
$1$-form, each $\omega_A$ is a closed $2$-form and $V$ is an
$nk$-dimensional integrable distribution on $M$, such that
\begin{enumerate}
\item $\eta_1\wedge\dots\wedge\eta_k\neq 0$,\quad $\eta_{A_{\rfloor
V}}=0,\quad\omega_{A_{\rfloor V\times V}}=0,$

\item $(\displaystyle {\cap_{A=1}^{k}} \ker\eta_A) \cap (\displaystyle
{\cap_{A=1}^{k}} \ker\omega_A)=\{0\}$, \quad  $\dim(\displaystyle
{\cap_{A=1}^{k}} \ker\omega_A)=k,$
\end{enumerate}
is called a {\rm $k$--cosymplectic structure} and the manifold $M$ a
{\rm $k$--cosymplectic manifold}.
\end{definition}

The canonical model for these geometrical structures is ${\r^k}\times
(T^1_k)^*M=J^1(M,\r^k)$. 
Let $J^1(M,{\r^k})$ be the $(k+(k+1)n)$-dimensional manifold of one jets
from $M$ to ${\r^k}$, with elements denoted by $j^1_{x,t}\sigma$. We
recall that one jets of mappings from $M$ to  $\r^k$ can be identified
with the manifold $J^1\pi$ of one jets of sections of the trivial bundle
$\pi: \r^k \times M \rightarrow M$.

$J^1 \pi$ is diffeomorphic to ${\r^k} \times (T^1_k)^*M $ via the
diffeomorphism  given by
$$
j^1_x\sigma\in   J^1\pi \rightarrow (\sigma (x),j^1_{x,0}\sigma_x)\in
\r^k \times (T^1_k)^*M\, ,$$ where $\sigma_x(\tilde{x})= \sigma
(\tilde{x})-\sigma(x)$ and $\tilde{x}$ denotes an arbitrary point in $M$.

Let   $\tau^*:\r^k \times (T^{1}_{k})^*M \rightarrow M$ denote the
canonical projection.  If $(x^i)$ are local coordinates on $U \subseteq M$
then the induced local coordinates $(t^A,x^i , p^A_i),\, 1\leq i \leq n,\,
1\leq A \leq k$, on $(\tau^*)^{-1}(U)\equiv {\r^k} \times (T^1_k)^*U$ are
given by
$$
t^A(j^1_x\sigma)=t^A,\qquad  x^i(j^1_x\sigma)=x^i(x),\qquad
   p^A_i(j^1_x\sigma)=d(\sigma^A_x)(x)(\frac{\displaystyle
\partial}{\displaystyle \partial x^i}_{\vert{x}} ) 
$$
where $\sigma^A_x=\pi_A\circ \sigma_x$.

An ${\r^k}$-valued $1$-form $\eta_0$ and an ${\r^k}$-valued $2$-form
$\omega_0$ on ${\r^k}\times (T^1_k)^*M$ are defined by
\begin{equation}\label{canon}
\eta_0=\displaystyle\sum_{A=1}^m(\eta_0)_A\,  {\hat{r}_A}
=\displaystyle\sum_{A=1}^k((\pi^1_A)^*dt)\,  {\hat{ r}_A},\quad \omega_0
=\displaystyle\sum_{A=1}^k (\omega_0)_A\, {\hat{r}_A}
=\displaystyle\sum_{A=1}^m(\pi^2_A)^*(\omega_M)\,  {\hat{r}_A}
\end{equation}
where $\pi^1_A:{\r^k} \times (T^1_k)^*M \rightarrow {\r}$ and $\pi^2_A:
{\r^k} \times (T^1_k)^*M \rightarrow T^*M$ are the projections defined by
$$
\pi^1_A((t^B),(p^B))=t^A \,,\quad  \pi^2_A((t^B),(p^B))=p^A\, ,
$$
and $\omega_M$ is the canonical symplectic form on $T^*M$.
In local coordinates we have
\begin{equation}\label{locexp}
(\eta_0)_A=dt^A, \quad (\omega_0)_A = \displaystyle \sum_{i=1}^m dx^i
\wedge dp^A_i\, \quad  1\leq A \leq k
\end{equation}

\noindent Moreover, let   $V=\ker T\mu^*$, where $\mu^* : {\r^k} \times (T^1_k)^*M
\rightarrow {\r^k} \times M$. Then locally
$$
 V=\langle\frac{\displaystyle\partial} {\displaystyle\partial p^1_{i}},
\dots, \frac{\displaystyle\partial}{\displaystyle\partial p^k_{i}}\rangle
\quad 1\leq A \leq k\, .
$$
and the  canonical $k$-cosymplectic structure on ${\r^k} \times
(T^1_k)^*M$  is   $((\eta_0)_A,(\omega_0)_A, V)$. Indeed a simple
computation in local coordinates shows that the forms
$((\eta_0)_A,(\omega_0)_A,V)$ satisfy  the conditions of Definition
\ref{deest} .

For any $k$-cosymplectic structure $(\eta_A,\omega_A,V)$ on $M$, there
exists a family of $k$ vector fields $(\xi_1,\dots,\xi_k)$
characterizated by the conditions
$$
\eta_A(\xi_B)=\delta_{AB},\quad \iota_{\xi_B}\omega_A=0,
$$
for all $1\leq A,B\leq k$. These vector fields are called the {\it Reeb
vector fields} associated to the $k$-cosymplectic structure.

\begin{thm}\cite{mmopm}
Let $(\eta_A,\omega_A,V,1\leq A\leq k)$ be a $k$-cosymplectic structure on
$M$. About every point of $M$ we can find a local coordinate system   $(t^A,x^i,p^A_i)$ such that
$$(\eta_0)_A=dt^A, \quad (\omega_0)_A = \displaystyle \sum_{i=1}^n dx^i
\wedge dp^A_i\,,\quad V=\langle\frac{\displaystyle\partial}
{\displaystyle\partial p^1_{i}}, \dots,
\frac{\displaystyle\partial}{\displaystyle\partial p^k_{i}}\rangle \quad
1\leq A \leq k\, ,
$$
and the Reeb vector fields are given by $\xi_A = \frac{\partial}{\partial
t^A}$.
\end{thm}

\subsection{Multisymplectic structures}

In $k$-symplectic geometry the model is the Whitney sum of $k$-copies of
the cotangent bundle of a manifold $M$. In multisymplectic geometry
\cite{GIMMsy,GIMMsy2,KijoSz,KijTul} one uses a completely different model.

   Let $E$ be an $m$-dimensional differentiable manifold and denote by
   $\bigwedge^kE$ the bundle of exterior $k$-forms on $E$ with canonical
   projection $\rho_k :\bigwedge^kE \to E$. Notice that
   $\bigwedge^1E=T^*E$.

\noindent    On $\bigwedge^kE$ there exists a canonical $k$-form $\Theta_E$ defined by
   $$(\Theta_E)_\alpha(v_1,\dots,v_k)=\alpha(T\rho_k(v_1),\dots,T\rho_k(v_k))$$
   for $\alpha\in \bigwedge^kE$ and
   $v_1,\dots,v_k\in T_\alpha(\bigwedge^kE) \, .$
   This is a direct extension of the construction of the canonical
   Liouville $1$-form on a cotangent bundle.

   Next, we define a $(k+1)$-form $\Omega_E$ by
   $$\Omega_E \, = \, -\, d\Theta_E \, .$$
   Taking bundle coordinates $(x^i,p_{i_1\ldots i_k})$, $1\leq i\leq m$,
   $1\leq i_1 < \dots <i_k\leq m$, on $\bigwedge^kE$, we have
   $$\Theta_E = p_{i_1\ldots i_k} dx^{i_1}\wedge \dots \wedge dx^{i_k}\,
,\quad
   \Omega_E = - dp_{i_1\ldots i_k} \wedge dx^{i_1}\wedge \dots \wedge
   dx^{i_k}. $$

       Assume that $E$ itself is
      fibered over some manifold $M$, with projection $\pi:E\to M$. For
      any $r$, with $0\leq r\leq k$, let $\bigwedge^k_rE$ denote the
      bundle over $E$ consisting of those exterior $k$-forms on $E$ which
      vanish whenever $r$ of its arguments are vertical tangent vectors
      with respect to $\pi$. Obviously, $\bigwedge^k_rE$ is a vector
      subbundle of $\bigwedge^kE$, and we will denote by $i_{k,r}:\bigwedge^k_rE
      \rightarrow \bigwedge^kE$ the natural inclusion.

         The restriction of $\Theta_E$ and $\Omega_E$ to $\bigwedge^k_rE$
         will be denoted by $\Theta_E^r$ and $\Omega_E^r$, respectively;
         that is
         $$\Theta_E^r = i_{k,r}^*\Theta_E ,\qquad \Omega_E^r = i_{k,r}^*\Omega_E  \, , $$
         and, clearly, $\Omega_E^r= -d\Theta_E^r$ .

         Based on  the properties of the $(k+1)$-forms $\Omega_E$ and
         $\Omega_E^r$, we introduce the following definition.

\begin{definition}
A closed $(k+1)$-form $\alpha$ on a manifold $N$ is called
{\rm multisymplectic} if it is non-degenerate in the sense that for a tangent
vector $X$ on $N$, $X \hook \alpha = 0 \, \mbox{ if and only if} \, X=0
\, .$ The pair $(N,\alpha)$ will then called a {\rm multisymplectic manifold}.
\end{definition}

Of course  the manifolds $(\bigwedge^kE,\Omega_E)$ and
$(\bigwedge^k_rE,\Omega_E^r)$, $0\leq r \leq k$, are multisymplectic.

To develop the  
multisymplectic   formalism of field theory we will   use
the  canonical multisymplectic manifold $(\bigwedge^n_2E,\Omega^2_E)$ and
the manifold $(\bigwedge^n_1E,\Omega^1_E)$. If $M$ is oriented with volume
form $\omega$ we can consider coordinates $(x^i,y^ A)$ on $E$ such that
$\omega=d^nx=dx^1\wedge\dots\wedge dx^n$.

 \noindent Elements of
$\bigwedge^n_1E$ and $\bigwedge^n_2E$ can be written, respectively,  as
   follows
   $$ p\, d^nx, \qquad  p\, d^nx + p^i_A dy^A\wedge d^{n-1}x_i$$
   where
$d^{n-1}x_i=\frac{\partial}{\partial x^i}\hook d^nx$.
      Then we take local coordinates
   $(x^i,y^A,p)$\,  on   $\bigwedge^n_1E$\,  and \, $(x^i,y^A,p, p^i_A)$\,  on
   $\bigwedge^n_2E$. Therefore the canonical
   multisymplectic $(n+1)$-form  $\Omega^2_E$ on $\bigwedge^n_2E$ is locally given by
\begin{equation}\label{theta2e}
\Omega^2_E= -dp \wedge d^nx - dp^i_A \wedge dy^A\wedge d^{n-1}x_i\,
\end{equation}
and  $\Theta^2_E=p\, d^nx + p^i_A dy^A\wedge d^{n-1}x_i $.

{\bf Remark} In \cite{Cant1} the authors have developed a geometrical
study of multisymplectic manifolds, exhibiting the complexity of a
classification. A characterization of multisymplectic manifolds which are
exterior bundles can be found in \cite{Cant2}.

\section{Relationships among the cotangent-like structures}

Here we show how the symplectic, $k$-symplectic, $m$-symplectic and
similar structures are related.  We also venture  further into the realm of
multisymplectic geometry by showing how the canonical $k$-symplectic
structure is induced from a special case of the multisymplectic structure
on $\cojetpi$.  We use here the definition of $\cojetpi$ given in \cite{CCI} rather than
the affine dual definition of $\cojetpi$ given in \cite{GIMMsy}.

\subsection{Relationships among $T^*M$, $(T^1_k)^*M$, and $LM$}

In Section 4.2 we have already seen the relationship between the
$k$-symplectic structure on $(T^1_k)^*M$  and the symplectic structure on
$T^*M$. 
   The relationship between the canonical symplectic structure on
   $T^*M$ and the soldering form on $LM$ can be found in
    \cite{No3}: if $\theta_0$ is the Liouville $1$-form on 
$T^*M$ and $\theta$  the soldering $1$-form on $LM$ then

$$
  (\theta_0)_{[ u , \alpha ] } (\bar{X}_{[ u, \alpha ]}) \, = \,
  \alpha  (\theta_u (X_u))
,\quad [u,s]\in T^*M\equiv LM\times_{Gl(n,\mathbb{R})}(\mathbb{R}^n)^*\, .
$$
In this equation $u$ is a point in $LM$,  $[u,\alpha]$ denotes a point
(equivalence class) in $T^*M$ thought of as the associated bundle
$LM\times_{GL(m,\mathbb{R})}(\mathbb{R}^n)^*$ and
 $$
    \bar{X}_{ [ u, \alpha ]}  \in T_{ [ u, \alpha ]} T^*M \, ,
\quad  X_u \in T_u(LM)
 $$
  are vectors that project to the same vector on $M$, and $\alpha\in \rn^*$ is non-zero.

\subsection{The multisymplectic  
form   and the canonical $k$-symplectic structure}

Now we shall describe the relationship between the canonical
multisymplectic form $\Omega_E^2$ on $\bigwedge^k_2E$ and the canonical
$k$-symplectic structure on $(T^1_k)^*M$ when $E$ is the trivial bundle
$E=\r^k \times M \rightarrow \mathbb{R}^k$. In  this case
$\bigwedge^k_2E$ is diffeomorphic to $ \r^k \times \mathbb{R}
\times (T^1_k)^*M$. Let us recall that $\bigwedge^k_2E$ is   the vector
bundle
$$\Lambda^k_2( \mathbb{R}^k\times M) = \{
 \alpha_{(t,x)} \in \Lambda^k(\r^k\times M) \, : \,
 v \hook w \hook \alpha_{(t,x)} =0 \, \forall v,w\in(V\pi)_{(t,x)} \}$$
where $V\pi$ is the vertical fiber bundle corresponding to $\pi$.
\bigskip

   We  define
$$
\begin{array}{cccc}
\Psi : & \Lambda^k_2(\mathbb{R}^k\times M) & \longrightarrow &
\mathbb{R}^k \times \mathbb{R} \times (T^1_k)^*M \\
 & \alpha_{(t,x)} & \rightarrow & (t,r,(\alpha^1)_x, \dots ,(\alpha^k)_x)
\end{array}
$$
where
$$
r = \alpha_{(t,x)} (\frac{\partial}{\partial t^1} (t,x), \dots,
\frac{\partial}{\partial t^k} (t,x) )
$$
and
$$
(\alpha^B)_x(-) = i^*_t ( \alpha_{(t,x)} (\frac{\partial}{\partial t^1}
(t,x), \dots,
 \frac{\partial}{\partial t^{B-1}} (t,x), - ,
 \frac{\partial}{\partial t^{B+1}} (t,x), \dots,
 \frac{\partial}{\partial t^k} (t,x))) \, \quad 1\leq B \leq k,
$$
where $i_t : M\rightarrow R^k \times M$ denotes the inclusion $x
\rightarrow (t,x)$.

   The inverse of $\Psi$
$$
\begin{array}{cccc}
\Psi^{-1} : & \mathbb{R}^k \times \mathbb{R} \times (T^1_k)^* M &
\longrightarrow &
 \Lambda^k_2(\mathbb{R}^k\times M) \\
 & (t,r,(\alpha^1)_x, \dots ,(\alpha^k)_x) & \mapsto & \alpha_{(t,x)}
\end{array}
$$
is given by
$$
\alpha_{(t,x)}= r (d^k t)_{(t,x)} + (pr^*_2)_{(t,x)}((\alpha^B)_x) \wedge
(d^{k-1} t^B)_{(t,x)}
$$
 where
$$ d^kt=dt^1\wedge \dots \wedge dt^k, \qquad
d^{k-1}t^B=\frac{\partial}{\partial t^B} \hook dt^k$$ and $pr_2
:\mathbb{R}^k \times M \rightarrow M$ is the canonical projection.

   Elements of $\bigwedge^k_2E$ can be written uniquely as
   $$p^B_i \, dx^i \wedge d^{k-1}t^B \, + \,p\,
d^kt$$ where $(x^i)$  are coordinates on $M$. Let us denote by
$(t_B,p,x^i,p^A_i)$ the corresponding coordinates on  $\bigwedge^k_2E
\equiv  \mathbb{R}^k \times \mathbb{R} \times (T^1_k)^*M$ . Locally
$\Psi$ is written as the identity.

The canonical $k$-form on $\bigwedge^k_2E\equiv
  \mathbb{R}^k \times \mathbb{R} \times (T^1_k)^*M $
is locally  given in this case by
\begin{equation}\label{locteta}
   \Theta^2_E = p^B_i \, dx^i \wedge d^{k-1}t^B \, + \,p\,
d^kt
\end{equation}
\noindent and the corresponding
 canonical multisymplectic $(k+1)$-form $\Omega^2_E = -d\Theta^2_E$
is locally given by
$$ \Omega^2_E =   dx^i \wedge dp^B_i \wedge d^{k-1}t^B - dp \wedge d^kt
$$
   Let $i:(T^1_k)^*M \rightarrow
  \mathbb{R}^k \times \mathbb{R} \times (T^1_k)^*M$ be the natural inclusion.
We define on
$(T^1_k)^*M $
the 1-forms $\lambda_B$, $1\leq B \leq k$, by
$$\lambda_B(-) = i^*(\Theta^2_E(\frac{\partial}{\partial t^1},
\dots, \frac{\partial}{\partial t^{B-1}}, \, - \,
,\frac{\partial}{\partial t^{B+1}},\dots, \frac{\partial}{\partial
t^k}),$$ \noindent and   from (\ref{locteta}) we  deduce $   \lambda_B =
p^B_i \, dx^i $. Hence $\lambda_B$
  is the Liouville
 form   on the $B$-th copy  $T^*M$ of $(T^1_k)^*M$.
To get this local expression  apply $\lambda_B$  to the partials
$\partial / \partial x^i$ and $\partial / \partial p^B_i$.

   Therefore the $2$ forms
$$\omega_B = -d\lambda_B = dx^i \wedge dp^B_i , \quad 1\leq B \leq k$$
\noindent define  the canonical $k$-symplectic structure on
$(T^1_k)^*M$, and $\omega_B$ can also be defined  as follows
\begin{equation}\label{relatomega}
\omega_B(-,-)) = i^*(\Omega^2_E(-,\frac{\partial}{\partial t^1}, \dots,
\frac{\partial}{\partial t^{B-1}}, \, - \, ,\frac{\partial}{\partial
t^{B+1}},\dots, \frac{\partial}{\partial t^k}),
\end{equation}
   The case $k=1$ gives us the canonical symplectic structure of
$T^*M$.

\begin{prop}
The relationship between the 2-forms of the canonical $k$-symplectic
structure on $(T^1_k)^*M $ and the  canonical multisymplectic form
 $\Omega^2_E$
is given by (\ref{relatomega}).
 \blob
\end{prop}

\section{ Field Theory on $k$-symplectic and $k$-cosymplectic manifolds }

Here we discuss the polysymplectic formalism \cite{gun} for Hamiltonian and Lagrangian field
theory using   $k$-symplectic manifolds.  We discuss the G\"{u}nther's formalism (autonomous case)
 using the
   $k$-symplectic structures and the $k$-tangent structures. The non
   autonomous case will be developed using the $k$-cosymplectic
   structures and the stable $k$-tangent structures \cite{mmm,mmopm}.


\subsection{$k$-vector fields}

Let $M$ be an arbitrary manifold and
$\tau : T^{1}_{k}M \longrightarrow M$
its $k$-tangent bundle .

\begin{definition}\label{kvector}
A section ${\bf X} : M \longrightarrow T^1_kM$ of the projection
$\tau$ will be called a  {\rm $k$-vector field} on $M$.
\end{definition}

Since $T^{1}_{k}M$ can be canonically identified with the
Whitney sum $T^{1}_{k}M \equiv TM \oplus \dots \oplus TM$
of $k$ copies of
$TM$,
  we deduce that a $k$-vector field ${\bf X}$ defines
a family of vector fields $X_{1}, \dots, X_{k}$ on $M$.

\begin{definition}\label{integsect}
An {\rm integral section} of the $k$-vector field ${\bf X}$ on $M$ is a map
  $\phi:U\subset \rk \rightarrow M$, where $U$ is an
open  subset of $\rk$
such that
$$
\phi_{*}(t)(\frac{\displaystyle\partial}{\displaystyle\partial t^A})
= X_{A}(\phi (t)) \, \quad \forall  t\in U,
\quad 1\leq A \leq k,$$
 or equivalently,    $\phi$ satisfies
\begin{equation}\label{integs}
X\circ\phi=\phi^{(1)},
\end{equation}
where  $\phi^{(1)}$ is the first  prolongation of $\phi$ defined by
$$
\begin{array}{rccl}\label{1prolong}
\phi^{(1)}: & U\subset \rk & \longrightarrow & T^1_kM \\ \medskip
 & t & \longrightarrow & \phi^{(1)}(t)=j^1_0\phi_t
\end{array}
$$
where $\phi_t (s)=\phi (s+t)$ for all $t,s \in \mathbb{R}$. If ${\bf X}$
has an integral section, ${\bf X}$ is said to be {\rm integrable}.
 \end{definition}

{\bf Remark}
Let us consider the trivial bundle
$\pi:E=R^k\times M \rightarrow R^k$.
  A jet field $\gamma$ on $\pi$ (see \cite{Saunders}) is
a section of the  projection
$
\pi_{1,0}:J^1\pi
\equiv \rk \times T^1_kM
\longrightarrow  E\equiv \rk \times M  \, .
$
    If we    identify each $k$-vector field ${\bf X}$ on $M$
with the jet field    $\gamma = (id_{\rk}, X)$ , that is $
\gamma(t,x)=(t,X_1(x),\dots , X_k(x))$, then  the  integral sections of
the jet field $\gamma$ correspond,   as defined by G\"{u}nther, to the {\em solutions}
of the $k$-vector field ${\bf X}$.

We remark that if $\phi$ is an integral section of a $k$-vector
field $(X_1,\dots,X_k)$ then each curve on $M$ defined by
$\phi_A=\phi\circ h_A$, where $h_A: \rne \rightarrow \rk$ is
the natural inclusion $h_A(t)=(0,\dots,t,\dots,0)$, is an integral
curve of the vector field $X_A$ on $M$, with $1\leq A\leq k$.
 We refer to  \cite{bar2,bar3} for a discussion on
the existence of integral sections.

\subsection{Hamiltonian formalism  and $k$-symplectic  structures}

In this section, following the ideas of G\"{u}nther \cite{gun}, we will
describe the Hamilton equations, for an autonomous Hamiltonian, in terms
of the geometry of $k$-symplectic structures, showing that the role
played by  symplectic manifolds  in classical mechanics  
is here played
by the $k$-symplectic manifolds.

Let $(M,\omega_A,V;1\leq A\leq k)$  be
 a $k$--symplectic manifold. Since $M$ is
a polysymplectic manifold let us consider the vector bundle morphism
defined by G\"{u}nther:
\begin{equation}\label{sharp}
\begin{array}{rccl}
\Omega^{\sharp}: & T^1_kM & \longrightarrow & T^*M  \\
\noalign{\medskip}& (X_1,\dots,X_k) & \longrightarrow &
\Omega^{\sharp}(X_1,\dots,X_k) =  \displaystyle \sum_{A=1}^k \, X_A \hook
\omega_A \, .
\end{array}
\end{equation}

\begin{definition} 
Let $H:M\longrightarrow \r$ be a function on $M$.
 Any $k$-vector field $(X_1,\dots,X_k)$ on $M$ such that
$$
 \Omega^{\sharp}(X_1,\dots,X_k)= dH
$$
will be called an {\rm evolution $k$-vector field} on $M$ associated
with the Hamiltonian function $H$.
\end{definition}

It should be noticed   that in general the solution to the above
equation is not unique.
Nevertheless, it can be proved \cite{mmopm} that there always exists an {\it evolution
$k$-vector field} associated with a Hamiltonian function $H$.

Let $(x^i,p^A_i)$ be a local coordinate system on $M$.  Then we have
\begin{prop}
If $(X_1,\dots,X_k)$ is an integrable evolution $k$-vector field
associated to $H$ then its integral sections
\[
\begin{array}{rccl}
\sigma: & \rk & \longrightarrow & M \\ \noalign{\medskip} & (t^B) &
\longrightarrow & (\sigma^i(t^B),\sigma^A_i(t^B)),
\end{array}
\]
are solutions of the classical local
 Hamilton equations associated with a regular multiple
integral variational problem \cite{Rund}:
$$\frac{\displaystyle \partial H}{\displaystyle\partial x^i}=
-\sum_{A=1}^k\frac{\displaystyle \partial\sigma^A_i} {\displaystyle
\partial t^A}, \quad \frac{\displaystyle \partial H} {\displaystyle
\partial p^A_i}= \frac{\displaystyle \partial\sigma^i}{\displaystyle
\partial t^A}, \quad 1\leq i\leq n, \, 1\leq\ A \leq k\, .
$$
\end{prop}

\subsection{Hamiltonian formalism and $k$-cosymplectic structures}

In this section we will describe the Hamilton equations for a
non-autonomous Hamiltonian in terms of the geometry of $k$-cosymplectic
structures, showing that the role played by  cosymplectic
 manifolds  in
classical mechanics (see \cite{ grad,albert,chlm}) is here
played by the $k$-cosymplectic manifolds.

Let $(M,\eta_A,\omega_A,V;1\leq A\leq k)$ be
a $k$--cosymplectic manifold. Let us consider the vector bundle morphism
defined by :
\begin{equation}\label{sharpco}
\begin{array}{rccl}
\Omega^{\sharp}: & T^1_kM & \longrightarrow & T^*M  \\
\noalign{\medskip}& (X_1,\dots,X_k) & \longrightarrow &
\Omega^{\sharp}(X_1,\dots,X_k) =  \displaystyle \sum_{A=1}^k \, X_A \hook
\omega_A +  \eta_A(X_A)\eta_A\, .
\end{array}
\end{equation}

\noindent  
 Let
$\xi_A$ the Reeb vector fields associated to the $k$-cosymplectic
structure $(\eta_A,\omega_A,V)$. Notice here that the hamiltonian  $H(t^A,x^i,p^A_i)$ is
non-autonomous.
\begin{definition}
Let $H:M\longrightarrow {\r}$ be a function on $M$.
 Any $k$-vector field $(X_1,\dots,X_k)$ on $M$ such that
$$
 \eta_A(X_B)=\delta^A_B,\quad \Omega^{\sharp}(X_1,\dots,X_k)=
 dH+\sum_{iA=1}^k(1-\xi_A(H))\eta_A
$$ 
 will be called an {\rm evolution $k$-vector field} on $M$ associated
with the Hamiltonian function $H$ for all $1\leq A,B\leq k$.
\end{definition}   

 It should be noticed that in general the solution to the above
equation is not unique. Nevertheless, it can be proved ~\cite{mmopm} that there always
exists an evolution $k$-vector field associated with a Hamiltonian
function $H$.

Let $(t^A,x^i,p^A_i)$ be a local coordinate system on $M$.  Then we have
\begin{prop}
If $(X_1,\dots,X_k)$ is an integrable evolution $k$-vector field
associated to $H$ then its integral sections
\[
\begin{array}{rccl}
\sigma: & \rk & \longrightarrow & M \\ \noalign{\medskip} & (t^B) &
\longrightarrow & (\sigma^A(t^B),\sigma^i(t^B),\sigma^A_i(t^B)),
 \end{array}
\]
satisfy $\sigma^A(t^1,\dots,t^k)=t^A$ and are solutions of the classical
local Hamilton equations associated with a regular multiple
integral variational problem \cite{Rund}:
$$\frac{\displaystyle \partial H}{\displaystyle\partial x^i}=
-\sum_{A=1}^k\frac{\displaystyle \partial\sigma^A_i} {\displaystyle
\partial t^A}, \quad \frac{\displaystyle \partial H} {\displaystyle
\partial p^A_i}= \frac{\displaystyle \partial\sigma^i}{\displaystyle
\partial t^A}, \quad 1\leq i\leq n, \, 1\leq\ A \leq k\, . \blob
$$
\end{prop}

\subsection{Second Order Partial Differential Equations on $T^1_kM$}

The idea of this subsection is to characterize the integrable $k$-vector
fields on $T^1_kM$ such that their integral sections are canonical
prolongations of maps from $\r^k$ to $M$.

 \begin{definition}\label{sode0}
A $k$-vector field on   $T^1_kM$, that is, a section $\xi\equiv
(\xi_1,\dots,\xi_k):T^1_kM\rightarrow T^1_k(T^1_kM)$ of the projection
$\tau_{T^1_kM}:T^1_k(T^1_kM)\rightarrow T^1_kM$, is  a {\rm Second Order
Partial Differential Equation (SOPDE)} if and only if it is  also a section
of the vector bundle $T^1_k\tau_M:T^1_k(T^1_kM)\rightarrow T^1_kM$, where
 $T^1_k(\tau_M)$ is defined
by $T^1_k(\tau_M)(j^1_0\sigma)=j^1_0(\tau_M \circ \sigma)$.
\end{definition}

Let $(x^i)$ be a coordinate system on $M$ and $(x^i,v^i_A)$  the induced
coordinate system on $T^1_kM$.  From the definition  we deduce that the
local expression of a SOPDE  $\xi$ is
\begin{equation}\label{localsode1}
\xi_A(x^i,v^i_A)=v^i_A\frac{\displaystyle \partial} {\displaystyle
\partial x^i}+
(\xi_A)^i_B \frac{\displaystyle\partial} {\displaystyle
\partial v^i_B},\quad 1\leq A \leq k   .
\end{equation}

We recall that the first prolongation  $\phi^{(1)}$   of
$\phi:U\subset \r^k \to M$ is  defined by
$$
\begin{array}{rccl}
\phi^{(1)}: & U\subset \r^k & \longrightarrow & T^1_kM) \\
\medskip
 & t & \longrightarrow & \phi^{(1)}(t)=j^1_0\phi_t
\end{array}
$$
where $\phi_t (s)=\phi (s+t)$ for all $t,s \in \mathbb{R}$. In local
coordinates:
\begin{equation}\label{localfi11}
\phi^{(1)}(t^1, \dots, t^k)=( \phi^i (t^1, \dots, t^k),
\frac{\displaystyle\partial\phi^i}{\displaystyle\partial t^A} (t^1,
\dots, t^k)), \qquad  1\leq A\leq k\, ,\,  1\leq i\leq n \, .
\end{equation}

\begin{prop}
Let $\xi$ an integrable  $k$-vector field on $T^1_kM$. The necessary and
sufficient condition for $\xi$ to be a Second Order Partial Differential
Equation (SOPDE) is that  its integral sections are first prolongations
$\phi^{(1)}$ of maps $\phi :\r^k \to
 M$. That is
 $$\xi_A(\phi^{(1)}(t))=\phi^{(1)}_*(t)(\frac{\partial}{\partial t_A}) (t)$$
 for all $A=1,\dots,k$.
 These maps $\phi$ will be called solutions of the SOPDE $\xi$.
\end{prop}

>From (\ref{localfi11})  and   (\ref{localsode1}) we have
 \begin{prop}\label{solsemi}
$\phi:\r^k \rightarrow M$ is a solution of the  SOPDE $\xi=(\xi_1,
\dots,\xi_k)$, locally given by (\ref{localsode1}), if and only if
$$
\frac{\displaystyle\partial\phi^i} {\displaystyle\partial
t^A}(t)=v^i_A(\phi^{(1)}(t)),\qquad \frac{\displaystyle\partial^2\phi^i}
{\displaystyle\partial t^A\partial t^B}(t) = (\zeta_A)^i_B(\phi^{(1)}(t)).
$$
\end{prop}

If $\xi:T^1_kM\rightarrow T^1_kT^1_kM$ is an integrable SOPDE then for all
integral sections $\sigma:U\subset\r^k\rightarrow T^1_kM$ we have $
(\tau_M\circ\sigma)^{(1)}=\sigma $ where $\tau_M:T^1_kM \rightarrow M$ is
the canonical projection.

Now we show how to characterize the SOPDEs using the canonical $k$-tangent
structure of $T^1_kM$.

\begin{definition}
The {\rm canonical vector field} $C$ on
$T^1_kM$ is the infinitesimal generator of the one parameter group
$$
\begin{array}{ccc}
\r \times (T^1_kM) & \longrightarrow &
 T^1_kM  \\
\noalign{\medskip} (s,(x^i,v^i_B)) & \longrightarrow & (x^i,e^s
\,v^i_B)\, .
\end{array}
$$
Thus $C$ is locally expressed as follows:
\begin{equation}\label{localc}
 C =
 \displaystyle \sum_{B} C_B =   \sum_{i,B} v^i_B
\frac{\displaystyle\partial}{\displaystyle \partial v_B^i},
\end{equation}
where each $C_B$ corresponds with the canonical vector field on the
$B$-th copy of $TM$ on $T^1_kM$.
\end{definition}

Let us remark that each vector field  $C_A$  on $T^1_kM$ can also be
defined using the $A$-lifts of vectors as follows:
$\,\, C_A((v_1)_q,\ldots,(v_k)_q) =((v_A)_q)^A(v))\,$.

 From (\ref{localJA}), (\ref{localsode1}) and (\ref{localc}) we deduce the following

\begin{prop}\label{pr235}
A $k$-vector field  $\xi=(\xi_1,\dots,\xi_k)$ on $T^1_kM$ is a SOPDE  if
and only if
$$
J^A(\xi_A)=C_A,\hspace{1cm}\forall\,1\leq A\leq k,
$$
where $(J^1,\dots,J^k)$ is the canonical   $k$-tangent  structure on
$T^1_kM$.
\end{prop}

\subsection{Lagrangian formalism and $k$-tangent structures}

Given a Lagrangian function of the form $L=L(x^i,v^i_A)$ one obtains, by
using a variational principle, the {\it generalized Euler-Lagrange
equations} for $L$:
\begin{equation}\label{lageq1}
\displaystyle \sum_{A=1}^k\frac{\displaystyle d}{\displaystyle d t^A}
(\frac{\displaystyle\partial L}{\displaystyle \partial v^i_A})-
\frac{\displaystyle \partial L}{\displaystyle \partial x^i}=0, \qquad
v^i_A     =\frac{\displaystyle \partial
 x^i}{\displaystyle \partial t^A}.
\end{equation}

In this section, following the ideas of G\"{u}nther \cite{gun},
we will describe the above equations (\ref{lageq1})
in terms of the geometry of $k$-tangent structures.
In classical mechanics the  symplectic structure of
Hamiltonian theory and the tangent structure of Lagrangian theory
play complementary roles \cite{cram1,cram2,grif1,grif2,klein}.
 Our purpose in this  section is to show that the
 $k$-symplectic structures and the $k$-tangent structures
play similarly complementary roles.

First of all, we note that such a $L$ can be considered as a
function $L:T^1_kM \rightarrow  \mathbb{R}$ with $M$ a manifold   with local
coordinates $(x^i)$. Next, we construct a $k$--symplectic structure
on the manifold $T^1_kM$, using its canonical  $k$--tangent structure for
each $1\leq A\leq k$.  We consider:
\begin{itemize}
\item  the vertical derivation $\imath_{J^A}$
of type $\imath_*$ defined by $J^A$,
which is defined by
\begin{gather*}
\iota_{J^A   }f=0   \\
(\iota_{J^A }\alpha)(X_1,\dots,X_p)=\displaystyle \sum_{j=1}^p
\alpha(X_1,\dots,J^A   X_j,\dots,X_p)\,,
\end{gather*}
\noindent for any function $f$  and
any  $p$-form $\alpha$ on $T^1_kM$;
\item the vertical differentation $d_{J^A }$   
of forms on $T^1_kM$ defined by
$$
d_{J^A   }=[\imath_{J^A   },d]=\imath_{J^A   }\circ d - d \circ \imath_{J^A   }\,,
$$
where $d$ denotes the usual exterior differentation.
\end{itemize}

Let us consider the $1$--forms $(\beta_L)_A = d_{J^A   }L\, ,\, 1\leq A
\leq k$. In a local coordinate system $(x^i,v^i_A)$ we have
\begin{equation}\label{betaloc}
(\beta_L)_A=\frac{\displaystyle \partial
 L}{\displaystyle\partial v^i_A     } dx^i, \, \, 1\leq A\leq k.
\end{equation}

\begin{definition}
A Lagrangian $L$ is called {\rm regular} if and
only if
\begin{equation}\label{det0}
det (\frac{\displaystyle\partial^2L}
{\displaystyle\partial v^i_A
\partial v^j_B})\neq 0,
\qquad   1 \leq i,j,
 \leq n,\quad 1\leq A,B \leq k  \, .
\end{equation}
\end{definition}

By  introducing the
following $2$--forms
$(\omega_L)_A = -d(\beta_L)_A \,, \,  1\leq A \leq k$,
 one can easily prove the following.
\begin{prop}
$L:T^1_kM\longrightarrow\r$ is a regular Lagrangian if and only if
$((\omega_L)_1, \dots,(\omega_L)_k,V)$ is a  $k$-symplectic structure on
$T^1_kM$,
where $V$ denotes the
vertical distribution of $\tau:T^1_kM
\rightarrow M$. \blob
\end{prop}

Let $L:T^1_kM\longrightarrow \r$ be a regular Lagrangian  and let us
consider the $k$--symplectic structure $((\omega_L)_1,\dots
,(\omega_L)_k, V)$ on $ T^1_kM$ defined by $L$.
Let $\Omega_L^{\sharp}$ be the morphism defined by this $k$--symplectic
structure
$$\Omega_L^{\sharp}:   T^1_k(T^1_kM)
\longrightarrow
  T^*(T^1_kM).$$
Thus, we can set the following equation:
\begin{equation}\label{lageq0}
  \Omega^{\sharp}_L(X_1,\dots,X_k)=
dE_L,
\end{equation}
where $E_L=C(L)-L$, and where  $C$ is the canonical vector field of the
vector bundle $\tau:T^1_kM \to M$.

\begin{prop}
 Let $L$ be  a regular Lagrangian.
 If  ${\bf \xi} =(\xi_1,\cdots,\xi_k)$ is a solution of (\ref{lageq0}) then it is a SOPDE.
 In addition if ${\bf \xi}$ is
integrable  then the solutions of $\xi$  are solutions of the
Euler-Lagrange equations (\ref{lageq1}).
\end{prop}

\proof It is a direct computation in local coordinates using
(\ref{localsode1}), (\ref{localc}) , (\ref{betaloc}) and (\ref{det0}).
\blob.

\noindent {\bf Remark} The {\it Legendre map}
defined by G\"{u}nther \cite{gun}
 \[
FL: T^1_kM \longrightarrow (T^1_k)^*M
\]
can be described here as follows: if $v_ x= (v_{1}, \dots , v_{k})_x \in
(T^1_kM)_q$ with $q\in M$ and $v_{A} \in T_{q}M$, then ${\cal FL}(v_x ) =
(\tilde{v^{1}}, \dots ,\tilde{v}^{k}) \in (T^1_kM)^{*}_x$, where
$\tilde{v}^{A} \in T^{*}_{x}M$ is given by
\[
\tilde{v}^{A}(z) = (\beta_L)_A(\bar{z}), \quad 1 \leq A \leq k,
\]
for any $z \in T_xM$, where $\bar{z} \in T_{v_x }(T_{k}^{1}M)$ with
$\tau_*(\bar{z})=z$.

 From (\ref{betaloc}) we deduce that  $  FL $ is locally given by
\begin{equation}\label{locfl}
(x^i,v^i_A)  \longrightarrow  (x^i, \frac{\displaystyle\partial
L}{\displaystyle\partial v^i_A     }).
\end{equation}
and from (\ref{betaloc}) and (\ref{locfl})  we deduce the following
\begin{lem}\label{le811}
For every $1\leq A\leq k$, we have $(\omega_L)_A= 
FL^*\omega_A$, where $\omega_1,\dots,\omega_k$ are the $2$-forms
of the canonical
$k$--sym\-plectic structure of $(T^1_k)^*M$. \blob
\end{lem}

Then from (\ref{det}) we get that

\begin{prop}
Let $L$ be a Lagrangian. The following conditions are equivalent:

1) $L$ is regular.

2) {\cal FL} is a local diffeomorphism.

3) $((\omega_L)_1, \dots,(\omega_L)_k,V)$ is a  $k$-symplectic structure
on
$T^1_kM$. \blob
\end{prop}

\subsection{Second order partial differential equations on $\r^k\times
T^1_kM$}

The idea of this subsection is to characterize the integrable $k$-vector
fields on $\r^k \times T^1_kM$ such that their integral sections are
canonical prolongations of maps from $\r^k$ to $M$.

\begin{definition}\label{sode2}
A $k$-vector field on   $\r^k \times T^1_kM$, that is, a section
$\xi\equiv (\xi_1,\dots,\xi_k):\r^k \times T^1_kM\rightarrow T^1_k(\r^k
\times T^1_kM)$ of the projection $\tau_{\r^k \times T^1_kM}:T^1_k(\r^k
\times T^1_kM)\rightarrow \r^k \times T^1_kM$, is  a {\rm Second Order Partial
Differential Equation (SOPDE)} if and only if:

1) $dt^A(\xi_B) = \delta^A_B$

2) $Tpr_2 \circ \xi_B \circ i_t$ is a SOPDE on $T^1_kM$, $\forall t \in
\r^k$, where $pr_2: \r^k \times T^1_kM \rightarrow T^1_kM$ is the
canonical projection and $i_t: T^1_kM \rightarrow \r^k \times T^1_kM$ is
the canonical inclusion.
\end{definition}

 Let $(x^i)$ be a coordinate system on $M$ and $(t^A,x^i,v^i_A)$  the induced
coordinate system on $\r^k \times T^1_kM$.  From (\ref{localfi2}) we
deduce that the local expression of a SOPDE  $\xi$ is
\begin{equation}\label{localsode2}
\xi_A(x^i,v^i_A)=\frac{\partial}{\partial t_A}+v^i_A\frac{\displaystyle
\partial} {\displaystyle
\partial x^i}+
(\xi_A)^i_B \frac{\displaystyle\partial} {\displaystyle
\partial v^i_B},\quad 1\leq A \leq k
\end{equation}
where $(\xi_A)^i_B $ are functions on $\r^k \times T^1_kM$.

\begin{definition}\label{de652}
 For $\phi:\r^k \rightarrow M$   a map, we define the {\rm first prolongation}
$\phi^{(1)}$ of $\phi$ as the map
$$
\begin{array}{rclcc}
\phi^{(1)}:\r^k & \longrightarrow & J^1\pi & \equiv & \r^k \times T^1_kM
,\\ t & \longrightarrow & j^1_t\phi & \equiv & (t,j^1_0\phi_t)
\end{array}
$$
In local coordinates:
\begin{equation}\label{localfi2}
\phi^{(1)}(t^1, \dots, t^k)=(t^1, \dots, t^k, \phi^i (t^1, \dots, t^k),
\frac{\displaystyle\partial\phi^i}{\displaystyle\partial t^A} (t^1,
\dots, t^k)), \qquad  1\leq A\leq k\, ,\,  1\leq i\leq n \, .
\end{equation}
\end{definition}

\begin{prop}
Let $\xi$ an integrable  $k$-vector field   on $\r^k \times T^1_kM$. The
necessary and sufficient condition for $\xi$ to be a Second Order Partial
Differential Equation (SOPDE) is that  its integral sections are first
prolongations $\phi^{(1)}$ of maps $\phi :\r^k \to
 M$. That is
 $$\xi_A(\phi^{(1)}(t))=\phi^{(1)}_*(t)(\frac{\partial}{\partial t_A}) (t)$$
 for all $A=1,\dots,k$.

 These maps $\phi$ will be called {\rm solutions} of the SOPDE $\xi$.
\end{prop}

>From (\ref{localfi2}) and (\ref{localsode2}) we have
 \begin{prop}\label{solsemi2}
$\phi:\r^k \rightarrow M$ is a solution of the  SOPDE $\xi$, locally
given by (\ref{localsode2}), if and only if
$$
\frac{\displaystyle\partial\phi^i} {\displaystyle\partial
t^A}(t)=v^i_A(\phi^{(1)}(t)),\qquad \frac{\displaystyle\partial^2\phi^i}
{\displaystyle\partial t^A\partial t^B}(t) = (\zeta_A)^i_B(\phi^{(1)}(t)).
$$
\end{prop}

If $\xi$ is an integrable SOPDE then for all integral sections
$\sigma:U\subset\r^k\rightarrow \r^k \times  T^1_kM$ we have $
(\tau_M\circ\sigma)^{(1)}=\sigma $ where $\tau_M:\r^k \times T^1_kM
\rightarrow M$ is the canonical projection.
Now we show how to characterize the SOPDEs on $\r^k \times T^1_kM$ using
the canonical $k$-tangent structure of $T^1_kM$.
Let us consider on $\r^k \times T^1_kM$ the tensor fields
$\hat{J^1},\dots,\hat{J^k}$ of type  $(1,1)$, defined as follows:
$$
\hat{J^A}=J^A-C_A\otimes dt^A, \quad 1\leq A \leq k\, .
$$
where we have transported the canonical $k$-tangent structure
$(J^1,\ldots,J^k)$ of $T^1_kM$ to  ${\r^k} \times T^1_kM$.
\begin{prop} \label{de651}
A $k$-vector field  $\xi=(\xi_1,\dots,\xi_k)$ on $\r^k \times T^1_kM$ is
a SOPDE if and only if
$$
\hat{J^A}(\xi_A)=0,\qquad \bar{\eta}_A(\xi_B)=\delta_{AB},
$$
for all  $1\leq A,B\leq k$.
\end{prop}

{\bf Remark:}  Let us consider the trivial bundles $\pi:E=\r^k\times M
\rightarrow \r^k$ and  $\pi_1:\r^k \times T^1_kM\rightarrow  \r^k$. We
identify each SOPDE $(\xi_1,\ldots,\xi_k)$ with the following {\it
semi-holonomic second order jet field}
$$\begin{array}{rcl}
J^1\pi\equiv \r^k \times T^1_kM & \rightarrow & J^1\pi_1 \equiv \r^k
\times T^1_k(T^1_kM) \\ (t^A,q^i,v^i_A) & \rightarrow &
(t^A,q^i,v^i_A,v^i_A,(\xi_A)_B^i)
\end{array}
$$

If the SOPDE $\xi$ on $\r^k \times T^1_kM$ is integrable, then its
integral sections are canonical prolongations of maps from $\r^k$ to $M$
and then $\xi$ defines a second-order jet field $\Gamma$ on $\pi$ whose
coordinate representation of the corresponding connection
$\tilde{\Gamma}$ is
$$\tilde{\Gamma} = dt^A\otimes \left(
\frac{\displaystyle \partial}{\displaystyle \partial t^A} + v^i_A
\frac{\displaystyle \partial}{\displaystyle \partial q^i} + (\xi_A)^i_B
\frac{\displaystyle \partial}{\displaystyle \partial v_B^i} \right)\, ,$$
 since
$(\xi_A)^i_B=(\xi_B)^i_A$ (see \cite{Saunders}).

The integrability of the SOPDE is equivalent to the condition given by
${\cal R}=0$, where ${\cal R}$ is the curvature tensor of the above
connection (see \cite{bar2} and \cite{Saunders}).

\subsection{Lagrangian formalism and stable $k$-tangent structures}

Given a nonautonomous Lagrangian ${\cal L}={\cal L}(t^A,q^i,v^i_A)$ one
 realizes that such an ${\cal L}$ can be considered as a
function ${\cal L}:{\r^k}\times T^1_kM \rightarrow \r$.

In this section we shall give a geometrical description of Euler Lagrange
equations (\ref{lageq1}) using a  $k$-cosymplectic structure on
${\r^k}\times T^1_kM $ associated to the regular  Lagrangian ${\cal L}$. This
$k$-cosymplectic structure shall be  constructed using  the {\it canonical}
 tensor fields $\tilde{J^A},\, 1\leq A \leq k$ of
type $(1,1)$ on ${\r^k}\times T^1_kM $
defined by
$$
\tilde{J^A}=\frac{\displaystyle\partial}{\displaystyle\partial t^A}\otimes
dt^A+ J^A =\frac{\displaystyle\partial}{\displaystyle\partial t^A}\otimes
dt^A+ \sum_{i =1}^{n}\frac{\displaystyle\partial}{\displaystyle\partial
v^i_A} \otimes dq^i \,,\qquad 1\leq A \leq k \, ,
$$
where we have transported the canonical $k$-tangent structure
$(J^1,\ldots,J^k)$ of $T^1_kM$ to  ${\r^k} \times T^1_kM$. The family
$(\tilde{J^A},dt^A,\frac{\partial}{\partial t_A})$ is called the {\it
canonical stable $k$-tangent structure} on ${\r^k}\times T^1_kM $.

For each $1\leq A \leq k$, we define:
\begin{itemize}
\item  the vertical derivation $\imath_{J^A}$ of forms on ${\r^k}\times
T^1_kM$ by
$$\imath_{\tilde{J^A}}f=0\,, \quad
(\imath_{\tilde{J^A}}\alpha)(X_1,\dots,X_p)=\displaystyle\sum_{j=1}^p
\alpha(X_1,\dots,\tilde{J^A}X_j,\dots,X_p)\,, $$

\noindent for any function $f$ and any  $p$-form $\alpha$ on
${\r^k}\times T^1_kM$;

\item the vertical differentation $d_{\tilde{J^A}}$ of forms on
${\r^k}\times T^1_kM$ by
$$
d_{\tilde{J^A}}=[\imath_{\tilde{J^A}},d]=\imath_{\tilde{J^A}}\circ d - d \circ
\imath_{\tilde{J^A}}\;,
$$
where $d$ denotes the usual exterior differentation.
\end{itemize}

Let us consider the $1$--forms
$$(\beta_{\cal L})_A = d_{\tilde{J^A}}{\cal L} -\xi_A({\cal L}) dt^A,
\quad 1\leq A \leq k \; .$$ In bundle coordinates $(t^A,q^i,v^i_A)$ we
have
\begin{equation}\label{locbeta}
(\beta_{\cal L})_A=\displaystyle \sum_{i=1}^n\frac{\displaystyle\partial
{\cal L}}{\displaystyle\partial v^i_A} dq^i, \, \, 1\leq A\leq k \;.
\end{equation}

\begin{definition}
A Lagrangian ${\cal L}$ is called {\rm regular} if and only if the Hessian matrix
\begin{equation}\label{det}
\left(\frac{\displaystyle\partial^2 {\cal L}} {\displaystyle\partial v^i_A
\partial v^j_B} \right)
\end{equation}
is non--singular.
\end{definition}

Now, we introduce the following $2$--forms
$$
(\omega_{\cal L})_A = -d(\beta_{\cal L})_A \,, \,  1\leq A \leq k \; .
$$
Using local coordinates one can easily prove the following proposition.

\begin{prop}
Let ${\cal L}:{\r^k}\times T^1_kM\longrightarrow R$ be a regular Lagrangian, and
$V_{1,0}$ the vertical distribution of the bundle $\pi_{1,0}:{\r^k}\times
T^1_kM \longrightarrow \r^k\times M$. Then, ${\cal L}$ is regular if and only if
$({\r^k}\times T^1_kM,\bar{\eta}_A,(\omega_{\cal L})_A,V_{1,0})$  is a
$k$--cosymplectic manifold. \blob
\end{prop}

Let ${\cal L}:{\r^k}\times T^1_kM\longrightarrow \r $ be a regular
Lagrangian and $(dt^A, (\omega_L)_A,V_{1,0})$ the associated
$k$-cosymplectic structure on $\rk \times T^1_kM$.
The equations
\begin{equation}\label{xileq}
dt^A((\xi_{\cal L})_B)=\delta^A_B, \quad (\xi_{\cal L})_A \hook (\omega_{\cal L})_B=0,\qquad
1\leq A,B \leq k\,  .
\end{equation}
define the Reeb vector fields $\{(\xi_{\cal L})_1,\dots,(\xi_{\cal L})_k\}$ on
${\r^k}\times T^1_kM$ which are locally given by
\begin{equation}\label{localxi1}
(\xi_{\cal L})_A=\frac{\displaystyle\partial}{\displaystyle\partial t^A}
+((\xi_L)_A)^i_B\frac{\displaystyle\partial} {\displaystyle\partial
v^i_B} \;,
\end{equation}
where the functions  $((\xi_{\cal L})_A)^i_B$ satisfy
\begin{equation}\label{xiconditions}
\frac{\displaystyle\partial^2 {\cal L}}{\displaystyle\partial t^A\partial v^j_C}
+\frac{\displaystyle\partial^2 {\cal L}}{\displaystyle\partial v^i_B
\displaystyle\partial v^j_C} ((\xi_{\cal L})_A)^i_B=0 \;,
\end{equation}
for all  $1\leq A,B,C\leq k$ and  $1\leq i,j\leq n$.

Since ${\cal L}$ is regular, from the local conditions (\ref{xiconditions}) we
can define, in a neighbourhood of each point of ${\r^k}\times T^1_kM$, a
$k$--vector field that satisfies (\ref{xileq}). Next one can construct a
global $k$--vector field $\xi_L$, which is a solution of (\ref{xileq}),
by using a partition of   unity.

Let ${\cal L}$ be a regular Lagrangian and let $\Omega_{\cal L}^{\sharp}$ be the
$\sharp$-morphism defined by the $k$-cosymplectic structure $(dt^A,
(\omega_{\cal L})_A,V_{1,0})$, as in (\ref{sharp}):
\begin{equation}\label{sharpl}
\begin{array}{rccl}
\Omega_{\cal L}^{\sharp}: & T^1_k({\r^k}\times T^1_kM)& \longrightarrow  &
T^*({\r^k}\times T^1_kM) \\ \noalign{\medskip}
 & (X_1,\dots,X_k) & \longrightarrow &
\Omega^{\sharp}_{\cal L}(X_1,\dots,X_k) =  \displaystyle \sum_{A=1}^k \, X_A
\hook (\omega_{\cal L})_A+ dt^A (X_A) dt^A  \; .
\end{array}
\end{equation}

A direct computation in local coordinates proves the following Proposition.

\begin{prop}
Let ${\cal L}$ be a regular Lagrangian and let $X =(X_1,\dots,X_k)$ be a
$k$-vector field such that
\begin{equation}\label{lageq}
\begin{array}{l}
dt^A(X_B)=\delta_{AB}, \quad 1\leq A,B\leq k\\
\noalign{\medskip}\Omega^{\sharp}_{\cal L}(X_1,\dots,X_k)=
dE_{\cal L}+\displaystyle\sum_{A=1}^k(1-(\xi_{\cal L})_A(E_{\cal L})) dt^A
\end{array}
\end{equation}
where $E_{\cal L}=C({\cal L})-{\cal L}$. Then $X =(X_1,\dots,X_k)$ is a SOPDE. In addition, if
$X =(X_1,\dots,X_k)$ is integrable then its solutions satisfy the
Euler-Lagrange equations (\ref{lageq1}).
\end{prop}

 In conclussion, we can consider  Eqs. (\ref{lageq}) as a {\it geometric
version} of the Euler-Lagrange field equations for a regular Lagrangian.

{\bf Remark}  We have given a geometric version of the Euler-Lagrange
equations for a non autonomous Lagrangian by constructing a $k$-cosymplectic
structure on $\r^k\times T^1_kM$ defined from the Lagrangian and the
canonical stable $k$-tangent structure on $\r^k\times T^1_kM $. We can
also construct this $k$-cosymplectic structure using the {\it Legendre
tranformation} ${\cal F}{\cal L}$ of ${\cal L}$ which is the map
\[
 {\cal FL}: \r^k\times T^1_kM \longrightarrow {\r^k}\times (T^1_k)^*M
\]
defined as follows:

\noindent If $(t,v) = (t^{1}, \dots, t^{k}, v_{1}, \ldots , v_{k}) \in
{\r^k}\times (T^1_kM)_x$ with $x\in M$ and $v_{A} \in T_xM$, then

$$  {\cal FL}(t,y) = (t^{1}, \ldots,t^k,p^1,\ldots p^k)
\in {\r^k}\times (T^1_kM)^{*}_x\, , \quad p^A \in T^*_xM \,
$$
is given by
\[
p^A(v_x) = (\beta_{\cal L})_A(\bar{v_x}), \quad 1 \leq A \leq k,
\]
for any $v_x \in T_xM$, where $\bar{v_x} \in T_v(T_{k}^{1}M)$ is any
tangent vector such that $d\tau_M(v)(\bar{v_x})=v_x$, with $\tau_M :
T^{1}_{k}M \longrightarrow M$ the canonical projection. In induced
coordinates we have
\begin{equation}\label{localfl}
{\cal FL}:(t^A,q^i,v^i_A)  \longrightarrow  (t^A,q^i,
\frac{\displaystyle\partial {\cal L}}{\displaystyle\partial v^i_A}).
\end{equation}

Now, from (\ref{locbeta}) and (\ref{localfl}) we deduce the following.

\begin{lem}
 
$
(\omega_L)_A = {\cal FL}^*((\omega_0)_A), \; \; \; dt^A =  
{\cal FL}^*((\eta_0)_A),
$
for all $A$.
\end{lem}

Then we have
\begin{prop}
The following conditions are equivalent:

1) ${\cal L}$ is regular.

2) ${\cal FL}$ is a local diffeomorphism.

3)$(dt^A,(\omega_{\cal L})_A,V_{1,0})$ is a $k$-cosymplectic structure on
$\r^k\times T^1_kM$.
\end{prop}
 .

\section{The  Cartan-Hamilton-Poincar\'{e}  Form on \jetpi\ and \lpie}

 In this section we further explore relationships between $n$-symplectic geometry on frame
bundles and multisymplectic geometry.  Since $m=n+k=\mbox{dim}(E)$ we will refer to the
$n$-symplectic geometry on $LE$ as $m$-symplectic geometry, and base the discussion on the
$n$-form on $\jetpi$ considered by Cartan, Hamilton and Poincar\'e. This form has various names
in the literature; here we will use the name Cartan-Hamilton-Poincar\'{e} (CHP) form.
Although this  $n$-form on $\jetpi$ has been in the literature for
many years, its definition on $\lpie$ is relatively recent. It appeared first in \cite{MN}, where
the
$n$-form was defined in terms of newly defined Cartan-Hamilton-Poincar\'{e} $1$-forms. These
Cartan-Hamilton-Poincar\'{e} $1$-forms play the role of an $m$-symplectic potential on $\lpie$ and
are discussed in Section 7.5. In Section 7.4 we give a new geometrical definition of $\Theta_L$
on
$\jetpi$. See also Section 9.4 where the Cartan-Hamilton-Poincar\'{e} $1$-forms are defined using
an
$m$-symplectic Legendre transformation.

\subsection{The  Cartan-Hamilton-Poincar\'{e} Form on \jetpi}
\label{pcformsect}

 One method  used to construct the Cartan-Hamilton-Poincar\'{e} Form on \jetpi\, is to first  
construct  a vector valued $m$-form $S_\omega$ on $J^1 \pi$ associated
with a volume form $\omega$ on $M$, as follows:

\noindent For each $1$-form $\sigma$ on
$J^1 \pi$  the vector valued $1$-form $S\sigma$ along $\pi_1: J^1 \pi
\rightarrow M$ is defined by
   $$\alpha ((S\sigma)(X)) = \sigma (S_\alpha (X))$$
for any vector field $X$ on $J^1 \pi$ and any $1$-form $\alpha$ on $M$.
Recall $S_\alpha$ was defined in Section \ref{vertendosect}.

   Now $S_\omega$ is defined according to the rule
$$S_\omega \, \hook \sigma = \imath_{\displaystyle S\sigma} \omega$$
where $\imath_{\displaystyle S\sigma} $ is the derivation of type $\imath_*$
corresponding to $S\sigma$, that is
$$\sigma(S_\omega(X_1,\dots ,X_m))=
(\imath_{\displaystyle S\sigma} \omega)(X_1,\dots ,X_m) =\displaystyle
\sum_{i=1}^n \omega((\pi_1)_*X_1,\cdots,S\sigma (X_i),\dots
,(\pi_1)_*X_m)$$ for any vector fields $X_1,\dots,X_m$ on $J^1\pi$. In
coordinates
\begin{equation}\label{SOMEGA}
S_\omega=(dy^A-y^A_j dx^j) \wedge \left( \basisx i \hook \omega \right)
\otimes \basisv Ai
\end{equation}

   If ${\cal L}_\pi : J^1 \pi \rightarrow \Lambda^n M$ is a Lagrangian
density, then ${\cal L}_\pi = {\cal L} \, \omega$ where ${\cal L} : J^1 \pi \rightarrow {\bf
R}$. The {\it
 Cartan-Hamilton-Poincar\'{e}
$n$-form of $\cal L$}  is defined by
\begin{equation}\label{omegalsaunders}
 \Theta_L =
 {\cal L} \, \omega + {S_\omega}^* d{\cal L} = {\cal L} \, \omega + d{\cal L} \circ S_\omega \, .
 \end{equation}

\noindent In coordinates
\begin{equation}\label{localomegal}
 \Theta_{\cal L} = {\cal L} \, \omega + \displaystyle \frac{\partial {\cal L}}{\partial y^A_i}
(dy^A - y^A_j\, dx^j)\wedge ( \displaystyle \frac{\partial L}{\partial
x^i} \hook \omega)
\end{equation}

\subsection{The tensors $S_\alpha$ and $S_\omega$ on $J^1 \pi$  viewed
 from $L_\pi E$}

   For each $1$-form $\alpha$ on $M$,
 we shall define on \lpie\  a
 tensor field $\tilde{S}_\alpha$, of type $(1,1)$  that
{\it projects} on the tensor $S_{\alpha}$ on \jetpi\ .
   Let $(B_i=B(\hat{r}_i),B_A= B(\hat{r}_A))$ be the standard
 vector fields of any torsion free
 linear connection on $\lambda : L_\pi E \rightarrow E$. In local
coordinates
 we have
\begin{equation}\label{b}
    B_i  =  v^s_i \, {\basisx s} + v^C_i \, {\basisy C} + V_i
   \quad , \quad
   B_A  =  v^C_A \, {\basisy C} + V_A
 \end{equation}

\noindent where $V_i, V_A$ are vertical with respect to $\lambda$.

   Now if $\alpha$ is an arbitrary  $1$-form on $M$  and
$(\pi \circ \lambda)^* \alpha$
 its pull-back to $L_\pi E$,
we consider on $L_\pi E$ the functions $((\pi \circ \lambda)^* \alpha)
(B_i)$ for each $1 \leq i \leq n$.
In coordinates, if
 $\alpha = \alpha_r \, dx^r$, then from (\ref{b})
 \begin{equation}\label{pib}
 ((\pi \circ \lambda)^* \alpha) (B_i) = \alpha_r \, dx^r
 \left( v^s_i \, {\basisx s} + v^C_i \, {\basisy C} + V_i \right) =
 \alpha_r \, v^r_i \, .
\end{equation}

Taken together the function  $\hat{\alpha}=(\alpha_av^a_i)\hat{r}_i$ is the $(\r^n)^*$-valued
tensorial
$0$-form on $LE$ corresponding to $\alpha$ on $M$.
\begin{definition}
   The vector-valued $1$-form $\tilde{S}_\alpha$
   on $L_\pi E$ is defined by
 $$
 \tilde{S}_\alpha = ((\pi \circ \lambda)^* \alpha) (B_i)  \,
 E^{*i}_B \otimes \theta^B \quad .
 $$
\end{definition}
 From (\ref{eteta})   and (\ref{pib})  we obtain that
in local coordinates

\begin{equation}\label{tildelocsomega}
\tilde{S}_\alpha = \alpha_j \, (dy^B - u^B_t \, dx^t) \otimes {\basisu Bj}
 \quad  .
\end{equation}

\begin{prop}
   The relationship between $\tilde{S}_\alpha$ on $L_\pi E$ and
 $S_\alpha$ on $J^1 \pi$ is given by
 $$
 \tilde{S}_\alpha \, \hook \, \rho_* = \rho_* \, \hook \, S_\alpha
 $$
 that is
 $$
 \rho_* (u) ( \tilde{S}_\alpha (u) (
X_u ) ) =
  S_\alpha (\rho (u)) ( \rho_* (u) (
X_u ) )
 $$
 for any $u \in L_\pi E$ and any  $X_u \in T_u ( L_\pi E)$.
\end{prop}

{\proof:} It is an immediate consequence of the local expressions of
$\tilde{S}_\alpha$ and $S_\alpha$ taking into account that $\rho^* y^B_t
= u^B_t$. \blob

   Now, proceeding analogously, we construct
a tensor field $\tilde{S}_\omega$ of type $(1,n)$ on \lpie\ using  the
tensor field $\tilde{S}_\omega$ on \lpie, associated with a volume form
$\omega$ on $M$. We then construct the {\it corresponding  Cartan-Hamilton-Poincar\'{e} 
form} on \lpie.

For each $1$-form $\sigma$ on \lpie\
the vector valued $1$-form $\tilde{S} \sigma$ along
$\pi \circ \lambda: \lpie  \rightarrow M$
is defined by
   \begin{equation}\label{ssigma}
\alpha ((\tilde{S}\sigma)(X)) = \sigma (\tilde{S}_\alpha (X))
\end{equation}
for any vector field $X$ on \lpie\ and any $1$-form $\alpha$ on  $M$.
    We shall compute the local expression of
 this $1$-form.   If we write
 $$
 \sigma = \sigma_i \, dx^i + \sigma_A \, dy^A + \sigma^j_i   \, du^i_j
 + \sigma^j_B \, du^B_j + \sigma^A_B \, du^B_A
 $$
 and we take $\alpha = dx^j$ then from (\ref{somega})
and (\ref{ssigma}) we obtain
\begin{gather*}
 dx^j(\tilde{S} \sigma ({\basisx i})) = -\sigma^j_B \, u^B_i,
   \quad
   dx^j (\tilde{S}\sigma ({\basisy A}))
    = \sigma^j_A,
 \\
 dx^j (\tilde{S} \sigma ({\basisu ij}))
   = dx^j (\tilde{S} \sigma ({\basisu Ai}))  =
                 dx^j (\tilde{S} \sigma ({\basisu AB})) =0
\end{gather*}
Therefore the local expression of $\tilde{S} \sigma$ is
 \begin{equation}\label{esetildesig}
 \tilde{S} \sigma \, = \, \sigma^j_B \, (dy^B - u^B_t \, dx^t)
 \otimes {\basisx j} \quad .
 \end{equation}

\begin{definition}
   The tensor field $\tilde{S}_\omega$ is defined according to the rule
$$\tilde{S}_\omega \, \hook \sigma = \imath_{\displaystyle \tilde{S}\sigma} \Omega$$
where $\imath_{\displaystyle {S}\sigma} $ is the derivation of type $\imath_*$
corresponding to $\tilde{S}\sigma$, that is
\begin{eqnarray}\label{exeseom}
\sigma(\tilde{S}_\omega(X_1,\dots ,X_n))& = &(\imath_{\displaystyle
S\sigma} \omega)(X_1,\dots ,X_n) \\ & =& \displaystyle \sum_{j=1}^n
\omega((\pi\circ \lambda)_* X_1,\dots,\tilde{S}\sigma (X_j),\dots
,(\pi\circ \lambda)_*X_n)
\end{eqnarray}
for any vector fields $X_1,\dots,X_n$ on \lpie\ and any $1$-form $\sigma$
on \lpie\ .
\end{definition}

 From (\ref{esetildesig}) and (\ref{exeseom}) we obtain that the local
expression of $\tilde{S}_\omega$ is
\begin{equation}\label{loceseom}
 \tilde{S}_\omega = (dy^A - u^A_t \, dx^t) \wedge
 \left( {\basisx i} \hook \omega \right) \otimes {\basisu Ai}
\end{equation}

 \begin{prop}
   The relationship between $\tilde{S}_\omega$ on $L_\pi E$ and
 $S_\omega$ on $J^1 \pi$ is given by
 $$
 \tilde{S}_\omega \, \hook \, \rho_* = \rho_* \, \hook \, S_\omega
 $$
 that is
 $$
 \rho_* (u) \left(
\tilde{S}_\omega (u) \left( (X_u)_1,
 \dots, (X_u)_n \right) \right) =
  S_\omega (\rho (u)) \left( \rho_* (u) \left( (X_u)_1 \right), \dots,
 \rho_* (u) \left( (X_u)_n \right) \right)
 $$
 for any $u \in L_\pi E$ and any  $(X_u)_1, \dots, (X_u)_n
 \in T_u ( L_\pi E)$.
\end{prop}

{\proof:} It is an immediate consequence of the local expressions
 (\ref{SOMEGA}) and (\ref{tildelocsomega}) of
$\tilde{S}_\omega$ and $S_\omega$ taking into account that $\rho^* y^B_t
= u^B_t$. \blob

\subsection {The Cartan-Hamilton-Poincar\'{e}  form $\Theta_L$ on $J^1 \pi$
viewed from $\lpie$ }

Using the tensor field $\tilde{S}_\omega$ we shall construct an $m$-form
on $L_\pi E$ that   projects to the corresponding Cartan-Hamilton-Poincar\'{e}
$m$-form on $J^1 \pi$.

\begin{definition}
A  Lagrangian on $\lpie$ is a function $L:\lpie\to\mathbb{R}$.
\end{definition}

\begin{definition}\cite{MN}
A Lagrangian on $\lpie$
is {\rm lifted} if it satisfies the auxiliary
conditions
\begin{equation}\label{lift}
\vertical ij(L)=0\ \,\qquad \vertical AB(L)=0
\end{equation}
\end{definition}

\rem Using (\ref{estar in lpie})
these conditions imply that $L$ is constant on the
fibers of $\rho:\lpie\to\jetpi$, and thus is the pull up of a function
${\cal L}$  on $\jetpi$, that is $\rho^*{\cal L} = L$.

\begin{definition}
   If $L : L_\pi E \rightarrow \mathbb{R}$ is a lifted Lagrangian on
 $L_\pi E$, then we define the {\rm Cartan-Hamilton-Poincar\'{e}  $m$-form} of $L$ by
$$
\theta_L =
 L \, \omega + {\tilde{S}^*_\omega} dL =
 L \, \omega + dL \circ \tilde{S}_\omega \, .
 $$
 \end{definition}
If $\omega = d^n x = dx^1 \wedge \dots
 \wedge dx^n$ then from (\ref{loceseom}) we obtain
that the local expression of $\theta_L$ is

 \begin{equation}\label{omegal}
 \theta_L = \left( L - u^A_i \, \frac{\partial L}{\partial u^A_i} \right)
 \, d^n x + \frac{\partial L}{\partial u^A_i} \, dy^A \wedge d^{n-1} x_i \, .
\end{equation}

\begin{prop}
 If $L$ is a lifted Lagrangian then the
corresponding
$m$-form satisfies $\rho^* \Theta_{\cal L} = \theta_L$,
where
 $\Theta_{\cal L}$ is the
Cartan-Hamilton-Poincar\'{e}
 $n$-form on $J^1 \pi$
corresponding to $\cal L$ .
\end{prop}

\proof It follows from the local expressions taking into account
 that
$\rho^* y^A_i = u^A_i$.  \blob

\subsection{The $m$-symplectic structure on $\lpie$ and the formulation of the
Cartan-Hamilton-Poincar\'{e}
$n$-form}

We consider next the definition  of the Cartan-Hamilton-Poincar\'e $1$-forms on $\lpie$ introduced
in \cite{MN,No5}. These $1$-forms combine  into an $\mathbb{R}^m$-valued $1$-form 
whose exterior derivative  plays the
role of a general $m$-symplectic structure on $\lpie$.

\begin{definition}\cite{MN} Let $ L: \lpie \to\mathbb{R}$ be a lifted Lagrangian
on \lpie\ , and $\tau(n)$ a positive function of $n=\dim M$. The {\rm
Cartan-Hamilton-Poincar\'{e} 1-forms} $\theta^\alpha_L$ on \lpie\ are
\begin{eqnarray}
\theta^i_{L}&=& \tn L \theta^i+\vertical
iA(L)\theta^A
           \label{theta i l defined}\\
\theta^A_{\lagrangian}&=&\theta^A
\end{eqnarray}
\label{definition of the chp 1-forms}
\end{definition}
where $\vertical iA $ are defined above in (\ref{estar in lpie}), and $\theta^i$ and $\theta^A$
are the components of the canonical soldering 1-form on \lpie.    

\rem The quantities $\vertical iA (\lag) $, referred to as the "covariant canonical momenta"
in \cite{MN}, are \underline{\em globally defined} on \lpie.  In local canonical coordinates
$(z^\alpha,\pi^\mu_\nu)$, these quantities have the local expressions
\begin{equation}
\vertical iA (\lag)=\pi^i_j p^j_B v^B_A\ \ ,\ \ p^j_B=\frac {\partial \lag}{\partial u^B_j}
\label{definition of Eia applied to L}
\end{equation}
and clearly are the frame components of the "canonical field momenta" $ p^j_B=\frac {\partial \lag}{\partial
u^B_j}$. For different values of
$\tau$ one can obtain the de Donder-Weyl theory~\cite{DW,Rund} and the Caratheodory
theory~\cite{Car,Rund} as special cases of the  formalism presented in reference \cite{MN}.  
The significance of these CHP 1-forms as regards other geometrical theories was also considered by 
MacLean and Norris.  In~\cite{MN} it was shown that
  one may   construct the CHP $n$-form on $\jetpi$  from the CHP 1-forms on \lpie.  In this regard
see also references~\cite{No3,FLN2}.
We now recall the construction of the Cartan-Hamilton-Poincar\'{e}  $n$-form on \jetpi\  from  
these  CHP 1-forms.

\begin{prop}\cite{MN} \label{quotient prop}
Let $(B_i,B_A)$ denote the standard horizontal vector fields of any
torsion free linear connection on  $\lambda:\lpie\to E$, and let
$\op{vol}$ denote the pull up to $\lpie$ of a fixed volume $n$-form
$\omega$ on $M$. Set $\op{vol}_i=B_i\hook\ \op{vol}$.  Then when
$\tn=\frac{1}{ n}$ the $n$-form
$$
\theta_L :=\theta^i_L \wedge \op{vol}_i
$$
passes to the quotient to define the CHP-$n$-form $\Theta_L$ on $\jetpi$ associated
with $vol=\omega$.

\label{chp forms pass to the quotient}
\end{prop}

   Next we shall show here that the Cartan-Hamilton-Poincar\'{e} $1$-forms
 can be obtained from the canonical
$m$-tangent structure
  $J^i, J^A$ on \lpie\ .
Let $\Lambda = f_1 \wedge \dots \wedge f_n$ be a fixed
contravariant volume on $M$, with $f_i$ locally written as
$f_i = \alpha^j_ i {\basisx j}$. Thus $\Lambda$ is a nowhere vanishing
$n$-vector on $M$, which is the covariant version of a volume
form on $M$. In coordinates
$$
\lambda = det(\alpha^i_j) \, {\basisx 1} \wedge \dots \wedge {\basisx n}
$$

   Now given an arbitrary point $u=(e_i,e_A)_e$ on $L_\pi E$ we
can define the $n$-vector
$$
[\tilde{e}_i] = \tilde{e}_1 \wedge \dots \wedge \tilde{e}_n
$$
where $\tilde{e}_i = (\pi \circ \lambda)_*(u)(e_i)$. $[\tilde{e}_i] $ is a
well-defined $n$-vector at $(\pi \circ \lambda)(u) = \pi (e) \in M$ since
the vectors \,$\tilde{e}_i$\, are linearly independent. In coordinates
$$
[\tilde{e}_i] = det(v^i_j) \, {\basisx 1} \wedge \dots \wedge {\basisx n}
$$

   We can now define a function $\sigma : L_\pi E \rightarrow \mathbb{R}$
relative to the fixed contravariant volume $\Lambda$ on $M$ by the
formula
$$
[\tilde{e}_i] = \sigma (u) \, \lambda (\pi (e))
$$

   Using the local expressions above it is easy to see that in local
coordinates on $L_\pi E$ one has
\begin{equation}\label{sig}
\sigma (u) = \frac{det(v^i_ j)(u)}{det(\alpha^i_j (\pi(e))}
\end{equation}

\begin{prop}
Let $L$ be a lifted Lagrangian on $L_\pi E$ and let $\sigma$ be the
function defined on $L_\pi E$ relative to a fixed contravariant volume
$\Lambda$ on $M$. Then the Cartan-Hamilton-Poincar\'{e} $1$-forms on $L_\pi E$
are given by the formula
$$
\theta^i_L = \frac{1}{\sigma} \, d_{\tilde{J}_i} (\sigma \, L)
$$
where
$$
\tilde{J}^i =  \frac{1}{n} (E^{*i}_j \otimes   \, \theta^j) \, + \,
E^{*i}_A \otimes \theta^A
$$
and $d_{\tilde{J}^i} = [\imath_{\tilde{J}^i},d]$ .
\end{prop}

\proof From (\ref{estar in lpie}) and (\ref{sig}) we obtain that
$$
E^{*i}_j (\sigma) = \sigma \, \delta^i_j \quad , \quad E^{*i}_aA (\sigma)
= 0
$$
and from (\ref{lift}) we obtain
$$
E^{*i}_j (\sigma \, L) = \sigma \, L \, \delta^i_j \quad , \quad
E^{*i}_A (\sigma \, L) = \sigma \, E^{*i}_A (L) \, .
$$

   Now from these last identities we have
$$
\begin{array}{lcr}
\displaystyle \frac{1}{\sigma} \, d_{\tilde{J}^i} (\sigma \, L) & = &
\displaystyle \frac{1}{\sigma} \, \left( d(\sigma \, L) \circ \tilde{J}^i \right)
  =\displaystyle \frac{1}{\sigma} \left( \frac{1}{n} E^{*i}_j (\sigma \, L) \theta^j
+ E^{*i}_A (\sigma \, L) \theta^A\right)  \\ \noalign{\medskip}
& = &\displaystyle \frac{1}{\sigma} \left( \frac{1}{n} \sigma \, L \,
\delta^i_j \theta^j + \sigma \, E^{*i}_A (L) \theta^A\right) =
 \displaystyle \frac{1}{n} \, L \, \theta^i + E^{*i}_A (L) \, \theta^A\,
 .\,\blob
\end{array}
$$
\rem To these three constructions of the Cartan-Hamilton-Poincar\'{e} $1$-forms on $\lpie$ we
add a fourth  in Section 9.4 where we show that the $\theta^\alpha_L$ are the pull-backs, under
a suitable defined
$m$-symplectic Legendre transformation, of the canonical $m$-symplectic structure on $LE$.

\section{ Multisymplectic formalism}

An alternative way to derive the field equations is to use the so-called
multisymplectic formalism, developed by the Tulczyjew school in Warsaw
(see \cite{ KijoSz,KijTul,Kijo,Snia}), and independently by Garc\'{\i}a and
P\'{e}rez-Rend\'{o}n \cite{GP1,GP2} and Goldschmidt and Sternberg \cite{GS}. This
approach was revised by Martin \cite{Mart1,Mart2} and Gotay et al
\cite{GIMMsy,GIMMsy2,Go1,Go2,Go3}, and more recently by Cantrijn et al \cite{Cant1,Cant2}.

\subsection{Lagrangian formalism}

Assume a Lagrangian ${\cal L}:\jetpi \to \mathbb{R}$ where \jetpi\ is the
$1$-jet prolongation of a fibered manifold $\pi:E \to M$. $M$ is
supposed to be oriented with volume form $\omega$. We take adapted
coordinates $(x^i,y^A,y^A_i)$ such that $\omega=dx^1\wedge \dots \wedge
dx^n=d^nx$.

 Denote $\Omega_{\cal L}=-d\Theta_{\cal L}$ where $\Theta_{\cal L}$ is the {\it
Cartan-Hamilton-Poincar\'{e}}
  $m$-form introduced in $7.1$ . From (\ref{omegalsaunders})   we have
  that in  local coordinates
      $$\Omega_{\cal L}= d(y^A_i \frac{\partial L}{\partial y^A_i}-{\cal L}) \wedge d^nx
      -d( \frac{\partial {\cal L}}{\partial y^A_i})\wedge dy^A\wedge d^{n-1}x^i$$
      where   $d^{n-1}x^i=
      \frac{\partial}{\partial x^i}\hook \omega$.

\begin{definition}
   $\Omega_{\cal L}$  is called the {\rm Cartan-Hamilton-Poincar\'{e}}
  $(n+1)$-form.
\end{definition}

   One can use this multisymplectic form to  re-express, in an intrinsic
   way, the {\it Euler-Lagrange equations}, which in coordinates take the
  classical form
\begin{equation}\label{lageqgim}
\displaystyle \sum_{i=1}^k\frac{\partial}{\partial x^i}
(\frac{\displaystyle\partial {\cal L}}{\displaystyle \partial
y^A_i})(x^i,\phi^B(x), \frac{\partial \phi^B}{\partial x^i}(x))-
\frac{\displaystyle
\partial {\cal L}}{\displaystyle
\partial y^A}(x^i,\phi^B(x), \frac{\partial \phi^B}{\partial x^i}(x))=0,
\end{equation}
for a (local) section $\phi$ of $\pi:E \to M$.

\begin{thm}
For a section $\phi$ of $\pi$ the following are equivalent:

\noindent (i) the Euler-Lagrange equations (\ref{lageqgim}) hold in
coordinates;

\noindent (ii) for any vector field $X$ on \jetpi\
\begin{equation}\label{modes1}
(j^1\phi)^*(X\hook \Omega_{\cal L})=0 \, .   \blob
\end{equation}
\end{thm}

   The proof can be found in \cite{GIMMsy}.

 $\Omega_{\cal L}$ is a multisymplectic form on \jetpi\ provided $L$ is
    regular, that is, the Hessian matrix
    $$( \frac{\partial^2 {\cal L}}{\partial y^A_i \partial y^B_j} ) $$
    is
    nonsingular.

 We can extend equations (\ref{modes1}) to sections $\tau$ of
$\jetpi\ \to M$, that is we consider sections $\tau$ such that
\begin{equation}\label{modes2}
\tau^*(X \hook \Omega_{\cal L})=0 \, ,
\end{equation}
 for any vector field $X$ on \jetpi\, .
If the Lagrangian ${\cal L}$ is regular then both problems (\ref{modes1}) and
(\ref{modes2}) are equivalent, that is, such a $\tau$ is automatically a
$1$-jet prolongation $\tau=j^1\phi$. Equation (\ref{modes2}) corresponds to
the so called de Donder  problem (see Binz {\it et al} \cite{Binz}.)

\subsection{Hamiltonian formalism}

  We have an exact
   sequence of vector bundles over $E$:
   $$0\to \, {\bigwedge}^n_1 E \,  \stackrel{i}{\longrightarrow} \,  {\bigwedge}^{n}_{2}
   E\,
   \stackrel{\mu}{\longrightarrow} \, J^1\pi^*\to 0$$
   where $J^1\pi^*$ is the quotient vector bundle
   $$J^1\pi^*=\frac{\displaystyle {\bigwedge}^n_2E}{\displaystyle {\bigwedge}^n_1E}\, ,$$
$i$ is the inclusion, and $\mu$ is the projection map.

 $J^1\pi^*$ is sometimes defined as the   affine dual bundle of \jetpi\,
    (see \cite{Saunders}).
  We have taken  local coordinates
   $(x^i,y^A,p)$\,  on   $\bigwedge^n_1E$\,  and \, $(x^i,y^A,p, p^i_A)$\,  on
   $\bigwedge^n_2E$,   and then $(x^i,y^A,p^i_A)$ can be taken  as local
   coordinates in $J^1\pi^*$.

   To develop a Hamiltonian theory, we need a Hamiltonian, in this case a
   section $H : J^1\pi^* \to \bigwedge^n_2E$ of the canonical projection
   $\mu$.
      In coordinates,
   we have
   $$H(x^i,y^A,p^i_A)=(x^i,y^A,-\hat{H},p^i_A)$$
   where $\hat{H}=\hat{H}(x^i,y^A,p^i_A)\in C^{\infty}(J^1\pi^*,\r)$.

   Take the pull-back $\Omega_H=H^*\Omega_E^2$ (we also have $\Theta_H=H^*\Theta_E^2$
   such that $\Omega_H=-d\Theta_H$), then from (\ref{theta2e}) we have
   $$ \Theta_H = -\hat{H} d^nx + p^i_A dy^A \wedge d^{n-1}x^i, \quad
      \Omega_H = d\hat{H} \wedge  d^nx - dp^i_A \wedge dy^A \wedge d^{n-1}x^i \, ,$$

      $\Omega_H$ is again a multisymplectic $(n+1)$-form. Now
        solutions of the
      {\it Hamilton equations}
      $$\frac{\partial \gamma^A}{\partial x^i}= -\frac{\partial \hat{H}}{\partial
      p^i_A} ,\qquad \sum_i \frac{\partial \gamma^i_A}{\partial
      x^i}=\frac{\partial \hat{H}}{\partial y^A} \, .$$
      are obtained by looking for sections
         \[
\begin{array}{ccccc}
\gamma & :& M &\longrightarrow  & J^1\pi^* \\
          &  &    (x^i)  & \mapsto &  (x^i,\gamma^A,\gamma^i_A)
          \end{array}
\]
     such that
      $$\gamma^*(Y\hook  \Omega_H)=0$$
      for any vector field $Y$ on $ J^1\pi^*$, see \cite{Cant1}\,.

      To relate both formalisms, we must use the Legendre transformation.
      For ${\cal L}$, we define a fibered mapping over $E$,
      $Leg : \jetpi \longrightarrow  \bigwedge^n_1E$,
      by
        $$[Leg(j^1_x\phi)](X_1,\dots ,X_n)\, =\,
         (\Theta_{\cal L})_{j^1_x\phi} (\tilde{X}_1,\dots ,\tilde{X}_n)$$
         for all $X_1,\dots ,X_n\in T_{\phi(x))}E$, where
         $\tilde{X}_1,\dots ,\tilde{X}_n\in T_{j^1_x\phi}(\jetpi)$ are such
         that they project on $X_1,\dots ,X_n$, respectively.

         In local coordinates
         $$Leg(x^i,y^A,y^A_i)\, = \, (x^i,y^A, {\cal L}-y^A_i\frac{\partial
         {\cal L}}{\partial y^A_i}, \frac{\partial {\cal L}}{\partial y^A_i})\, .$$
         If we compose $Leg: \jetpi \to \bigwedge^n_1E$ with
         $\mu :  \bigwedge^n_1E \to J^1\pi^*$, we obtain the reduced
         Legendre transformation
         $$\begin{array}{ccccc}
leg & :& \jetpi &\longrightarrow  & J^1\pi^* \\
          &  &    (x^i,y^A,y^A_i)  & \mapsto &  (x^i,y^A,\frac{\partial {\cal L}}{\partial y^A_i} )
          \end{array}
$$
which extends the usual one in mechanics, and the Legendre map defined by
G\"{u}nther. (see remark in  Section $6.5$).

A direct computation shows that $leg^* \Theta^2_E=\Theta_{\cal L}, \quad
leg^*\Omega^2_E = \Omega_{\cal L}$.

 It is clear that $leg:\jetpi \to J^1\pi^*$
is a local diffeomorphism if and only if ${\cal L}$ is regular.
  If ${\cal L}$ is regular, then we can define a (local) section $H$ as follows
   $H=Leg \circ leg^{-1}$
   $$\jetpi \longrightarrow {\bigwedge}^n_2E$$.
   \begin{prop}
The following assertions are equivalents:

1) ${\cal L}$ is regular.

2) $\Omega_{\cal L}$ is multisymplectic, and

3) $leg : J^1\pi \to J^1\pi^*$ is a local diffeomorphism.

   \end{prop}

\subsection{Ehresmann connections and the Lagrangian and Hamiltonian formalisms}

   A different geometric version of the field equations was  
given recently, based on Ehresmann connection  \cite{Cant2}\,.

   In mechanics we look for curves and their linear approximations; that
   is, we look for tangent vectors. In Field Theory, we
   look for sections, and their linear approximations
   are just horizontal subspaces of Ehresmann connections in the fibration
   $\pi_1: \jetpi \to M$.

  A connection in $\pi_1$ (in the sense of
Ehresmann \cite{Roux,Lib1}) is defined by a complementary distribution
{\bf H} of $V\pi_1$, i.e., we have the following Withney sum of vector
bundles over $E$:
\[
T(\jetpi) = \hbox{{\bf H}} \oplus V\pi_1 \; .
\]

As  is well-known, we can characterize a connection in $\pi_1$ as a
(1,1)-tensor field $\Gamma$ on $\jetpi$ such that
\begin{itemize}
\item $\Gamma^2 = Id$, and

\item  the eigenspace at the point $z \in \jetpi$ corresponding to the
eigenvalue $-1$ is the vertical subspace $(V\pi_1)_{z}$.
\end{itemize}
In other words, $\Gamma$ is an almost product structure on $\jetpi$ whose
eigenvector bundle corresponding to the eigenvalue $-1$ is just the
vertical subbundle $V\pi_1$.

We denote by
$$
 \hbox{\bf h} = \frac{1}{2}(Id + \Gamma) \; , \; \hbox{{\bf v}} =
\frac{1}{2}(Id - \Gamma) \; ,
$$
 the horizontal and vertical projectors,
respectively. Hence, the horizontal distribution is given by $\hbox{{\bf
H}} = Im \; \hbox{\bf h}$ and $Im \; \hbox{{\bf v}} = V\pi_1$.

We say that $\Gamma$ is {\it flat} if the horizontal distribution is
integrable. In such a case, from the Frobenius theorem, there exists a
horizontal local section $\gamma$ of $\pi_1$ passing through each point of
$\jetpi$. Let us recall that a local section $\gamma$ of $\pi_1 : \jetpi
\to M$ is called {\it horizontal} if it is an integral submanifold of the
horizontal distribution.

Suppose that $\hbox{\bf h}$ is locally expressed in fibered coordinates
$(x^{i}, y^A, y^A_i)$ as follows:
 \begin{equation}\label{projector}
  {\bf h} = dx^i\otimes
[\frac{\partial}{\partial x^i} +\Gamma^A_i \frac{\partial}{\partial y^A}
+ \Gamma^A_{ji}\frac{\partial}{\partial y^A_j}]
\end{equation}

 \noindent A direct computation in local coordinates shows that the equation
   $$\imath_{\bf h} \Omega_{\cal L}= (n-1)\Omega_{\cal L}$$
   may be  considered as the geometric version of the field equations,
   where
   ${\bf h}$ is the horizontal projector  of the  Ehresmann connection in $\jetpi
   \to M$.
   Indeed, from (\ref{projector}) and the local expression of $\Omega_{\cal L}$
   we deduce that  $\imath_{\bf h} \Omega_{\cal L} = (n-1)\Omega_L$ if and only
   if
   \begin{equation}\label{iago1}
   \frac{\partial {\cal L}}{\partial y^A}-\frac{\partial^2 {\cal L}}{\partial y^A_i \partial x^i}
   -\Gamma^B_i \frac{\partial^2 {\cal L}}{\partial y^A_i \partial
   y^B}-\Gamma^B_{ij}\frac{\partial^2 {\cal L}}{\partial y^A_i \partial y^B_j}
      +(\Gamma^B_i -y^B_i) \frac{\partial^2 {\cal L}}{\partial y^A \partial
   y^B_i}=0\, ,
       \end{equation}
   \begin{equation}\label{iago2}
   (\Gamma^B_j-y^B_j)\frac{\partial^2 {\cal L}}{\partial y^A_i \partial y^B_j}=0\,
   .
   \end{equation}
If $L$ is regular, (\ref{iago2}) implies $\Gamma_j^B=y^B_j$, for all
$B,j$, and then (\ref{iago1}) becomes
\begin{equation}\label{iago3}
  \frac{\partial {\cal L}}{\partial y^A}-\frac{\partial^2 L}{\partial y^A_i
\partial x^i}
   -y^B_i \frac{\partial^2 {\cal L}}{\partial y^A_i \partial y^B}
   - \Gamma^B_{ji} \frac{\partial^2 L}{\partial y^A_i \partial
   y^B_j}=0\, ,
   \end{equation}

\noindent  Hence, if $\Gamma$ is flat and  $\gamma: M \to \jetpi$ is a
 a horizontal local section  locally given by
$\gamma(x^{i}) = (x^{i}, \gamma^A,\gamma^A_i)$,   then taking into account
that $\gamma_*(T_xM)={\bf H}_{\gamma(x)}$ we obtain
\begin{equation}
\Gamma^A_{i} \, = \, y^A_i \, =\, \frac{\partial \gamma^A}{\partial x^{i}}
= \gamma^A_{i}\,, \qquad
  \Gamma^{A}_{ji} \, = \, \frac{\partial
\gamma^A_j}{\partial x^{i}}\,  = \, \frac{\partial^2 \gamma^A}{\partial
x^{i}\partial x^j} \, .
\end{equation}
 This implies that $\gamma$ is a $1$-jet prolongation, i.
e. $\gamma=j^1\phi$ and, $\phi$ is a solution of (\ref{iago3}), that is,
$\phi$ is solution of the Euler-Lagrange equations
 (\ref{lageqgim}).

      Again, we can look for Ehresmann connections in the fibration $J^1\pi^* \to M$.
      Indeed, if $\tilde{h}$ is the horizontal projector of such a
      connection, we deduce that
   $$\imath_{\tilde{h}} \Omega_H = (n-1) \Omega_H$$
   if and only if
   $${\bigwedge}^A_i = -\frac{\partial H}{\partial p^i_A}, \qquad \sum_i
   {\bigwedge}^A_{ii} = \frac{\partial H}{\partial y^A}\, ,$$
   where
   $$ \tilde{h}= dx^i\otimes [\, \frac{\partial}{\partial x^i} + {\bigwedge}^A_i \frac{\partial}{\partial y^A}
   +{\bigwedge}_{ji}^A \frac{\partial}{\partial p^j_A} ]$$

\noindent    Therefore, if $\tilde{h}$ is flat, and $\gamma$ is an integral section
   of $\tilde{h}$, we deduce that $\gamma$ satisfies the Hamilton
   equations for $H$.

\subsection{Polysymplectic formalism}

   An alternative formalism for Classical Field Theories is the
   so-called polysymplectic approach 
(see \cite{GMS4,GMS1,GMS2,igor1,igor11,igor2,Sarda,Sarda2,Sarda3,SZ}). 
The geometric ingredients
   are almost the same as in multisimplectic theory, except that we
   consider vector-valued Cartan-Hamilton-Poincar\'{e} forms.

   We start with a fibred bundle $\pi:E \to M$ as above, and introduce the
   following  spaces

   \begin{itemize}

   \item The Legendre bundle
   $$\Pi = \bigwedge^nM\otimes_E V^*\pi \otimes_E TM$$
   where $V^*\pi$ is the dual vector bundle of the vertical bundle $V\pi$.

   \item The homogeneus Legendre bundle
   $$Z_E=T^*E\wedge (\bigwedge^{n-1}M) \, .$$
      \end{itemize}
   $Z_E$ (resp. $\Pi$) will play the role of $\bigwedge^m_2E$ (resp.
   $J^1\pi^*$) in multisymplectic formalism. Accordingly, we introduce
   coordinates $(x^i,y^A,p,p^i_A)$ on $Z_E$, and $(x^i,y^A,p^i_A)$ on
   $\Pi$. Moreover, there exists a canonical embedding
   $\theta :\Pi \to \bigwedge^{n+1}E\bigotimes_E TM$ defined
   by
   $\theta=-p^i_A \, dy^A \wedge \omega \otimes \frac{\partial}{\partial
   x^i}$.
   \begin{definition}
The polysymplectic form on $\Pi$ is the unique $TM$-valued $(n+2)$-form
$\Omega$ such that the relation
$$\imath_\phi \Omega = -d(\phi \hook \theta)$$
holds for any $1$-form $\phi$ on $M$.
 \end{definition}
   A direct computation shows that $\Omega$ has the following local
   expression
   $$\Omega = dp^i_A \wedge dy^A \wedge \omega \otimes \frac{\partial}{\partial
   x^i} \, .$$
A covariant Hamiltonian is given by a Hamiltonian form, that is, a
section $H$ of the canonical projection $Z_E \to \Pi$, as in the
multisymplectic settings. The field equations are provided by a
connection $\gamma$  in the fibration $\Pi \to M$ such that $\gamma \hook
\Omega$ is closed, and $\gamma$ is then called a Hamilton connection (see
\cite{GMS1} for details).

The Cartan-Hamilton-Poincar\'{e} $m$-form $\Theta_{\cal L}$ defines the Legendre
transformation
$$ {\cal FL } : J^1\pi \longrightarrow Z_E$$
by
$$ {\cal FL}(x^i,y^A,y^A_i)=(x^i,y^A,{\cal L}- y^A_i \frac{\partial
{\cal L}}{\partial y^A_i},\frac{\partial {\cal L}}{\partial y^A_i})\, .$$
   On the other hand, notice that $Z_E$ is canonically embedded into
   $\bigwedge^nM$, so that it inherits the restriction $\Xi_E$
   of the canonical multisymplectic form $\Omega_E$, say
   $$\Xi_E =\Omega_{E \rfloor Z_E} \, .$$

   Let $Z_{\cal L}={\cal FL}(J^1\pi)$ and assume that it is embedded into
   $Z_E$. Therefore we have an $n$-form $\Xi_{\cal L}$ on $Z_{\cal L}$ which is just
   the restriction of $\Xi_E$.
   Of course we have
   $$\Theta_{\cal L}= {\cal FL}^*(\Xi_{\cal L}) \, .$$
   The Legendre morphism $ {\cal FL}$ permits then to transport
   sections from the fibration $\jetpi \to M$ to $Z_{\cal L} \to M$, and conversely:

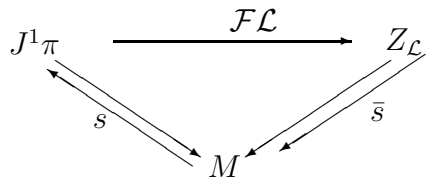
\begin{picture}(300,80)(0,0)
\put(90,65){\makebox(0,0){$J^1\pi$}} 
\put(230,65){\makebox(0,0){$Z_{\cal L}$}}
\put(162,18){\makebox(0,0){$M$}} 
\put(160,69){ ${\cal FL}$}
\put(120,65){\vector(1,0){90}} 
\put(150,18){\vector(-3,2){55}} 
\put(238,60){\vector(-3,-2){55}} 
\put(115,35){\makebox(0,0){$s$}}

\put(98,58){\vector(3,-2){55}}
\put(225,58){\vector(-3,-2){55}} 
\put(115,35){$$} 
\put(217,35){$\bar{s}$}
\end{picture}
\bigskip

\noindent such that, if $s$ is a solution of the equation $s^*(X\hook
   d\Theta_{\cal L})=0$ for all vector fields on $\jetpi$, then $  {\cal FL}
   \circ s$ is a solution of the equation $\gamma^*(\bar{X}\hook d\Xi_{\cal L} )=0$,
   for all vector fieds $\bar{X}$ on $Z_{\cal L}$, and conversely (see \cite{GMS1} ).

   In \cite{GMS1} is also analyzed the case of singular Lagrangians in
   order to compare the Hamiltonian and Lagrangian formalism.

\section{$n$-symplectic geometry  }

$n$-symplectic geometry on frames bundles was originally developed as a generalization of
Hamiltonian mechanics.  The theory has, however, turned out to be a covering theory of
both symplectic and multisymplectic geometries in the sense that these latter structures
can be derived from $n$-symplectic structures on appropriate frames bundles
\cite{No3,FLN2}.  In this
section we compare the $n$-symplectic geometry to $k$-symplectic/polysymplectic
geometry and to multisymplectic geometry as well.  Moveover we present a recent
extension  of the  algebraic structures on an $n$-symplectic manifold to a general
$n$-symplectic manifold.

\subsection{The structure equations of $n$-symplectic geometry}

The difference between $n$-symplectic and $k$-symplectic/polysymplectic
geometry lies not in the properties of the canonical 2-form -- they are
essentially the same.  Instead the real difference lies in the structure
equations, the specification of $LM$, and the algebraic structures based on the $m$-symplectic
Poisson bracket.

In $n$-symplectic geometry, one works with the soldering form on the frame
bundle $LM$.  The differential of the soldering form is a family of 2-forms
that, together with the right grouping of the fundamental vertical vector fields,
makes $LM$ a $m$-symplectic manifold.  However in $n$-symplectic geometry we
prefer to think of $d\theta$ as a vector valued 2-form -- as a single unit
rather than a collection.

Recall the structure equation of $m$-symplectic geometry for first order observables:
\begin{equation}\label{eq. de est.}
d\hat f^i =- X_{\hat f} \hook d\theta^i
\end{equation}
So we have vector-valued observables ($\hat f^i$) and scalar-valued vector
fields ($X_{\hat f}$), whereas the polysymplectic formalism
has scalar observables and
vector-valued vector fields.

In the polysymplectic formalism there exist corresponding vector fields for all
functions, but these vector fields are not unique.  Contrastingly, in the
first order $m$-symplectic formalism the vector fields are unique, but
only exist for a special class of functions (see Section 9.4).  This uniqueness allows for
the definition of Poisson brackets, which are not available in the
polysymplectic formalism.

The   $m$-symplectic   formalism  extends to allow higher order
observables. For example, in the second order symmetric case we have:
\def\sup{\rule{0 em}{1 em}^}
\begin{equation}
d\hat f^{ij} = -2 X_{\hat f}\sup{(i} \hook d\theta\sup{j)}
\end{equation}
Now we obtain vector-valued vector fields from an
$\r^n\otimes_s\r^n$-valued function.  In fact, we remark that the trace
$\sum_{i=j} f^{ij}$ will satisfy the polysymplectic equation with the
vector field $-2 X_{\hat f}^{i}$. In this second order case the vector
fields are no longer unique, but this does not impede the definition of
Poisson brackets.

For the $p$-th order case in $m$-symplectic geometry we have
\begin{equation}
d\hat f^{i_1\ldots i_p}
 = -p! X_{\hat f}\sup{(i_1\ldots i_{p-1}} \hook d\theta\sup{i_p)}
\qquad\mbox{or}\qquad d\hat f^{i_1\ldots i_p}
 = -p! X_{\hat f}\sup{[i_1\ldots i_{p-1}} \hook d\theta\sup{i_p]}
\end{equation}
for the symmetric and anti-symmetric cases respectively.
The Poisson bracket of a $p$-th and a $q$-th order observable is a $(p+q-1)$-th
order observable.  The full algebra is developed in \cite{No1}.
There is nothing in the polysymplectic formalism to compare   to this in general.

 It has been shown recently that the $n$-symplectic Poisson
brackets defined on frame bundles extends to Poisson brackets on a general polysymplectic
manifold. We present in the next sections  a summary of the general results shown by
Norris \cite{No5} for a general $n$-symplectic (polysymplectic) manifold.

\subsection{General $n$-symplectic geometry}\label{definition general m-symplectic}

Let $P$ be an $N$-dimensional manifold, and let $(\hat{r}_\alpha)$ denote
the standard basis of $\r^n$, with $1\leq n\leq N$.  We suppose  there
exists on
$P$ a   {\bf general $n$-symplectic structure}, namely an
$\r^n$-valued 2-form
$\hat{\omega} = \omega^\alpha \otimes\hat{r}_\alpha$    that 
satisfies the
following two conditions:
\begin{eqnarray}
&(C-1)&\hfill\qquad d\omega^\alpha=0\quad\forall\ \ \alpha=1,2,\dots,n
\label{closed}\\
 &(C-2)&\qquad X\hook\hat{\omega} = 0 \quad\Leftrightarrow\quad X=0
\label{non-degenerate} 
\end{eqnarray}

\begin{definition}
The pair $(P ,\hat{\omega})$ is a general $n$-symplectic manifold.
\label{definition of n-symplectic manifold}
\end{definition}

\rem In references~\cite{No1,No2,No3,No4,FLN1,FLN2,MN} the term $n$-symplectic
structure   refers
to the two-form that is the exterior derivative of the $\r^n$-valued soldering 1-form  on
frame bundles or subbundles of frame bundles. As outlined earlier in this paper  G\"unther~\cite{gun}
was perhaps the first to consider a manifold with a non-degenerate
$\r^n$-valued 2-form, and he used the terms 
  {\bf\it polysymplectic structure} and {\bf\it polysymplectic manifold} for the non-dengenerate 2-form 
and manifold, respectively.  In addition, when one adds two extra conditions to conditions
C-1 and C-2 one arrives at a {\bf\it
$k$-symplectic manifold}. 
 Specifically, if     $P$ is required to support  an $np$-dimensional distribution $V$
such that
\begin{eqnarray*}
&(C-3)&\hfill\qquad N = p(n + 1)
\label{first extra k-symplectic   assumptions}\\
 &(C-4)&\qquad \hat\omega \vert_{V\times V}=0
\label{second extra k-symplectic   assumptions} 
\end{eqnarray*}
then $P$ is a {\em  k-symplectic manifold}  as defined by both de Leon, Salgado, et
 al.~\cite{mc1} and also
by  Awane~\cite{aw1}.  To make this identification one needs to make the   notational
changes $n
\longrightarrow k$ and
$p
\longrightarrow n$ in the above discussion. Thus all $k$-symplectic manifolds are
$n$-symplectic, but not conversely.  The   example
$(LE,d\hat\theta)$ of an
$m$-symplectic manifold introduced in Section $4.4$   is   also a
$k$-symplectic manifold.  On the other hand the important example of the adapted frame
bundle \lpie\   is $m$-symplectic, but not $k$-symplectic.  The
problem is that   the k-symplectic dimensional requirement  
$N=p(m+1)$  cannot be satisifed on
\lpie.
 
We will use the name {\it general $m$-symplectic structure} for the structure in definition
\ref{definition general m-symplectic} in order to emphasis the geometrical and algebraic
developments that  the $m$-symplectic   approach  provides.  Howewer, the definition of a general
$m$-symplectic structure is identical with the definition of a polysymplectic structure.

\bigskip

\subsection{Canonical coordinates}

Awane  \cite{aw1} has proved a generalized Darboux theorem for $k$-symplectic geometry. 
Thus in the neighborhood of each point $u\in P$ one can find canonical (or Darboux)
coordinates
$(\pi^\alpha_a,z^b)$,
$\alpha,\beta=1,2,\dots k$ and $a,b=1,2,\dots n$. With respect to such canonical
coordinates 
$\hat{\omega}$ takes the form
\begin{equation}
\hat{\omega} = (d\pi^\alpha_a\wedge dz^a)\otimes  \hat{r}_\alpha
\label{canonical form}
\end{equation}
Hence we have the following locally defined equations:
\begin{equation}
d\pi^\alpha_a = -\basisz {a}\hook \omega^\alpha\ ,\qquad
dz^a  = \basispi {\alpha}{a} \hook \omega^\alpha\ ,\qquad ( \Sigma_\alpha\kern-12pt /  \kern10pt)
\label{local form of structure equations}
\end{equation}

\rem The $n$-symplectic approach used   to characterize algebras of observables requires the
existence of such canonical coordinates.  From the results in \cite{No1} one knows that not all
functions are {\em allowable
$n$-symplectic} observables, even in the canonical case of frame bundles. Thus, for
example, whether or not there exist pairs
$(\hat f^{\multialpha p},X_{\hat f}^{\multialpha {p-1}})$, $p=1,2,\dots$ that satisfy equation
(\ref{sym structure equation}) below for a general $m$-symplectic manifold is an existence
question.  The formulas
(\ref{local form of structure equations}) will provide local examples of rank 1 solutions
of the $n$-symplectic structure equations (\ref{sym structure equation})
 when either the geometry is specialized to $(k=n)$-symplectic geometry where a Darboux
theorem holds, or when canonical coordinates are simply known to exist. 
Fortunately in the case of   
the adapted frame bundle
\lpie,  canonical coordinates are known to exist.
\bigskip

\example On the bundle of linear frames $\lambda:LE\to E$ one can introduce canonical coordinates in the
  $(z^\alpha,\pi^\alpha_\beta)$ as in ection $2.5$.    With respect to such a coordinate
system on $LE$ the soldering 1-form $\hat\theta$ has the local coordinate expression
\begin{equation}
\hat\theta = (\pi^\alpha_\beta dz^\beta)\otimes  \hat{r}_\alpha
\label{soldering form in canonical coordinates}
\end{equation}
The   $m$-symplectic 2-form $d\hat\theta$   clearly has   the canonical form
(\ref{canonical form})
 in such a coordinate system.\bigskip

\subsection{The Symmetric Poisson Algebra Defined by $\hat{\omega}$}

	In this section we generalize the algebraic structures of $n$-symplectic geometry on frame
bundles to a general $n$--symplectic manifold. Throughout this section we let $(P,\hat\omega)$ be
a   general 
$n$-symplectic manifold as defined above.   It is   convenient to introduce  the
multi-index notation
 $$
\hat{  r}_{\alpha_1 \alpha_2 \dots \alpha_{n-\mu}}   = 
 \hat{r }_{\alpha_1}\otimes_s \hat{  r}_{\alpha_2}\otimes_s\cdots \otimes_s \hat{ 
r}_{\alpha_{n-\mu}}  \ \ ,\qquad 0\leq \mu\leq n-1
$$
 In addition  
round brackets around indices ($\alpha\beta\gamma$) denote symmetrization over the
enclosed indices.

\begin{definition}  For each $p\geq 1$   let $SHF^p$ denote the set of all
$(\otimes_s)^p\r^n$-valued functions $\hat f=(\hat f^{\multialpha p}) =(\hat
f^{(\multialpha p)})$     on $P$  that satisfy the equations
\begin{equation}
d\hat f^{\multialpha p} = - p! X_{\hat f}^{(\multialpha
{p-1}}\hook \omega^{\alpha_p)}
\label{sym structure equation}
\end{equation}
for some set of vector fields $(X_{\hat f}^{\multialpha {p-1}})$.  We then set
\begin{equation}
SHF = \oplus_{p\geq 1} SHF^p
\end{equation}
 $\hat f\in SHF^p$  is a \underline{symmetric Hamiltonian function of rank p}.

\end{definition}
\bigskip

\example The locally defined functions $\hat f$ that satisfy  (\ref{sym structure equation}) 
for the canonical $m$-symplectic manifold $(LE,d\hat\theta)$ were given in
reference~\cite{No1}. In particular, contrary to the situation in symplectic geometry, not all 
$(\otimes_s)^p\r^m$-valued functions on $LE$ are compatible with equation (\ref{sym
structure equation}). The $p=1,2$
cases will clarify the structure.  Let
$ST^p(LE)$ denote the vector space of symmetric $(\otimes_s)^p\r^m$-valued
GL(m)-tensorial functions     on $LE$ that correspond uniquely to symmetric rank $p$
contravariant tensor fields on $E$.  Similarly let $C^\infty(E,(\otimes_s)^p\r^m)$ denote
the set of smooth $(\otimes_s)^p\r^m$-valued functions on $LE$ that are constant on fibers
of $LE$.   Then
 \begin{eqnarray}
 SHF^1& =&ST^1(LE) +C^\infty(E,\r^m)  \label{the rank 1 algebra} \\
  SHF^2&=&ST^2(LE) +T^1(LE)\otimes_s C^\infty(E,\r^m)+ C^\infty(E,\r^m\otimes_s \r^m)
\label{the rank 2 algebra}\\ \nonumber
\end{eqnarray}
\noindent For example, if $\hat f= (\hat f^{\alpha })\in SHF^1$ and 
$\hat f= (\hat f^{\alpha\beta})\in SHF^2$, then in canonical coordinates
$(\pi^\alpha_\beta,z^\gamma)$ the functions $\hat f^{\alpha}$ and $\hat f^{\alpha\beta}$ have the general forms
\begin{equation}
\hat f^{\alpha } = {A}^{a } \pi^\alpha_a   + {B}^{\alpha } \ ,\qquad\hat f^{\alpha\beta} = A^{\mu\nu}
\pi^\alpha_\mu\pi^\beta_\nu + B^{\mu(\alpha} \pi_\mu^{\beta)} + C^{\alpha\beta} 
\label{generic rank 1 and 2 fhat}
\end{equation}
where  ${A}^{a } $, $ {B}^{\alpha }$, 
$A^{\mu\nu}=A^{(\mu\nu)}$, $B^{\mu\nu}$ and $C^{\mu\nu}=C^{(\mu\nu)}$ are all constant on the fibers
of $\lambda:LE\to E$ and hence are pull-ups of functions defined on $E$.  
\bigskip

  The analogous results for the general $n$-symplectic form given in (\ref{canonical
form}) above are straight forward to work out in canonical coordinates.  For the
$p=1$ and $p=2$ symmetric cases, one finds:
\begin{equation}
 \hat f^{\alpha } = {\cal A}^{a } \pi^\alpha_a   + {\cal B}^{\alpha } \ ,\qquad
\hat f^{\alpha\beta} = {\cal A}^{ab} \pi^\alpha_a\pi^\beta_b +
{\cal B}^{a(\alpha} \pi_a^{\beta)} + {\cal C}^{\alpha\beta} 
\label{rank 2 example in darboux coordinates}
\end{equation}
where now all coefficients are functions of the coordinates $z^a$.

     Although $\hat{\omega}$ is non-degenerate in the sense given in equation
(\ref{non-degenerate}) above, because of the symmetrization on the right-hand-side in
(\ref{sym structure equation}) the relationship between $\hat f$ and $(X_{\hat
f}^{\multialpha {p-1}})$ is not unique unless $p=1$. Given a pair $(\hat f^{\multialpha
p},X_{\hat f}^{\multialpha {p-1}})$ that satisfies (\ref{sym structure equation}) one can
always add to $X_{\hat f}^{\multialpha {p-1}}$ vector fields $Y^{\multialpha {p-1}}$ that
satisfy the kernel equation
\begin{equation}
Y^{(\multialpha {p-1}}\hook \hat{\omega}^{\alpha_p)} = 0
\label{kernel condition}
\end{equation}
to obtain a new pair $(\hat f^{\multialpha p},\bar{X}_{\hat f}^{\multialpha {p-1}})$ that also satisfies
(\ref{sym structure equation}), where
\[
\bar{X}_{\hat f}^{\multialpha {p-1}} =  {X}_{\hat f}^{\multialpha {p-1}} + Y^{\multialpha
{p-1}}
\]
Hence we associate with $\hat f\in SHF^p$ an equivalence class of
$(\otimes_s)^{p-1}\r^n$-valued vector fields, which we denote by $\vectorclass f =
\vectorclassindicesinside f{\multialpha {p-1}}{\rhat {\multialpha {p-1}}}$.    We will
see below that even though we obtain equivalence classes of Hamiltonian vector fields
rather than vector fields, the geometry still carries natural   algebraic structures.

\begin{definition}  For each $p\geq 1$   let $SHV^p$ denote the vector space of all 
equivalence classes of 
$(\otimes_s)^{p-1}\r^n$-valued
vector fields $\vectorclass f = \vectorclassindicesinside f{\multialpha {p-1}}{\rhat {\multialpha
{p-1}}}$ on $P$  that satisfy the equations (\ref{sym structure equation})
 for some $\hat f= \hat f^{\multialpha p} \rhat {\multialpha p} \in SHF^p$.  
  We then set
\begin{equation}
SHV = \oplus_{p\geq 1} SHV^p
\end{equation}
 $\vectorclass f$   will be referred to as the   \underline{generalized rank p Hamiltonian vector field
defined by $\hat f$}.

\end{definition}

\example 
The Hamiltonian vector field $\xhatf$ for the rank 1 element in (\ref{generic rank 1 and 2 fhat}) is unique, and has
the form
\begin{equation}
\xhatf = A^\alpha\basisz \alpha - (\frac {\partial A^\beta}{\partial z^\gamma}\pi_\beta^\alpha + \frac {\partial
B^\alpha}{\partial z^\gamma})\basispi \alpha \gamma
\label{hamiltonian vector field for general rank 1}
\end{equation}
The  equivalence class
of  $\r^m$-valued Hamiltonian vector fields corresponding to the rank 2 element in
(\ref{generic rank 1 and 2 fhat}) on $LE$ has representatives of the form
\begin{equation}
\xhatf^{\alpha} =(A^{\mu\nu}\pi^\alpha_\mu  + B^{\nu\alpha})\basisz {\nu} -\frac 12\left(\popo
{A^{\mu\beta}}{z^\gamma} \pi_\mu^\alpha\pi^\nu_\beta +  \popo {B^{\mu(\alpha}}{z^\gamma}\pi_\mu^{\nu)}+ 
\popo {C^{\alpha\nu}}{z^\gamma}\right)\basispi {\nu}{\gamma} +Y^{\alpha\nu}_\gamma\basispi {\nu}{\gamma}
\label{hamiltonian vector field for general rank 2}
\end{equation}
where $Y^{\alpha\beta}_\gamma$ are functions subject to the constraint
\[
Y^{(\alpha\beta)}_\gamma = 0
\]
but are otherwise completely arbitrary.  The fact that $Y^\alpha=Y^{\alpha\mu}_\nu\basispi \mu\nu$ is purely
vertical on $\lambda:LE\to E$ follows from (\ref{kernel condition}).

    For the $n$-symplectic rank 2 symmetric observable given above in (\ref{rank 2
example in darboux coordinates}), one can check easily that the local coordinate form of a
representative $X^\alpha_{\hat f}$ of the equivalence class of Hamiltonian vector fields
$\vectorclass f^\alpha$ that satisfies (\ref{sym structure equation}) has the form
\begin{equation}
X^\alpha=({\cal A}^{ab}\pi_a^\alpha + {\cal B}^{b\alpha})\basisz b -\frac 12 \left(  
\popo{{\cal A}^{ab}}{z^d} \pi_a^\alpha\pi^\sigma_b +  \popo {{\cal
B}^{a(\alpha}}{z^d}\pi_a^{\sigma)}+ 
\popo {{\cal C}^{\alpha\sigma}}{z^d}\right)\basispi {\sigma}{d} +Y^{\alpha }  
\end{equation}

\subsubsection{Poisson Brackets}

	We show that the $n$-symmetric Poisson brackets defined  on frame bundles can also be defined in
a general $n$-symplectic manifold.

\begin{definition}\label{symmetric algebra} For $p,q\geq 1$ define a  map
$\{\ ,\ \}:SHF^p\times SHF^q\to SHF^{p+q-1}$ as follows.  For
 $\hat f=f^{\multialpha p}\multiRbasisalpha p \in SHF^p$
and
$\hat g =g^{\multibeta q}\multiRbasisbeta q\in
SHF^q$  
\begin{equation}
\pb f g^{\alpha_1 \alpha_2\dots \alpha_{p+q-1}} :=p!X_{\hat f}^{(\alpha_1 \alpha_2 \dots
\alpha_{p-1}}\left(\hat g^{\alpha_p
\alpha_{p+1} \dots \alpha_{p+q-1})}\right)
\label{symmetric bracket defined}
 \end{equation}
where $\xhatf^{\alpha_1 \alpha_2 \dots \alpha_{p-1}}$ is any set of representatives of the
equivalence class $\vectorclass f $.  

\end{definition}

We need to make certain that $\pb f g $ is well-defined.  Suppose  we have two representatives
$  {X}_{\hat f}^{\multialpha {p-1}} $ and 
$\bar{X}_{\hat f}^{\multialpha {p-1}} =  {X}_{\hat f}^{\multialpha {p-1}} + Y^{\multialpha
{p-1}}$ of $\vectorclass f$.  Then it follows easily from (\ref{kernel condition}) that
\[
\bar{X}_{\hat f}^{(\alpha_1 \alpha_2 \dots
\alpha_{p-1}}\left(\hat g^{\alpha_p
\alpha_{p+1} \dots \alpha_{p+q-1})}\right)
=
\xhatf^{(\alpha_1 \alpha_2 \dots
\alpha_{p-1}}\left(\hat g^{\alpha_p
\alpha_{p+1} \dots \alpha_{p+q-1})}\right)
\]
Hence the bracket is independent of choice of representatives.  That $\pb fg $ actually is in 
$SHF^{p+q-1}$ will follow from   Corollary (\ref{the corollary}) below.

\begin{definition}
Let 
$ \vectorclass f=\vectorclassindicesinside f{\alpha_1 \alpha_2 \dots \alpha_{p-1}}{{\hat
r}_{\alpha_1\alpha_2\dots
\alpha_{p-1}}}$  
 and $ \vectorclass g=\vectorclassindicesinside g{\alpha_1 \alpha_2 \dots
\alpha_{p-1}}{{\hat r}_{\alpha_1\alpha_2\dots \alpha_{p-1}}}
$ 
denote the equivalence classes of vector-valued   vector fields determined by
$\hat f\in SHF^p$ and $\hat g\in SHF^q$, respectively.  Define a bracket
$\lbrack\!\lbrack\ ,\  \rbrack\!\rbrack:SHV^p\times SHV^q\to SHV^{p+q-1}$ by
\begin{eqnarray}
\lbrack\!\lbrack \vectorclass f, \vectorclass g\rbrack\!\rbrack 
   & = &
\lbrack\!\lbrack[\xhatf^{(\alpha_1 \alpha_2 \dots \alpha_{p-1}}\ ,\ \xhatg^{\alpha_p \alpha_{p+1} \dots
\alpha_{p+q-2})}]
   {\hat r}_{\alpha_1\alpha_2\dots \alpha_{p+q-2}}\rbrack\!\rbrack
\label{bracket for SHV defined}
\end{eqnarray}
where the "inside" bracket on the right-hand side is the ordinary Lie bracket of vector fields
calculated using arbitrary representatives.  (Notice  the  symmetrization over
all the upper indices in this equation.)  
\end{definition}

We again need to show that this bracket is well-defined.  This is shown 
in the following lemma, in which we
will need the formula
\begin{equation}
L_{X^{(J}} \omega^{\alpha)}=0
\label{basic fact}
\end{equation}
which follows easily from (\ref{sym structure equation}) and the formula $L_X\omega = X\hook d\omega +
d(X\hook \omega)$.  In (\ref{basic fact})   $J$ denotes the multiindex $\multialpha {p-1}$, and $X^J$
denotes a representative  of a rank p Hamiltonian vector field satisfying equations (\ref{sym structure
equation}).  The next lemma shows that the  bracket  defined in (\ref{bracket for SHV defined}) is (i)
independent of choice of representatives, and (ii)  closes on the set of equivalence classes of
vector-valued Hamiltonian vector fields.  The proof of the lemma can be found in
\cite{No5}, which is quite similar to the proof of the analogous result in symplectic geometry.

\begin{lem} Let $ \vectorclass f $  
 and $ \vectorclass g $ 
denote the equivalence classes of vector-valued   vector fields determined by
$\hat f\in SHF^p$ and $\hat g\in SHF^q$, respectively. Then 
\begin{equation}\label{eq95}
\lbrack\!\lbrack \vectorclass f,\vectorclass g \rbrack\!\rbrack = \frac {(p+q-1)!}{p!\ q!} 
\vectorclass {\pb fg}
\end{equation}

\end{lem}

\begin{cor}
\label{the corollary}
\[
\pb fg \in SHF^{p+q-1}
\]
\end{cor}

\begin{thm} $(SHV,\lbrack\!\lbrack\ ,\ \rbrack\!\rbrack)$ is a Lie Algebra.
\end{thm}

\proof The bracket defined in (\ref{bracket for SHV defined}) is clearly anti-symmetric.  To check the
Jacobi identity we note that we only need check it for arbitrary representatives, and we may use the very
definition (\ref{bracket for SHV defined}) for the calculation.  Since the "inside" bracket on the
right-hand-side in (\ref{bracket for SHV defined}) is the ordinary Lie bracket for vector fields, we
see that the bracket defined in (\ref{bracket for SHV defined}) also must obey the identiy of
Jacobi.\blob
\bigskip  

We can now show that $SHF$ is a Poisson algebra under the bracket defined in (\ref{symmetric bracket
defined}).\bigskip

\begin{thm}
$(SHF, \npbblank)$ is a Poisson algebra over the commutative algebra $(SHF,\otimes_s)$.
\end{thm}
 
\proof The symmetrized tensor product ${}\otimes_s{}$
 makes $SHF$ into a commutative algebra.
If we now consider again elements $\hat f\in SHF^p$, $\hat g\in SHF^q$ and $\hat h\in SHF^r$, then by using
definition (\ref{symmetric bracket defined})   one may show that
\begin{equation}
\pbnorighthat f {\hat g\otimes_s \hat h}=
\pb f g \otimes_s \hat h + \hat g\otimes_s\pb f h\ \ \ .
 \end{equation}
\noindent Thus the  bracket defined in (\ref{symmetric bracket defined}) acts as a derivation on the
commutative algebra. \blob

 \bigskip

\example In the canonical case $P=LE$ the Poisson brackets just defined have a well-known
interpretation. As mentioned above the homogeneous elements in $SHF^p$  make up the space
$ST^p(LE)$, the symmetric rank p GL(m)-tensorial functions that correspond to symmetric rank~p
contravariant tensor fields on $E$. Then   $ST=\oplus_{p\geq 1} ST^p\subset SHF$, and the bracket
$\npbblank:ST^p\times ST^q\to ST^{p+q-1}$ has been shown~\cite{No4} to be the frame bundle
version of the  \underline{Schouten-Nijenhuis} bracket  of the corresponding symmetric tensor
fields on $E$.  

 There is also a Schouten-Nijenhuis bracket for \underline{anti-symmetric} contravariant tensor fields
on $E$, and as one might expect this bracket also extends to $LE$.  This leads to a graded
$m$-symplectic Poisson algebra of anti-symmetric tensor-valued functions on $LE$~\cite{No3}.

\subsection{The Legendre Transformation in $m$-symplectic theory   on $\lpie$}

One can define the CHP 1-forms,  defined above in Definition \ref{definition of the chp 1-forms},
using a frame bundle version of the Legendre transformation.  Given a lifted Lagrangian $\lag
:\lpie\to \r$ we obtain a mapping
$\phi_{\lag }:\lpie\to LE$ given by
 
\begin{equation}
\phi_{\lag }(u)=\phi_{\lag }(e,e_i,e_A)=\left(e,\frac{1}{\tau\lag (u)}e_i,
e_A-\frac{1}{\tau\lag (u)}\vertical aA(\lag )(u)e_a\right)
\label{frame legendre transformation defined}
\end{equation}
 The condition that this mapping end up in $LE$  is that the Lagrangian
be {\bf non-zero}, and {\em for the rest of this paper we will assume this condition}.   We   refer to
this mapping as the {\it $m$-symplectic Legendre transformation}.  Our goal is to prove Theorem
(\ref{pull back of soldering form is chp form}), namely that $\hat\theta_{\lag}
=\phi_{\lag}^*(\hat\theta)$ where
$\hat\theta$ is the canonical soldering 1-form on the image  $\qsubl$  of $\phi_{\lag}$.   

To clarify the meaning of the Legendre transformation (\ref{frame legendre
transformation defined}) we introduce a new manifold $\tilde P $ as follows. Let $J $ denote the
subgroup of $GL(n)$ consisting of matrices of the form 
\[
\left(
\begin{array}{cc}
I & \xi  \\
0 & I\\
\end{array}
  \right)\qquad \xi\in\r^{n\times k}
\]
   Define $\tilde P$  by 
\begin{equation}
\tilde P = \lpie \cdot J  = \{ (e_i,e_A+\xi^j_A e_j )\ | \ (e_i , e_A)\in\lpie \ ,\
\xi\in\r^{n\times k}\}
\label{definition of P}
\end{equation}
    
\bigskip

 We collect together the pertinent results that are proved in \cite{MN,No5} and that lead up to
Theorem  (\ref{pull back of soldering form is chp form})

\begin{lem} $\tilde P$  is a open dense submanifold of the bundle  of  frames $LE $ of $E$.
\label{mike's lemma}
 \end{lem}

\begin{lem}   There is a canonical diffeomorphism from $\tilde P$  to the product manifold 
$ \lpie \times\r^{m\times k} $.  
\end{lem}

Using this fact one can the prove the following lemma. We let $Q_L$ denote the range of the
Legendre transformation.

\begin{lem}  If the Lagrangian \lag\  is non-zero, then the Legendre transformation $\phi_{\lag}:\lpie\to \qsubl$  
is a diffeomorphism.

\end{lem}

  These facts taken together lead to the following fundamental theorem:

\begin{thm}\label{pull back of soldering form is chp form}
Let \lag\  be the pull-up to $\lpie$ of a non-zero Lagrangian   on \jetpi, and    let 
$\phi_{\lag} $ denote the $m$-symplectic Legendre transformation defined above in
(\ref{frame legendre transformation defined}).  Then
\begin{equation}
\hat\theta_{\lag} =\phi_{\lag}^*(\hat\theta)
\label{theta L is pull back of theta}
\end{equation}
\end{thm}

\proof The proof is a direct calculation using the definition (\ref{frame legendre transformation
defined}).

\rem This theorem has an obvious analogue in symplectic mechanics, where the symplectic form on the
velocity phase space $TE$ is, for a regular Lagrangian, the pull back under the Legendre transformation of
the canonical 1-form on $T^*M$.  There is also a similar theorem in {\em multisymplectic geometry} where the CHP
m-form on  $\jetpi$  is known~\cite{GIMMsy} to be the pull back of the canonical multisymplectic
m-form on
\cojet.\bigskip

Now \qsubl, being a submanifold of $LE$, supports the restriction $\hat\theta\vert_{\qsubl} $ of
the  $\r^m$-valued  soldering 1-form 
$\hat\theta $.  It is easy to verify that the closed $\r^m$-valued 2-form  
$d\hat\theta\vert_{\qsubl} $ is also non-degenerate, and hence
$(\qsubl,d(\hat\theta\vert_{\qsubl}))$ is an $m$-symplectic manifold. Using the fact that $\qsubl$ 
and \lpie\  are diffeomorphic under the Legendre transformation, we obtain the following corollary
to Theorem \ref{pull back of soldering form is chp form}.

 \begin{cor}   $(\lpie, d\hat\theta_{\lag})$ {\it is an m-symplectic manifold}.
 \label{the fundamental corollary}
 \end{cor}

To find the {\em allowable observables} of this theory one can set up \cite{No5} the equations of
$m$-symplectic reduction to find the subset of $m$-symplectic observables on $LE$ that reduce to the
submanifold
\qsubl.

\subsection{The Hamilton-Jacobi and Euler-Lagrange   equations in $m$-symplectic theory on
$\lpie$}

Working out the local coordinate form of the CHP-1-forms, given in 
Definition \ref{definition of the chp 1-forms}, in  Lagrangian coordinates one finds 
\begin{eqnarray}
\theta^i_{L} &=& -H^i_j dx^j+P^i_A dy^A \label{theta i l}\\
\theta^A_{L} &=& P^A_j dx^j + P^A_B dy^B 
\end{eqnarray}
where 
\begin{eqnarray}
H^i_j &=& u^i_k(p^k_B u^B_j -\tau(n) L \delta^k_j) \label{covariant hamiltonian}\\
P^i_B &=& u^i_k p^k_B \label{covariant canonical momentum}\\
P^A_j &=& -u^A_B u^B_j   \\
P^A_B & =& u^A_B \label{P AB defined}
\end{eqnarray}

The $H^i_j$ are the components  of the {\bf covariant Hamiltonian}, and
  the $P^i_B$ are the components of the {\bf covariant canonical momentum} \cite{MN}.
Defining symbols $h^k_j$ by the formula
\begin{equation}
h^k_j=p^k_B u^B_j -\tau(n) L \delta^k_j
\label{little hij defined}
\end{equation}
the covariant Hamiltonian (\ref{covariant hamiltonian}) can be
expressed as $H^i_j=u^i_k h^k_j$.  Setting $\tau(n)=1$ one finds that
$h^i_j$ has the form of \car's Hamiltonian  tensor \cite{Rund,Car}.
Similarly, setting $\tau=\frac 1n$ one finds that $h=h^i_i$ yields
the Hamiltonian in the de Donder-Weyl theory~\cite{Rund,DW}.

\subsubsection{The $m$-symplectic Hamilton-Jacobi Equation on $\lpie$}\label{hamilton jacobi}

The \car-Rund and de Donder-Weyl Hamilton-Jacobi equations occur as special cases of a
general Hamilton-Jacobi equation that can be set up on $\lpie$.
Proceeding by analogy with the time independent
Hamilton-Jacobi theory we seek
Lagrangian submanifolds of $\lpie$.
However, since the dimension of $\lpie$ is in general not twice the dimension
of $E$, a new definition is needed.  For our purposes here we will consider
$m=n+k$ dimensional submanifolds of $\lpie$ that arise as sections of $\lambda$.
In particular we consider  sections $\sigma:E\to\lpie$ that
satisfy
\begin{equation}
\sigma^*(d\theta^\alpha_{L})=0
\label{lagrangian submanifold condition}
\end{equation}
These are  the {\em $m$-symplectic Hamilton-Jacobi equations} \cite{MN}.

Since  $\sigma^*(d\theta^\alpha_{L})=d\left(\sigma^*(\theta^\alpha_{L})\right)$
the condition (\ref{lagrangian submanifold condition}) asserts that the 1-forms
$\sigma^*(\theta^\alpha_{L})$ are locally exact, and we express this as
\begin{equation}
\sigma^*(\theta^\alpha_{L})=dS^\alpha
\label{definition of s functions}
\end{equation}
in terms of $m=n+k$ new functions $S^\alpha$ defined on open subsets of $E$.
For convenience we will denote   objects on $\lpie$ pulled back
to $E$ using $\sigma$ with an over-tilde. Thus, for example,
$
\tilde H^i_j=H^i_j\circ \sigma$ and $ \tilde P^i_A=P^i_A\circ\sigma
$.
Then we get from (\ref{covariant hamiltonian})--(\ref{P AB defined}) 
and (\ref{definition of s functions})  
\begin{eqnarray}
\mbox{\bf(a)}\ \ \tilde H^i_j&=&-\frac{\partial S^i}{\partial x^j}\ ,\hskip1truein
\mbox{\bf(b)}\ \ \tilde P^i_A=\frac{\partial S^i}{\partial y^A}  
\label{h and p in terms of S}\\
\mbox{\bf(a)}\ \ \tilde u^A_B\tilde u^B_j&=&-\frac{\partial S^A}{\partial x^j}\ ,\hskip1truein
\mbox{\bf(b)}\ \ \tilde u^A_B=\frac{\partial S^A}{\partial y^B}  
\label{extra hj equations}
\end{eqnarray}
Recalling that $H^i_j=P^i_Bu^B_j-\tau(n) L u^i_j$ and $P^i_A$
 are functions of the coordinates $x^i$, $y^A$, $u^i_j$ and $u^A_i$, equations 
(\ref{h and p in terms of S}) can be combined into the single equation
\begin{equation}
 H^i_j(x^a,y^B,u^a_b,u^B_a,\frac{\partial S^i}{\partial y^B})\circ\sigma
=-\frac{\partial S^i}{ \partial x^j}
\label{generalized hj equation}
\end{equation}
Similarly combining equations (\ref{extra hj equations}) we obtain
\begin{equation*}
\frac{dS^A}{dx^j}=0
\end{equation*}

\noindent  We next consider special cases of these  {\bf $m$-symplectic Hamilton-Jacobi equations}.

\subsubsection{The Theory of \car\ and Rund} 

We note from (\ref{covariant hamiltonian}), (\ref{covariant canonical momentum}), 
and (\ref{little hij defined})  that $H^i_j=u^i_kh^k_j$ and
$P^i_A=u^i_kp^k_A$, where the matrix of functions $(u^i_j)$ is $\Gln[n]$-valued.
Using the notation $P^i_j=-H^i_j$ and $\tilde u^i_j=u^i_j\circ\sigma$
 we may rewrite (\ref{covariant hamiltonian}) and (\ref{covariant canonical momentum})
 in the form
\begin{equation}
\tilde P^i_j=-\tilde u^i_k\tilde h^k_j\ ,\hskip1truein 
\tilde P^i_A=\tilde u^i_k\tilde p^k_A\ 
\label{rund's equations}
\end{equation}
If we take $t(n)=1$  then these equations are the equations defining
the {\em canonical momenta} in Rund's canonical formalism
 for \car's
geodesic field theory (see equations (1.22), page 389 in \cite{Rund}, with the obvious
change in notation).
In this situation equation (\ref{generalized hj equation}) can be identified
with the Rund's Hamilton-Jacobi equation  for \car's 
theory (see equation (3.29) on page 240 in \cite{Rund}).
  We recall \cite{Rund} that one can
derive the Euler-Lagrange field equations from this Hamilton-Jacobi
equation.

 In (\ref{rund's equations}) we have the result that
the arbitrary non-singular matrix-valued functions $(\tilde u^i_j)$ that occur 
in Rund's canonical
formalism for \car's theory have a geometrical interpretation in the present
setting.  Specifically they correspond to the coordinates for  linear
frames for $M$. These defining relations are derived from Rund's
{\bf transversality condition}, and  we now show that this condition has the elegant reformulation
  as the kernel of $(\theta^i_{\lagrangian})$.

We will say that a vector $X$ at $e\in E$ is transverse to a solution surface through $e$
that is defined by a given Lagrangian $L$, if $X=d\lambda(\hat X)$,
where $\hat X\in T_u(\lpie)$ satisfies $\hat X\hook\theta^i_{L}=0$,
for some $u\in\lambda^{-1}(e)$.  $\hat X$ thus satisfies the equations
\begin{equation*}
\begin{array}{l}
0=-H^i_j X^j+P^i_A X^A=u^i_k\left(-h^k_j X^j+p^k_A X^A  \right) \\
\\
\quad X^j=\hat X(x^j)\ ,\quad X^A=\hat X(y^A)
\end{array}
\end{equation*}
from which we infer
\begin{equation}
0=-h^k_j X^j+p^k_A X^A 
\label{rund's transversality condition} 
\end{equation}
This is Rund's transversality condition  for the theory of \car\
when we take $\tau (n)=1$ (see equation (1.10), page 388 in~\cite{Rund}).
 The canonical momenta $P^i_j$ and $P^i_A$ are
defined by Rund to be solutions of
\begin{equation}
0=P^i_j X^j+P^i_A X^A
\label{defining relation}
\end{equation}
when $(X^j,X^A)$ satisfy (\ref{rund's transversality condition}).
Rund's solutions of these equations are given in (\ref{rund's equations}).
Looking at (\ref{rund's equations}),  (\ref{rund's transversality condition}) and 
(\ref{defining relation}) we see that the introduction of the
$u^i_j$ in (\ref{rund's equations}) amounts to the introduction
of the $\Gln[n]$ freedom for linear frames for $M$.

\subsubsection{de Donder-Weyl Theory} Returning to (\ref{generalized hj equation})
let us reduce this equation by making several assumptions. We suppose that
$\lagrangian$ is regular (in the usual sense on $\jetpi$), that  
the section $\sigma$ is such that $\tilde u^i_j=\delta^i_j$, and we make the
choice $\tau (n)=\frac{1}{n}$.  Now
 summing $i=j$ in (\ref{generalized hj equation}) we obtain
\begin{equation*}
\tilde   h(x^i,y^B,\frac{\partial S^i}{\partial y^B})=-\frac{\partial S^i}{\partial x^i}
\end{equation*}
where $\tilde h=\tilde p^i_A \tilde u^A_i-\tilde{L}$.
 This equation is the  Hamilton-Jacobi equation of the de Donder-Weyl theory,
as presented by Rund (see equation (2.31) on page 224 in~\cite{Rund}).
  We recall~\cite{Rund} that one can  derive in this case also the Euler-Lagrange 
field equations from the de Donder-Weyl Hamilton-Jacobi equation.

\subsection{Hamilton  Equations in $m$-symplectic geometry}

The structure of equations (\ref{theta i l}) - (\ref{covariant canonical momentum})
suggests that one should be able to derive generalized Hamilton equations if
the canonical momenta $p^i_A=\frac{\partial L}{\partial u^A_i}$ can
be introduced as part of a local coordinate system on $\lpie$.  Part of the  original philosophy
used in developing $m$-symplectic geometry in reference~\cite{No1} was to
switch from scalar equations to tensor equations, motivated by the fact
that the soldering 1-form is vector-valued.  In particular, the basic
structure equation (\ref{eq. de est.}) in $m$-symplectic geometry is tensor-valued. We show next
that 
\begin{equation}
u^*(\eta\hook d\theta^i_L)=0\ 
\label{new ham equation}
\end{equation}
where $u:M\to\lpie$ is a section of $\pi\circ\lambda$,
and $\eta$ is any vector field on $\lpie$, yields generalized
canonical equations that contain known canonical equations as special cases.
 We consider here only $d\theta^i_{L}$ since 
by Proposition (\ref{quotient prop}) it alone is needed to construct the CHP-$m$-form
on $\jetpi$.

We need the following definition in order to  introduce
the canonical momenta as part of a  coordinate system on $\lpie$.

\begin{definition} A Lagrangian $L$ on $\lpie$ is {\bf regular} if
the $(n+k)\times (n+k)$ matrix
\begin{equation*}
\left(\vertical iA\circ\vertical jB(L)    \right)
\end{equation*}
is non-singular.
\end{definition}
Working out the terms of this matrix in Lagrangian coordinates using (\ref{estar in lpie})  we
obtain
\begin{equation*}
\vertical iA\circ\vertical jB(L)    
=   u^j_a u^i_b v^E_B v^D_A \left(\frac{\partial^2L}{\partial u^E_a\partial u^D_b}  \right)
\end{equation*}
It is clear that this definition is equivalent to the
standard definition of regularity on $\jetpi$. 

 We now consider the
transformation of coordinates from the set $(x^i,y^A,u^i_j,u^A_k,u^A_B)$
to the new set $(\bar x^i,\bar y^A,\bar u^i_j,p^j_A,\bar u^A_B)$ where
\begin{equation*}
\bar x^i=x^i\ ,\quad \bar y^A= y^A\  ,\quad \bar u^i_j=u^i_j\ ,\quad\bar u^A_B=u^A_B\ ,\quad 
p^i_A=\frac{\partial L}{\partial u^A_i}
\end{equation*}
  Computing the Jacobian
one finds that the new barred functions will be a proper coordinate system
whenever the Lagrangian is regular. For  the remainder of  this section we shall assume
 that $L$ has this property, despite the fact that 
 many important examples (see~\cite{GIMMsy, GIMMsy2})  have non-regular Lagrangians.
    Moreover, for simplicity we will drop
the bars on the new coordinates.

In the generalized canonical equation (\ref{new ham equation}) we now take
$\eta=\frac{\partial}{\partial p^i_A}$. We find the result
\begin{equation*}
0=\left(\frac{\partial H^j_k}{\partial p^i_A}\circ u  \right)+
(u^j_i\circ u)\left(\frac{\partial (y^A\circ u)}{\partial x^k}  \right)
\end{equation*}
Using $H^j_k=u^j_i h^i_k$ and the fact that $(u^i_j)$ is a non-singular
matrix valued function, this last equation reduces to

\begin{equation*}
\frac{\partial h^j_k}{\partial p^i_A}\circ u=\frac{\partial(y^A\circ u)}{\partial x^k}\delta^j_i
\end{equation*}
This is our first set of  {\bf $m$-symplectic Hamilton equations}.  Notice that by
summing $j=k$ in this equation we obtain
\begin{equation}
\frac{\partial h}{\partial p^i_A}\circ u=\frac{\partial(y^A\circ u)}{\partial x^i}
\label{first  de Donder Weyl hamilton equation}
\end{equation}
Upon setting  $\tau(n)=\frac{1}{n}$ we obtain half of the de Donder-Weyl
canonical equations.  Under suitable but complicated conditions these equations,
with $\tau(n)=1$,  will also reproduce part of Rund's canonical equations for
the theory of \car.

In the generalized canonical equation (\ref{new ham equation}) we now take
$\eta=\basisy A$. We find 
\begin{equation*}
0=u^*\left(d(u^i_kp^k_A)+u^i_k\frac{\partial h^k_j}{\partial y^A}dx^j\right)			  
 \end{equation*}
Using an ``over bar''  notation to denote 
 objects pulled back to $M$ by $u$ we may write this as
\begin{equation}
\frac{\partial}{\partial x^j}\left(\bar u^i_k\bar p^k_A\right)=
-\bar u^i_k\left(\frac{\partial h^k_j}{\partial y^A}\right)\circ u
\label{second set of hamilton equations}
\end{equation}
 This is our second
set of {\bf $m$-symplectic Hamilton  equations}.

Notice that what is non-standard in  (\ref{second set of hamilton equations}) 
is the appearance of the  derivatives of the functions $\bar u^i_j=u^i_j\circ u$.
If, however, the section $u:M\to \lpie$ is such that the $\bar u^i_j$ are constants,
 then these equations reduce to
\begin{equation*}
\frac{\partial(\bar p^k_A)}{ \partial x^j}=-\frac{\partial h^k_j}{\partial y^A}\circ u
\end{equation*}
Setting  $\tau(n)=\frac{1}{n}$ and summing $k=j$ in this equation we obtain
\begin{equation*}
\frac{\partial(\bar p^i_A)}{ \partial x^i}=-\frac{\partial h}{\partial y^A}\circ u
\end{equation*}
These equations, together with equations (\ref{first  de Donder Weyl hamilton equation})
when $\tau(n)=\frac{1}{n}$, are the complete canonical equations in the de Donder-Weyl theory.

\section*{Acknowledgments}

This work was Partially supported by grants DGICYT (Spain) PB97-1257,
  PGC2000-2191-E  
 and   PGIDT01PXI20704PR.


\begin{thebibliography}{99}

\bibitem{GS} H. Goldschmidt, S. Sternberg:
The Hamilton-Cartan formalism in the calculus of variations. 
{\sl Ann. Inst. Fourier} {\bf 23} (1973), 203-267.

\bibitem{b} M. Bruckheimer:
Ph.D. dissertation, University of Southampton, 1960.

\bibitem{cgc} R.S. Clark, D.S. Goel:
An almost cotangent manifolds.
{\sl J. Differential Geom.} {\bf 9} (1974), 109-122.

\bibitem{gun} Ch. G\"{u}nther: The polysymplectic Hamiltonian
formalism in field theory and calculus of variations I: The local
case. {\sl J. Differential Geom.} {\bf 25} (1987), 23-53.

\bibitem{mc1} M. de Le\'{o}n, I. M\'{e}ndez, M. Salgado:
$p$-Almost cotangent structures.
{\sl Boll. Unione Mat.
Ital.} (7) {\bf 7-a} (1993), 97-107.

\bibitem{mc2} M. de Le\'{o}n, I. M\'{e}ndez, M. Salgado:
Regular $p$-almost cotangent structures.
{\sl J. Korean Math. Soc.} {\bf25}, (1988), No.2, 273-287.

\bibitem{aw1} A. Awane: $k$-symplectic structures,
{\sl J. Math. Phys.} {\bf 33} (1992), 4046-4052.

\bibitem{aw2} A. Awane: $G$-spaces $k$-symplectic homog\`enes,
{\sl J. Geom. Phys.} {\bf 13} (1994), 139-157.

\bibitem{No1} L.K. Norris:
{Generalized symplectic geometry on the frame bundle of a manifold},
Lecture given at the {AMS
Summer Research Institute on Differential Geometry, 1990}, at U.C.L.A.

\bibitem{No2} L.K. Norris:
Generalized symplectic geometry on the frame bundle of a manifold,
{\sl Proc. Symp. Pure Math.}  {\bf 54}, Part 2
(Amer. Math. Soc., Providence RI, 1993), 435-465.
\bibitem{No3} L.K. Norris: Symplectic geometry on $T^*M$ derived from
$n$-symplectic geometry on $LM$.
{\sl J.~Geom.\ Phys.} {\bf 13} (1994), 51-78.
\bibitem{No4} L.K. Norris: Schouten-Nijenhuis Brackets.
{\sl J. Math.\ Phys.} {\bf  38} (1997), 2694-2709.


\bibitem{cb} R.S. Clark, M. Bruckheimer:
Sur les estructures presque tangents.
{\sl C. R. Acad. Sci. Paris S\'er . I Math.} {\bf 251} (1960), 627-629.

\bibitem{e} H.A. Eliopoulos:
Structures presque tangents sur les vari\'{e}t\'{e}s diff\'{e}rentiables.
{\sl  C. R. Acad. Sci. Paris S\'er . I Math.} {\bf 255} (1962), 1563-1565.

\bibitem{mt1} M. de Le\'{o}n, I. M\'{e}ndez, M. Salgado: 
$p$-almost tangent structures.
{\sl Rend.   Circ. Mat. 
Palermo} Serie II {\bf XXXVII} (1988), 282-294.

\bibitem{mt2} M. de Le\'{o}n, I. M\'{e}ndez, M. Salgado:
Integrable $p$--almost tangent structures and tangent
bundles of $p^1$-ve\-lo\-ci\-ties.
{\sl Acta Math. Hungar.}, Vol. 58 (1-2) (1991), 45-54.

\bibitem{Saunders} D.J. Saunders:
{\sl The Geometry of Jet Bundles}. Cambridge University Press, Cambridge, 1989.

\bibitem{bc} F. Brickell, R.S. Clark:
Integrable almost tangent structures. {\sl J. Differential. Geom.} {\bf 9} (1974), 557-563.

\bibitem{cgt} R. S. Clark, D.S. Goel:
On the geometry of an almost tangent structure.
{\sl Tensor (N. S.)} {\bf 24} (1972), 243-252.

\bibitem{ct} M. Crampin,  G. Thompson:
Affine bundles and integrable almost tangent structures.
{\sl  Math. Proc. Cambridge Philos. Soc.} {\bf 101} (1987), 61-67.

\bibitem{cram1} M. Crampin: Tangent bundle geometry for Lagrangian dynamics.
{\sl J. Phys. A: Math. Gen.} {\bf 16} (1983), 3755--3772.

\bibitem{cram2} M. Crampin: Defining Euler-Lagrange fields in terms of almost tangent structures.
{\sl  Phys. Lett. A} {\bf 95} (1983), 466-468.

\bibitem{grif1} J. Grifone:
Structure presque-tangente et connexions, I.
{\sl Ann. Inst. Fourier} {\bf 22} (1972), 287-334.

\bibitem{grif2} J. Grifone:
Structure presque-tangente et connexions, II.
{\sl Ann. Inst. Fourier} {\bf 22} (1972), 291-338.

\bibitem{klein} J. Klein:
Espaces variationelles et m\'{e}canique.
{\sl  Ann. Inst. Fourier} {\bf 12} (1962),1-124.

\bibitem{ts} G. Thompson, U. Schwardmann:
Almost tangent and cotangent structures in the large.
{\sl  Trans. Amer. Math. Soc.} {\bf 327} (1991), 313-328.

\bibitem{mor} A. Morimoto: Liftings of some types of tensor fields and
connections to tangent $p^r$-velocities.
{\sl Nagoya Math. J.} {\bf 40} (1970), 13-31.


\bibitem{FLN2} R.O. Fulp, J.K. Lawson, L.K. Norris:
Generalized symplectic geometry as a covering theory for the Hamiltonian
theories of classical particles and fields.
{\sl J. Geom. Phys.} {\bf 20} (1996), 195-206.

\bibitem{Lawson} J.K. Lawson: {\sl Generalized symplectic geometry for classical
fields and spinors}, Ph.D. dissertation, Dept. Math., North Carolina State
Univ., Raleigh, 1994.

\bibitem{KN} S. Kobayashi, K. Nomizu:
{\sl Foundations of differential geometry}, Vol. I.
Interscience, New York, 1963.

\bibitem{t} G. Thompson:
Integrable almost cotangent structures and Legendrian bundles.
{\sl Math. Proc. Cambridge Philos. Soc.} {\bf 101} (1987), 61-67.

 \bibitem{No5} L.K. Norris:
The n-symplectic Algebra og Observables in Covariant Lagrangian field Theory. To appear
in {\sl J. Math. Phys.}

\bibitem{mmm} M. de Le\'{o}n, E. Merino, M. Salgado:
$k$-cosymplectic manifolds and Lagrangian field theories. 
{\sl J. Math. Phys.} {\bf 42} (2001), .

\bibitem{GIMMsy} M. Gotay, J. Isenberg, J. Marsden:
{\sl Momentum Maps and Classical Relativistic Fields, Part I: Covariant
Field Theory}, 1997, MSRI Preprint.

\bibitem{GIMMsy2} M. Gotay, J. Isenberg, J. Marsden:
{\sl Momentum Maps and Classical Relativistic Fields, Part II: Canonical
Analisys of Field Theories}, 1999, MSRI Preprint.

 
\bibitem{KijoSz} J. Kijowski, W. Szczyrba:
Multisymplectic manifolds and the geometrical construction of the Poisson brackets in the classical field
theory. G\'{e}om\'{e}trie symplectique et physique math\'{e}matique (Colloq.
Internat. C.N.R.S., Aix-en-Provence, 1974), pp. 347-349. 

\bibitem{KijTul} J. Kijowski, W. M.  Tulczyjew:
{\sl A symplectic framework for field theories}.
Lecture Notes in Physics, 107. Springer-Verlag, New York, 1979.

 

\bibitem{Cant1} F. Cantrijn, A. Ibort, M. de Le\'{o}n:
On the geometry of multisymplectic
manifolds. 
{\sl J. Austral. Math. Soc. Ser. A} {\bf 66} (1999), 303-330.

\bibitem{Cant2} F. Cantrijn,  A. Ibort, M. de Le\'{o}n:
Hamiltonian structures on multisymplectic manifolds.  
{\sl Rend. Sem. Mat. Univ. Politec. Torino}, {\bf 54} (1996), 225-236.

\bibitem{CCI} J.F. Carinena, M. Crampin, A. Ibort:
On the multisymplectic formalism for first order field theories.
{\sl Diff. Geom. Appl.} {\bf 1} (1991), 345-374.


\bibitem{mmopm} M. de Le\'{o}n, E. Merino, J Oubina, P. Rodrigues, M. Salgado:
Hamiltonian systems on $k$-cosymplectic manifolds,
{\sl J. Math. Phys,}, ${\bf 39}$ , 91997), 876-893. 

\bibitem{bar2} A. Echevarr{\'\i}a-Enr{\'\i}quez, M.C. Mu\~{n}oz-Lecanda, N. Rom\'{a}n-Roy: 
Multivector fields and connections: Setting Lagrangian equations in field theories.
{\sl J. Math. Phys.} {\bf 39} (1998), 4578-4603.

\bibitem{bar3} A. Echevarr{\'\i}a-Enr{\'\i}quez, M.C. Mu\~{n}oz-Lecanda, N. Rom\'{a}n-Roy: 
Multivector field formulation of Hamiltonian
field theories: equations and symmetries.
{\sl J. Phys. A: Math. Gen.} {\bf 32} (1999), 8461-8484.

\bibitem{Rund} H. Rund:
{\sl The Hamilton-Jacobi Theory in the Calculus of Variations},  
Reprinted Edition, Robert E. Krieger Publishing Co. Inc.,
Huntington, New York, 1973.

\bibitem{grad}
F. Cantrijn, A. Ibort, M. de Le\'on:  Gradient vector fields on cosymplectic manifolds,
{\sl J. Phys. A: Math. gen.} ${\bf 25}$ (1992), 175-188,

\bibitem{albert} A. Albert:
Le th\'{e}or\`{e}me de r\'{e}duction de Marsden-Weinstein en
g\'{e}om\'{e}trie cosymplectique et de contact. 
{\sl J. Geom. Phys.} {\bf 6} (1989), 627-649.

\bibitem{chlm} D. Chinea, M. de Le\'{o}n, J.C. Marrero:
The constraint algorithm for time-dependent Lagrangians.
{\sl J. Math.  Phys .}, {\bf 35} (1994), 3410-3447.

\bibitem{MN} M. McLean, L.K. Norris:
Covariant Field Theory
on Frame Bundles of Fibered Manifolds.
{\sl J. Math. Phys.} 41 (2000), 6808-6823.

\bibitem{DW} T. de Donder, 
``Th\'eorie invariantive du calcul des variations, nouvelle \'edit.,''
Gauthier-Villars, Paris (1935); H. Weyl, ``Geodesic fields in the calculus of variations
for multiple integrals,'' { Ann. Math. (2)}, {\bf 36},  607-629 (1935).

\bibitem{Car} C. \car, 
``\"Uber die Variationsrechnung bei mehrfachen Integralen,''
 {\sl  Acta Szeged Sect. Scient. Mathem.}, {\bf 4}, 193-216 (1929).

\bibitem{Kijo} J. Kijowski:
A finite-dimensional canonical formalism in the classical
field theory. 
{\sl Comm. Math. Phys.} {\bf 30} (1973), 99-128.

\bibitem{Snia} J. Sniatycki:
On the geometric structure of classical field
theory in Lagrangian formulation. 
{\sl Math. Proc. Cambridge Philos. Soc.} {\bf 68} (1970), 475-484.

 \bibitem{GP1} P.L. Garc\'{\i}a, A. P\'{e}rez-Rend\'{o}n:
Symplectic approach to the theory of quantized fields, I. 
{\sl Comm. Math. Phys.} {\bf 13} (1969) 24-44.

\bibitem{GP2} P.L.  Garc\'{\i}a, P. L.; A. P\'{e}rez-Rend\'{o}n:
Symplectic approach to the theory of quantized fields, II. 
{\sl Arch. Ratio. Mech. Anal.} {\bf 43} (1971), 101-124.


\bibitem{Mart1}G. Martin:
Dynamical structures for $k$-vector fields. 
{\sl Internat. J. Theoret. Phys.} {\bf 27} (1988), 571-585.

\bibitem{Mart2} G. Martin:
A Darboux theorem for multi-symplectic manifolds.
{\sl Lett. Math. Phys.} 16 (1988), 133-138.

\bibitem{Go1} M.J. Gotay:
An exterior differential systems approach to the Cartan
form. In: {\sl Symplectic geometry and mathematical physics (Aix-en-Provence,
1990)}. Progr. Math., 99, Birkh\"{a}user Boston, Boston, MA, 1991,
pp. 160-188.

\bibitem{Go2} M.J. Gotay:
A multisymplectic framework for classical field theory and the calculus of variations, I.
Covariant Hamiltonian formalism. In: {\sl Mechanics, analysis and geometry: 200
years after Lagrange}. North-Holland Delta Ser., North-Holland,
Amsterdam, 1991, pp. 203-235.
\bibitem{Go3} M.J. Gotay:
A multisymplectic framework for classical field theory and the calculus of variations, II.
Space $+$ time decomposition. 
{\sl Differential Geom. App.} {\bf 1} (1991), 375-390.
\bibitem{Binz} E. Binz, J. Sniatycki, H. Fischer:
{\sl Geometry of classical fields}. 
North-Holland Mathematics Studies, {\bf 154}.
North-Holland Publishing Co., Amsterdam, 1988.

\bibitem{Roux} A. Roux:
{\sl Jets et connexions}. Publ. Dep. de Math\'ematiques, Lyon, 1975.

\bibitem{Lib1} P. Libermann:
Parall\'elismes.
{\sl J. Differential Geom.} {\bf 8} (1973), 511-539.

\bibitem{GMS4} G. Giachetta, L. Mangiarottti, G. Sardanashvily:
{\sl New Lagrangian and Hamiltonian Methods in Field Theory}.
World Scientific, Singapore, 1997.

\bibitem{GMS1} G. Giachetta, L. Mangiarotti, G. Sardanashvily:
Covariant Hamilton equations for field theory. 
{\sl J. Phys. A: Math. Gen.} {\bf 32} (1999), 6629-6642.

\bibitem{GMS2} G. Giachetta, L. Mangiarotti, G. Sardanashvily:
BRST-extended polysymplectic Hamiltonian formalism for field theory. 
{\sl Nuovo Cimento B} (12) 114 (1999), 939-955.

\bibitem{igor1} I.V. Kanatchikov:
Novel algebraic structures from the polysymplectic form in field theory.
In: {\sl Physical Applications and Mathematical Aspects of Geometry, Groups and Algebra},
Vol. 2, H.A. Doebner, W. Scherer, C. Schulte eds., World Scientific,
Singapore 1997, pp. 894-899.
\bibitem{igor11} I. I.V. Kanatchikov:
On field theoretic generalizations of a Poisson algebra.
{\sl Rep. Math. Phys.} {\bf 40} (1997), 4225-234.
\bibitem{igor2} I. I.V. Kanatchikov:
Canonical structure of Classical Field Theory in the polymomentum phase space.
{\sl Rep. Math. Phys.} {\bf 41} (1998), 49-90.

\bibitem{Sarda0} G. Sardanishvily:
{\sl Gauge theory in jet manifolds}.
Hadronic Press, Inc, Palm Harbor, 1993.

\bibitem{Sarda} G. Sardanashvily:
{\sl Generalized Hamiltonian formalism for field theory. Constraint systems}. 
World Scientific, Singapore, 1995.
\bibitem{Sarda2} G. Sardanashvily:
Stress-energy-momentum tensors in constraint field theories. 
{\sl J. Math. Phys.} {\bf 38} (1997), 847-866.
\bibitem{Sarda3} G. Sardanashvily:
SUSY-extended field theory. 
{\sl Internat. J. Modern Phys. A} {\bf 15} (2000), 3095-3112.
\bibitem{SZ} G. Sardanashvily, O. Zakharov, On application of the Hamilton
formalism in fibred manifolds to field theory. 
{\sl Differential Geom. App.} {\bf  3} (1993), 245-263.

\bibitem{FLN1} R.O. Fulp, J. K. Lawson and L. K. Norris:
Geometric prequantization on the spin bundle based on $n$-symplectic
geometry: The Dirac equation.
{\sl  Int. J. Theor. Phys.} {\bf 33} (1994) 1011-1028.













 

\bibitem{aw3} A. Awane, M. Goze:
{\sl Pfaffian systems, $k$-symplectic systems}. 
Kluwer Academic Publishers, Dordrecht, 2000.

 
 
 

 
 

 
 

 

 

 
 

 
\bibitem{bar1} A. Echevarr{\'\i}a-Enr{\'\i}quez, M.C. Mu\~{n}oz-Lecanda, N. Rom\'{a}n-Roy: 
Geometry of Lagrangian first-order classical field theories.
{\sl Fortschr. Phys.} {\bf 44} 3 (1996), 235-280.

 
\bibitem{bar4} A. Echevarr{\'\i}a-Enr{\'\i}quez, M.C. Mu\~{n}oz-Lecanda, N. Rom\'{a}n-Roy: 
Geometry of multisymplectic hamiltonian first-order field theories.
{\sl J. Math. Phys.} {\bf 41} (2000), 7402-7444.

\bibitem{bar5} A. Echevarr{\'\i}a-Enr{\'\i}quez, M.C. Mu\~{n}oz-Lecanda, N. Rom\'{a}n-Roy: 
A geometrical analysis of the field equations in field theory.
{\sl arXiv:math-ph/0105018}.

 

 
 

 
 
 

 
 
 

 

 

\bibitem{hrabak1} S.P.Hrabak:
On a Multisymplectic Formulation of the Classical BRST symmetry for First Order Field Theories Part I: Algebraic Structures.
arXiv.org/math-ph/9901012 

\bibitem{hrabak2} S.P.Hrabak:
On a Multisymplectic Formulation of the Classical BRST Symmetry for First Order Field Theories Part II: Geometric Structures.
arXiv.org/math-ph/9901013

 
 

 

 

 

 

 

 

 

 

 

 

 

\bibitem{LR1} M. de Le\'on, P.R. Rodrigues:
Formalisme hamiltonien symplectique sur les fibr\'es
tangents d'ordre sup\'erieur.
{\sl C. R. Acad. Sci. Paris}, s\'erie II, {\bf 301}(1985), 103-106.

\bibitem{LR11} M. de Le\'on, P.R. Rodrigues:
{\sl Generalized Classical Mechanics and Field Theory}.
North-Holland Mathematics Stu\-dies, 112, North-Holland, Amsterdam, 1985.

\bibitem{LR2} M. de Le\'on, P.R. Rodrigues:
A contribution to the global formulation of the higher
order Poincar\'e-Cartan form.
{\sl Lett.   Math. Phy.} {\bf 14} 4 (1987), 353-362.

\bibitem{LR3} M. de Le\'on, P.R. Rodrigues:
$n^k$-Almost Tangent Structures and the Hamiltonization
of Higher Order Field Theories.
{\sl J. Math. Phys.} {\bf 30} (1989), 1351-1353.

 
 
\bibitem{MS1} L. Mangiarotti, G. Sardanashvily:
{\sl Connections  in classical and quantum field theory}.
World Scientific, Singapore .

\bibitem{mascho} J.E. Marsden, S. Shkoller:
Multisymplectic geometry, covariant hamiltonians and water wawes.
{\sl Math. Proc. Cambridge. Philos.  Soc.} {\bf 125} (1999), 553-575.
 
 

 

 

 

\bibitem{Pu} M. Puta: Some remarks on the $k$-symplectic manifolds.
{\sl Tensor, N.S.} {\bf 47} (1988), 109-115.

 
 
 
\bibitem{saunders1} D.J. Saunders:
An alternative approach to the Cartan form in Lagrangian field theories.
{\sl J. Phys. A: Math. Gen.} {\bf 20} (1987), 339-349.

\bibitem{saunders2} D.J. Saunders:
Jet fields, connections ans second order differential equations.
{\sl J. Phys. A: Math. Gen.} {\bf 20} (1987), 3261-3270.

 

 
\bibitem{Snia2} J. Sniatycki:
The Cauchy data space formulation of classical field theory. 
{\sl Rep. Math. Phys.} {\bf 19} (1984), 407-422.


 

 

\bibitem{Wo} N. Woodhouse, {\sl Geometric Quantization}, 2nd ed. ( Oxford Press, Oford,
1992).

\bibitem{Yano} K. Yano, S. Ishihara:
Horizontal lifts from manifolds to its tangent bundle.
{\sl J. Math. and Mech.} {\bf 16} (1967), 1015-1030.

\bibitem{Yano2} K. Yano. S. Ishihara:
{\sl Tangent and cotangent bundles}.
Marcel Dekker, New York, 1973.

\end{thebibliography}
\end{document}